\documentclass[preprint]{aa}  
\usepackage{graphicx}
\usepackage{verbatim}
\usepackage{subcaption}
\usepackage{txfonts}
\usepackage[]{hyperref}

\begin{document} 
   \title{The search for radio emission from exoplanets using LOFAR beam-formed observations: Jupiter as an exoplanet}
   \author{Jake D. Turner\inst{1,2,3},
           Jean-Mathias Grie{\ss}meier\inst{1,4},
           Philippe Zarka\inst{4,5},
           Iaroslavna Vasylieva\inst{6}
          }
   \institute{Laboratoire de Physique et Chimie de l'Environnement et de l’Espace (LPC2E) Université d’Orléans/CNRS, Orléans, France \\
              \email{jt6an@virginia.edu}
         \and
           Department of Astronomy, University of Virginia, Charlottesville, VA, USA  
          \and
          Department of Astronomy, Cornell University, Ithaca, NY, USA
        \and
            Station de Radioastronomie de Nan\c{c}ay, Observatoire de Paris, PSL Research University, CNRS, Univ. Orl\'{e}ans, OSUC, 18330 Nan\c{c}ay, France
        \and
            LESIA, Observatoire de Paris, CNRS, PSL, Meudon, France
        \and
        Institute of Radio Astronomy, National Academy of Sciences of Ukraine, Kharkov, Ukraine
             }
   \date{}
 
  \abstract
   {The magnetized Solar System planets are strong radio emitters and theoretical studies suggest that the radio emission from nearby exoplanets in close-in orbits could reach intensity levels $10^{3}-10^{7}$ times higher than Jupiter's decametric emission. Detection of exoplanets in the radio domain would open up a brand new field of research, however, currently there are no confirmed detections at radio frequencies.
   }
   {We investigate the radio emission from Jupiter, scaled such that it mimics emission coming from an exoplanet, with low-frequency beam-formed observations using LOFAR. The goals are to define a set of observables that can be used as a guideline in the search for exoplanetary radio emission and to measure effectively the sensitivity limit for LOFAR beam-formed observations.}
   {We observe ``Jupiter as an exoplanet'' by dividing a LOFAR observation of Jupiter by a down-scaling factor and adding this observation to beam-formed data of the ``sky background''. Then we run this artificial dataset through our total intensity (Stokes-I) and circular polarization (Stokes-V) processing and post-processing pipelines and determine up to which down-scaling factor Jupiter is still detected in the dataset. }
   {We find that exoplanetary radio bursts can be detected at 5 pc if the circularly polarized flux is $10^5$ times stronger than the typical level of Jupiter's radio bursts during active emission events ($\sim4\times10^{5}$ Jy). Equivalently, circularly polarized radio bursts can be detected up to a distance of 20 pc (encompassing the known exoplanets 55 Cnc, Tau Bo\"{o}tis, and Upsilon Andromedae) assuming the level of emission is $10^{5}$ times stronger than the peak flux of Jupiter's decametric burst emission ($\sim6\times10^{6}$ Jy). 
   }
  {}

   \keywords{Planets and satellites: magnetic fields -- Radio continuum: planetary systems -- Magnetic fields --  Astronomical instrumentation, methods and techniques --  Planet-star interactions}
   
   \titlerunning{Jupiter as an Exoplanet}
   \authorrunning{J.D. Turner,
           J.-M. Grie{\ss}meier,
           P. Zarka,
           I. Vasylieva}
   \maketitle
%

\section{Introduction} 

The detection and characterization of exoplanetary radio emission would constitute a new and important field of exoplanet science. For example, the detection of planetary auoral radio emission is probably the only method to unambiguously detect exoplanetary magnetic fields (\citealt{G2015}).
To date, no confirmed radio detection has been achieved, even though a certain number of observations have been conducted over the past few decades (e.g. \citealt{Winglee1986}; \citealt{Bastian2000}; \citealt{Ryabov2004}; \citealt{George2007};  \citealt{Lazio2007}; \citealt{Smith2009}; \citealt{Lecavelier2009,Lecavelier2011}; \citealt{Lazio2010a,Lazio2010b}; \citealt{Stroe2012}; \citealt{Hallinan2013}; \citealt{Lecavelier2013}; \citealt{Sirothia2014}; \citealt{Murphy2015}; \citealt{Lynch2017}; \citealt{Turner2017pre8}; \citealt{Lynch2018}; \citealt{OGorman2018}). A summary of all the observational campaigns can be found in \citet[Table 2]{Griessmeier17PREVIII}. In parallel to observational studies, a number of theoretical studies has been published (e.g. \citealt{Zarka1997pre4}; \citealt{Farrell1999,Farrell2004}; \citealt{Zarka2001}; \citealt{Lazio2004}; \citealt{Stevens2005}; \citealt{Griessmeier05AA,Gr2007}; \citealt{Jardine2008}; \citealt{Vidotto2010r,Vidotto2015}; \citealt{Hess2011}; \citealt{Nichols2011,Nichols2012}; \citealt{See2015}; \citealt{Nichols2016}); an overview is given, e.g., in recent review articles such as \citet[][]{Zarka11PREVII,Zarka2015SKA,G2015,Griessmeier17PREVIII}.\\
\indent Starting with \citet{Zarka1997pre4} and \citet{Farrell1999}, a number of articles have attempted to estimate the radio flux density that can be expected for different types of exoplanets. Of course, such estimates have to be taken carefully. For example, \citet{Gr2007} give uncertainties of approximately one order of magnitude for the flux density and an uncertainty of a factor of 2-3 for the maximum emission frequency for the planet Tau Bo\"{o}tis b. The uncertainties are even larger when different models are compared. Still, such numbers can be used to determine whether the detection of exoplanetary auroral radio emission seems realistic with a given radio telescope and observational setup. Indeed, according to most recent estimates, emission frequencies are compatible with the frequencies at which some radio telescopes of latest generation operate, and estimated flux densities are close to the sensitivity of these instruments. In particular, \cite{Griessmeier17PREVIII} find that the flux densities of 15 exoplanets are above the theoretical detection limit of LOFAR as given by \citet{Turner2017pre8}.\\
\indent With such encouraging radio predictions, radio observations of exoplanets are undertaken by most low-frequency radio telescopes. For these observations, different observing modes and strategies can be used. In the following, we will differentiate between (a) imaging observations and (b) beam-formed observations. Many recent observations (e.g. \citealt{Hallinan2013}; \citealt{Sirothia2014}; \citealt{Lynch2017}) have been recorded in the form of interferometric images using an array of distributed antennas or dishes (e.g. GMRT, LOFAR). Interferometric observations have the advantage of a higher robustness against localized (i.e. site-specific) Radio Frequency Interference (RFI), and are equally sensitive to continuous and moderately bursty signals (i.e. longer than the shortest time constant in imaging pipelines, typically a few seconds; e.g. \citealt{Offringa2014}). They are computationally expensive, but offer the possibility to exclude a bad antenna or dish from the analysis even during offline processing.
Beam-formed observations have the advantage of a higher time resolution, which can be used to localize and excise short and sporadic RFI more precisely. They cannot reliably detect continuous or slowly varying emission, but excel at the detection of short bursty signals. Compared to imaging observations, only a handful of pixels have to be analyzed, which reduces the computational cost: Typical observations use 1 ON-beam and 1 to 3 simultaneous OFF-beams, see e.g. \citet{Zarka1997pre4} or \citet{Turner2017pre8}. 

For both imaging and beam-formed observations, the determination of a minimum detectable flux density is not straightforward in the case of a bursty signal. The reason for this is that the upper limit depends on the detection method. In this work, we present a detection tool that integrates the processing steps described in \citealt{Turner2017pre8} (RFI-mitigation, normalization by the time-frequency (t-f) response function, t-f integration) and a series of sensitive observables based on the work of \citet{Vasylieva2015}. In order to test, validate, and quantify the sensitivity reached with this tool, we apply it to a LOFAR observation of Jupiter's magnetospheric radio emission in which the signal from Jupiter is attenuated. The idea is simple: we observe Jupiter, divide its signal by a fixed factor before adding it to an observation of ``sky background'', thereby creating an artificial dataset best described as ``Jupiter as an exoplanet''. We then run our pipeline and check whether the (attenuated) radio signal from Jupiter is detected. The maximum factor by which we can divide Jupiter's signal and still achieve a detection gives the sensitivity of our setup (i.e. the combination of the telescope and the processing chain). This method is mainly designed for use with beam-formed data, but an extension to radio imaging observations is under preparation and will be described elsewhere (Loh et al. in prep).

Finally, the instantaneous flux density of Jupiter was obtained from a well-calibrated observation using the Nancay Decameter Array (NDA; \citealt{Boischot1980,Lamy2017}) simultaneous to our LOFAR observation of Jupiter. The NDA observation is used to convert the sensitivity of our setup into physical units.

\section{Observations}

\begin{figure*}[ptbh!]
\center
\vspace{-0.5em}

    \begin{subfigure}[c]{0.73\textwidth}
        \centering
        \caption{}
        \includegraphics[width=\textwidth]{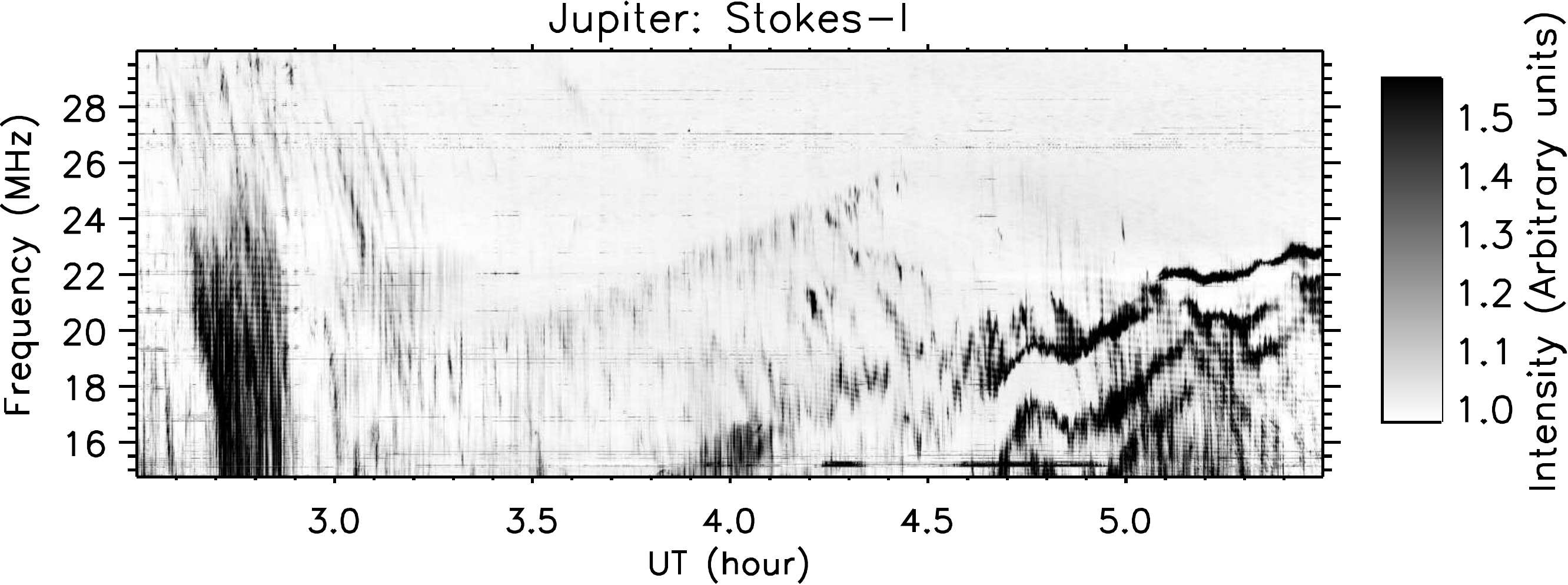}
        \label{}
    \end{subfigure}%
    
    \vspace{-0.5em}
    \begin{subfigure}[c]{0.73\textwidth}
        \centering
        \caption{}
        \includegraphics[width=\textwidth]{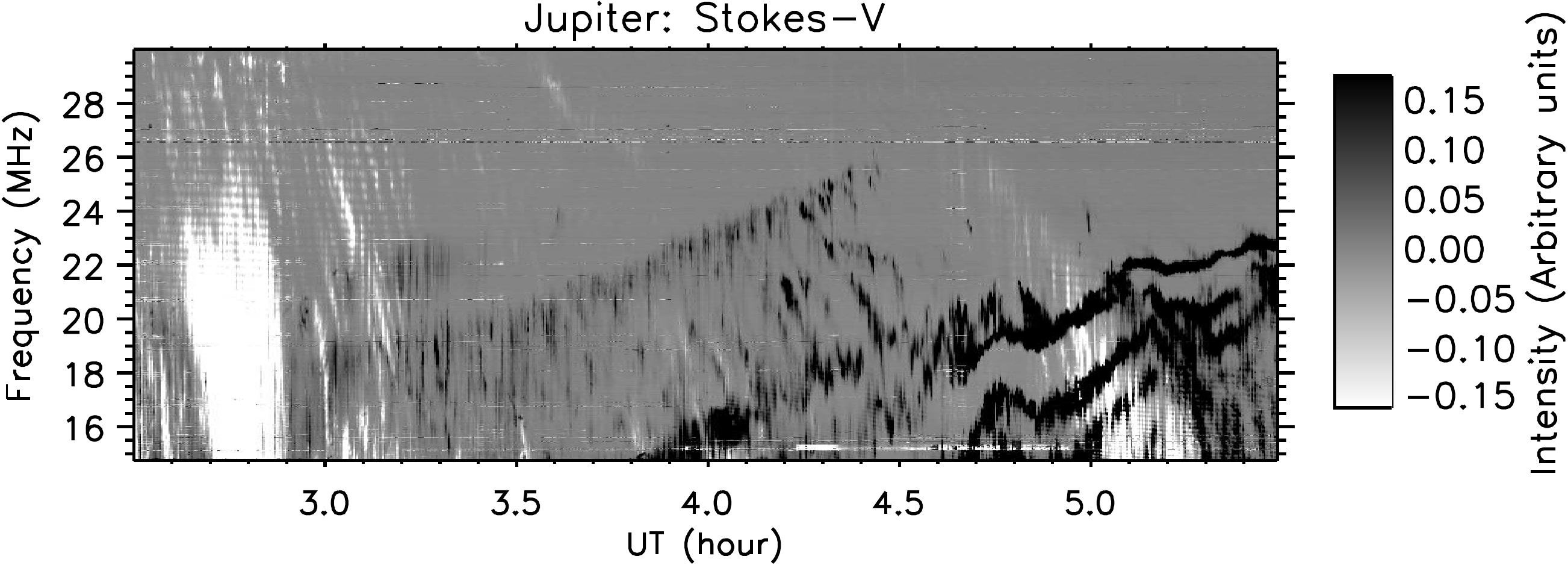}
        \label{}
    \end{subfigure}%
    
    \vspace{-0.5em}
       
        \begin{subfigure}[c]{0.73\textwidth}
        \centering
        \caption{}
        \includegraphics[width=\textwidth]{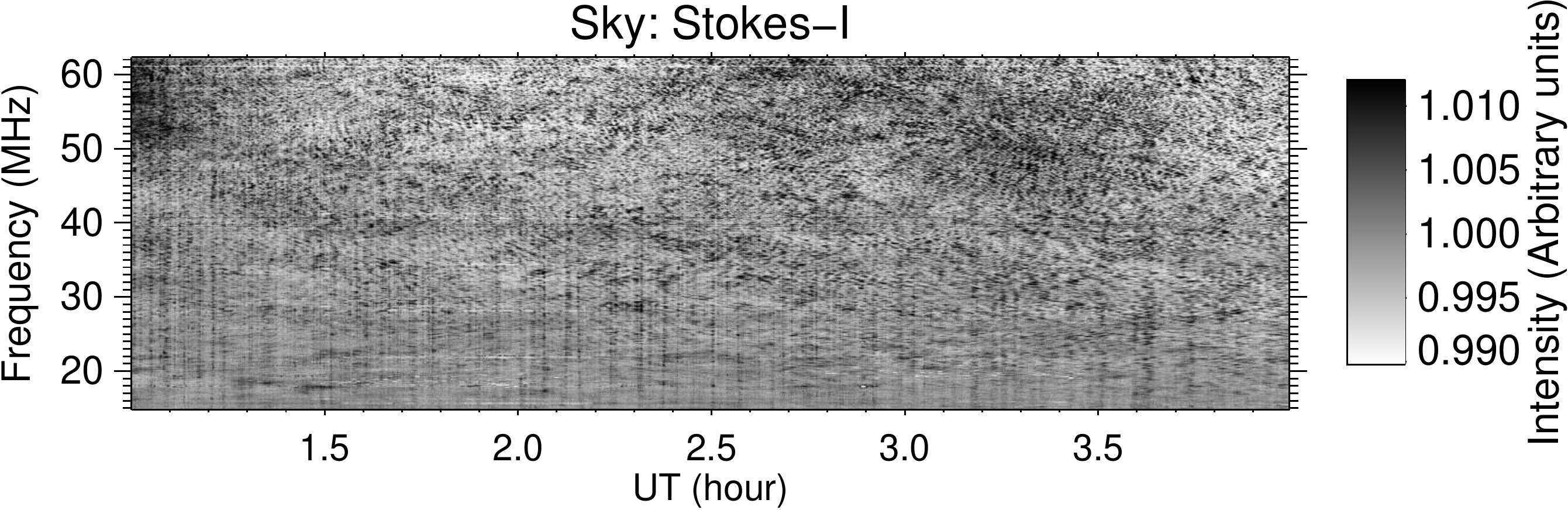}
        \label{}
    \end{subfigure}%
    
    \vspace{-0.5em}
    
          \begin{subfigure}[c]{0.73\textwidth}
        \centering
        \caption{}
        \includegraphics[width=\textwidth]{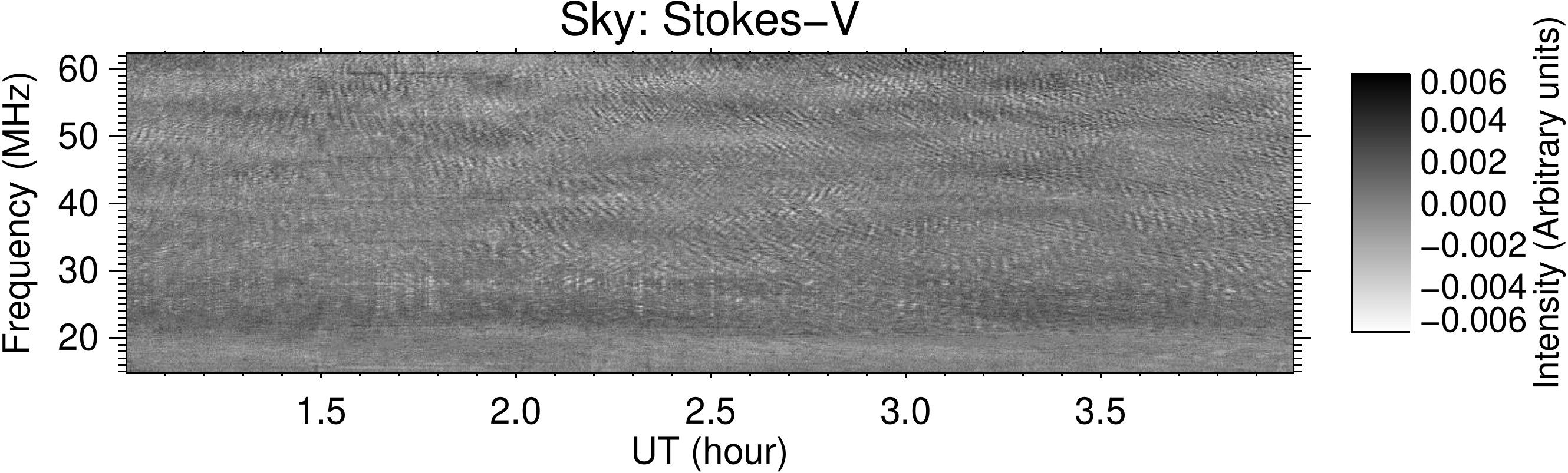}
        \label{}
    \end{subfigure}%
    
    \vspace{-0.5em}

\caption{
Dynamic spectrum with LOFAR LBA of Jupiter in Obs~$\#$1 in Stokes-I (\textbf{a}) and Stokes-V (\textbf{b}) and of the OFF-beam 1 in Obs~$\#$2 in Stokes-I (\textbf{c}) and Stokes-V (\textbf{d}).  In Obs~$\#$1, we only show the data from 15 to 30 MHz because there was no emission of Jupiter occurring above 30 MHz. The Stokes-I observations (panels \textbf{a} and \textbf{c}) are divided by an average value of the background at each frequency, whereas the Stokes-V observations (panels \textbf{b} and \textbf{d}) are subtracted by an average background. As seen in panel \textbf{b}, the emission around 2.5-3 UT has negative (right-handed) circular polarization and most of the other emission has positive (left-handed) circular polarization. 
}
\label{fig:DynSpec}
\end{figure*}

For this study, we use four different sets of Low-Frequency Array (LOFAR; \citealt{vanHaarlem2013}) Low Band Antenna (LBA) beam-formed observations in the frequency range 15--62 MHz. The detailed setup and the summary of all observations (date, time, and beam directions) can be found in Table \ref{tb:setup} and \ref{tb:obs}, respectively. In this paper, we focus on the total intensity (Stokes-I) and circular polarization (Stokes-V) components of the emission. All observations were intentionally scheduled during night time hours to mitigate strong contamination by RFI. The first observation (hereafter Obs~$\#$1) was taken on February 11, 2017 from 02:30 to 5:30 UT and the ON-beam was pointed at Jupiter. The dynamic spectrum of this beam can be found in Figs. \ref{fig:DynSpec}a and \ref{fig:DynSpec}b. The structure of the Jupiter emission is very complex and the analysis of this structure (e.g. \citealt{Burke1955}; \citealt{Carr1983}; \citealt{Zarka1998}; \citealt{Kaiser1993}; \citealt{Lecacheux2004}; \citealt{Imai2015}; \citealt{Marques2017}) is beyond the scope of this study. As expected, Jupiter's emission is only seen below 40 MHz in the observation (\citealt{Marques17}). 


\begin{table*}[!pht]
\centering
\caption{Summary of LOFAR LBA beam-formed observations}
\begin{tabular}{ccccc}
\hline 
\hline
 Parameter              &  Obs $\#$1            & Obs $\#$2             & Obs $\#$3             & Obs $\#$4\\ 
\hline
\hline
LOFAR OBS ID            &L568467                   & L570725                  & L569123     & L547645         \\
Date (UT)               &February 11, 2017  &February 18, 2017  & February 26, 2017 &September 28, 2016\\
Time (UT)               &02:30-05:30         & 01:12-04:12        & 01:16--04:16      &23:00--04:00       \\
Target                  &Jupiter            &Tau Bo\"{o}tis          &Tau Bo\"{o}tis          &Upsilon Andromedae\\
ON-beam RA (2000)       &13:27:49.42        &13:47:15.74         & 13:47:15.74               & 01:36:47.84               \\
ON-beam DEC (2000)      &-07:39:01.70       &+17:27:24.90         &+17:27:24.90            & +41:24:19.60                 \\
OFF-beam 1 RA (2000)    &13:25:51.27        &13:54:44.95       &13:54:44.95      &01:40:00             \\
OFF-beam 1 DEC (2000)   &-09:35:11.94       &+16:49:29.20       &+16:49:29.20       &+38:00:00              \\
OFF-beam 2 RA (2000)    & 13:35:55.97        &13:58:10.366      &13:58:10.366       &01:30:00                   \\ 
OFF-beam 2 DEC (2000)   & -09:05:16.10      &+19:00:01.37       &+19:00:01.37       &+48:00:00                  \\
\hline
\end{tabular}
\label{tb:obs}
\end{table*}


Due to its anisotropic beaming, Jupiter's emission is visible from Earth only $\sim$10\% of the time. It does not, however, occur randomly, but depends on the geometrical position of the Earth, Jupiter, and Jupiter's satellite Io, as expressed by Io's orbital phase and the CML (Central Meridian Longitude = the observer's Jovicentric longitude). Statistical studies have identified times when the probability of detecting Jupiter's decametric emission from Earth is > 50\%, (\citealt{Marques17}), and for a specific geometry (so-called Io-B emission), the occurrence rate reaches 94\% (i.e.  nearly permanent emission) (\citealt{Zarka17PREVIII}). To determine a good time window for Obs~$\#$1, we made use of the Io-phase/CML diagrams provided by Nan\c{c}ay Radio observatory\footnote{https://realtime.obs-nancay.fr/dam/data\_dam\_affiche/\linebreak[1]data\_dam\_affiche.php?init=1\&lang=en\&planete=jupiter}.

\begin{table}[!ht]
\centering
\caption{Setup of the LOFAR LBA beam-formed observations}
\begin{tabular}{ccc}
\hline 
\hline
Parameter   & Value  & Units   \\ 
\hline
\hline 
Array Setup                     & Core \\
 $\#$ of Stations \tablefootmark{a}                & 24       \\
 Lower Frequency &           14.7   &MHz    \\
 Upper Frequency &           62.4   &MHz       \\
 Channel Bandwidth  ($b$)      & 3.05  & kHz    \\
 $\#$ of Subbands                       & 244           \\
 Channels per Subband           &64             \\
 Time Resolution ($\tau_{r}$)         & 10.5 &msec  \\
 Angular Resolution\tablefootmark{b} & 9.2  &  arcmin\\
 Raw Sensitivity\tablefootmark{c} ($\Delta S$)          & 208  & Jy   \\ 
 Polarizations                  & IQUV   \\
\hline
\end{tabular}
\tablefoot{\\
\tablefoottext{a}{The data streams from all stations are combined during the observations. Subsequent data processing deals with this combined data stream.} \\
\tablefoottext{b}{The angular resolution was calculated at 45 MHz (\citealt{vanHaarlem2013}).}\\
\tablefoottext{c}{The theoretical (thermal noise) sensitivity ($\Delta S$) was calculated using the sensitivity equation $\Delta S= S_{sys}/(N\sqrt{n_{pol}\tau_{r} b })$, where $S_{sys}$ is the system equivalent flux density (SEFD) and equal to 40 kJy (obtained from LOFAR calibration data; \citealt{vanHaarlem2013}), $N$ is the number of stations used, $n_{pol}$ is the number of polarizations (2), $b$ is the channel bandwidth, and $\tau_{r}$ is the time resolution. $\Delta S$ was calculated with the values given in this table.}
}
\label{tb:setup}
\end{table}

Two OFF-beams were obtained simultaneously with the ON-beam, however the OFF-beams show strong contamination by emission from Jupiter despite being located $\sim2$ degrees away from Jupiter. Therefore, a second observation to obtain "clean" OFF-beams was taken on February 18, 2017 from 01:12 to 4:12 UT (hereafter Obs~$\#$2). Obs~$\#$2 will be used as the ``sky background'' to which we will add the attenuated Jupiter signal. Two OFF-beams were obtained at beam positions chosen such that no significant low-frequency point sources were located within the beam. For this we used the TGSS survey \citep{Intema2017} at 150 MHz. The dynamic spectrum of one of the OFF-beams can be found in Figs. \ref{fig:DynSpec}c and \ref{fig:DynSpec}d.

While most of the analysis was done using Obs~$\#$2 for the ``sky background'', we also used two other dates of observations with two OFF-beams to verify our results. The third dataset was taken on February 26, 2017 from 01:16 to 04:16 UT (hereafter Obs~$\#$3) and was pointed at the same OFF-beam positions as Obs~$\#$2. This date had far worse RFI conditions than Obs~$\#$2 and also had noticeable large-scale scintillation due to a disturbed ionosphere. The fourth dataset was taken on September 28, 2016 from 23:00 to 04:00 UT (hereafter Obs~$\#$4; Table \ref{tb:obs}). Obs~$\#$4 was comparable in quality to Obs~$\#$2 (no large scale scintillation patterns) and RFI conditions but was pointed at a different part of the sky. 


\section{``Jupiter as an exoplanet''} 

\subsection{Scaling Jupiter's signal} \label{sec:scaleJup}

We add the Jupiter signal, multiplied by a factor $\alpha$ ($<<1$), to the sky plus instrumental background of a typical exoplanet observation, and then try to detect it with our two-step processing pipeline (Sect. \ref{sec:pipeline}). As we will test below the post-processing in 10 MHz bands, we use the Jupiter signal of Figs. \ref{fig:DynSpec}a and \ref{fig:DynSpec}b detected in the band 15--25 MHz. In order to test our pipeline across the entire LOFAR-LBA range, we need to be able to add the attenuated Jupiter signal to any 10 MHz band in the range 10-90 MHz. Having no absolute calibration available in the LOFAR-LBA range, we proceed in two steps: (i) the Jupiter signal detected by LOFAR in Obs~$\#$1 ($I_{J1}$) is expressed in terms of the sky background in the band of observation 15--25 MHz ($I_{S1}$), i.e. the ratio ($I_{J1}/I_{S1}$) is computed as in the following section (Sect. \ref{sec:JupSig}), and it is then transferred to an arbitrary 10 MHz band in the sky background in Obs~$\#$2 ($\text{I}_{S2}$);~(ii) the flux density of the Jupiter emission is computed from simultaneous calibrated observations performed at the NDA. These two steps are detailed below.


For step (i), we add the dynamic spectrum of the Jupiter observation in the range 15--25 MHz to the dynamic spectrum of the sky background in an arbitrary 10 MHz band of an exoplanet observation (with the same observational setup; Table \ref{tb:setup}) to get a test Stokes-I dynamic spectrum $I_{sim}$ following
      \begin{align}
           I_{sim}  =&  I_{S2} + \alpha I_{J2},  \\ 
                    =&  I_{S2} \left( 1 + \alpha \frac{ I_{J1} }{ I_{S1} } \frac{ S_{S1} }{S_{S2} } \right), \label{eq:transfer_I}
      \end{align}
where $I_{J2}$ is the Jupiter signal as it would have been observed in the test frequency band, $I_{J1}/I_{S1}$ and $I_{S2}$ are derived from the Stokes-I observational data, $\alpha$ ($<<$1) is the variable down-scaling parameter, and the ratio $S_{S1}/S_{S2}$ can be computed as the ratio of the SEFD in the band 15--25 MHz and in the test frequency band. Similarly, the test dynamic spectrum for Stokes-V $V_{sim}$ is 
\begin{align}
    V_{sim} =& V_{S2} + \alpha V_{J1} \left(\frac{I_{S2}}{I_{S1}}\right) \left(\frac{S_{S1}}{S_{S2}}\right), \label{eq:transfer_V}
\end{align}
where $V_{S2}$ is the Stokes-V sky background in Obs~$\#$2 and $V_{J1}$ is the Stokes-V Jupiter signal from 15-25 MHz. The full derivation of equations \eqref{eq:transfer_I} and \eqref{eq:transfer_V} can be found in Appendix \ref{app:Jupiter}. From the LOFAR calibration data (\citealt{vanHaarlem2013}), we approximate that the SEFD on an LBA station is 40 kJy in the range 30-70 MHz and that it increases approximately as $\lambda^2$ below 30 MHz (mainly due to the steep increase of the sky background). Thus, when transferring the Jupiter signal from the range 15--25 MHz ($\lambda=12-20$ m) to a test frequency band above 30 MHz, equations \eqref{eq:transfer_I} and \eqref{eq:transfer_V} can be simply rewritten
\begin{align}  
      I_{sim}  =  I_{S2} \left( 1 + \alpha \frac{ I_{J1} }{ I_{S1} } \left[\frac{ max(\lambda_{J1},10 m) }{max(\lambda_{S2}, 10 m) }\right]^2 \frac{N_{2} }{ N_{1} } \right), \label{eq:apply_LOFAR_I} \\
      V_{sim} = V_{S2} + \alpha V_{J1}  \frac{ I_{S2} }{ I_{S1} }  \left[\frac{ max(\lambda_{J1},10 m) }{max(\lambda_{S2}, 10 m) }\right]^2  \frac{N_{2}}{N_{1}}  \label{eq:apply_LOFAR_V}  
\end{align} 
with $N_{2}$ and $N_{1}$ the number of LBA stations involved in each observation.


Note that equations \eqref{eq:transfer_I} and \eqref{eq:transfer_V} can be used to add the signal (of Jupiter or other) observed with one telescope in a given frequency range to the background recorded with another telescope in another frequency range, as long as the SEFD of the two telescopes in their respective spectral ranges are known. Equations \eqref{eq:apply_LOFAR_I} and \eqref{eq:apply_LOFAR_V} are the application for the considered LOFAR-LBA observations. The Jupiter signal thus transferred retains its absolute intensity (e.g. in Jy).

\begin{figure}[htb!]
\center

    \begin{subfigure}[c]{0.47\textwidth}
        \centering
        \caption{}
        \includegraphics[width=\textwidth]{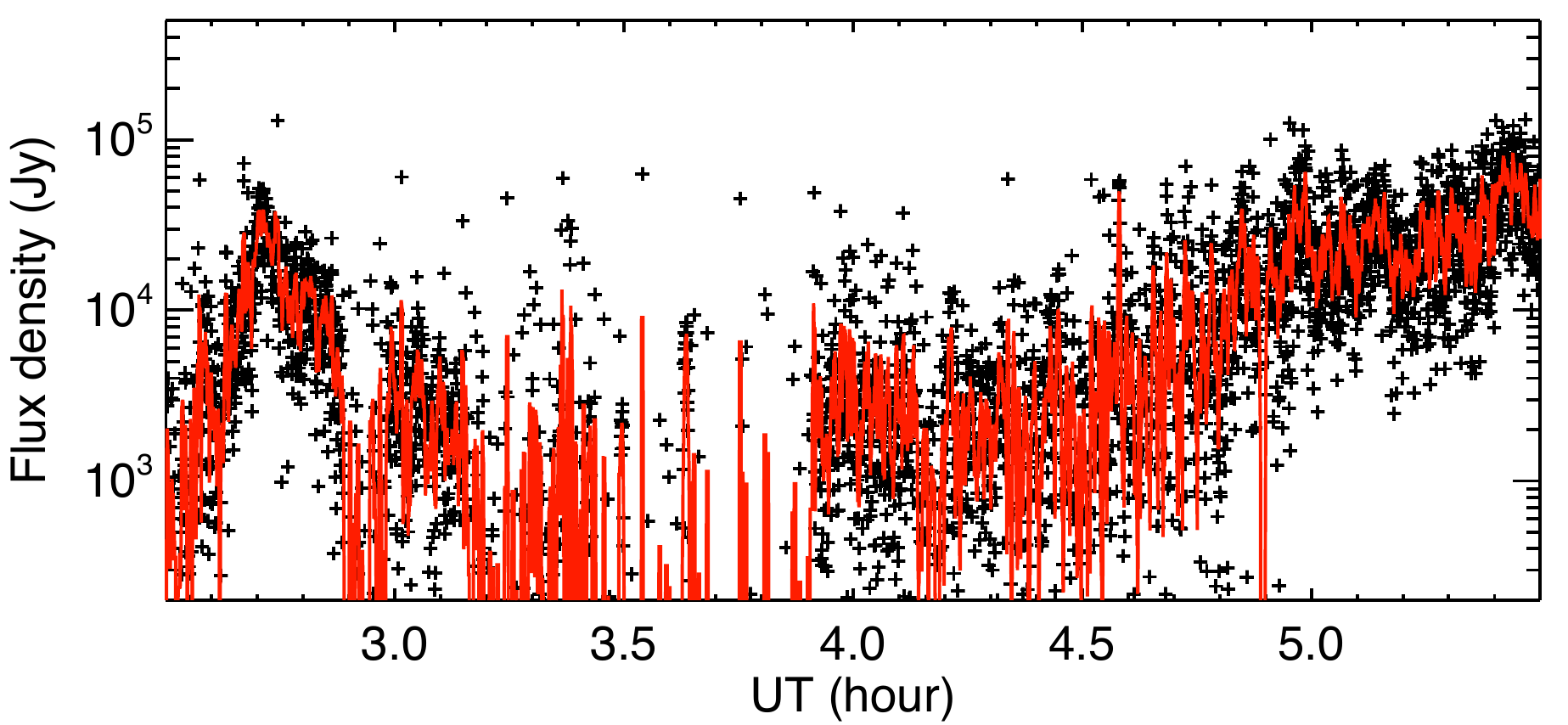}
        \label{}
    \end{subfigure}%
  \vspace{-0.5em}
    \begin{subfigure}[c]{0.47\textwidth}
        \centering
        \caption{}
        \includegraphics[width=\textwidth]{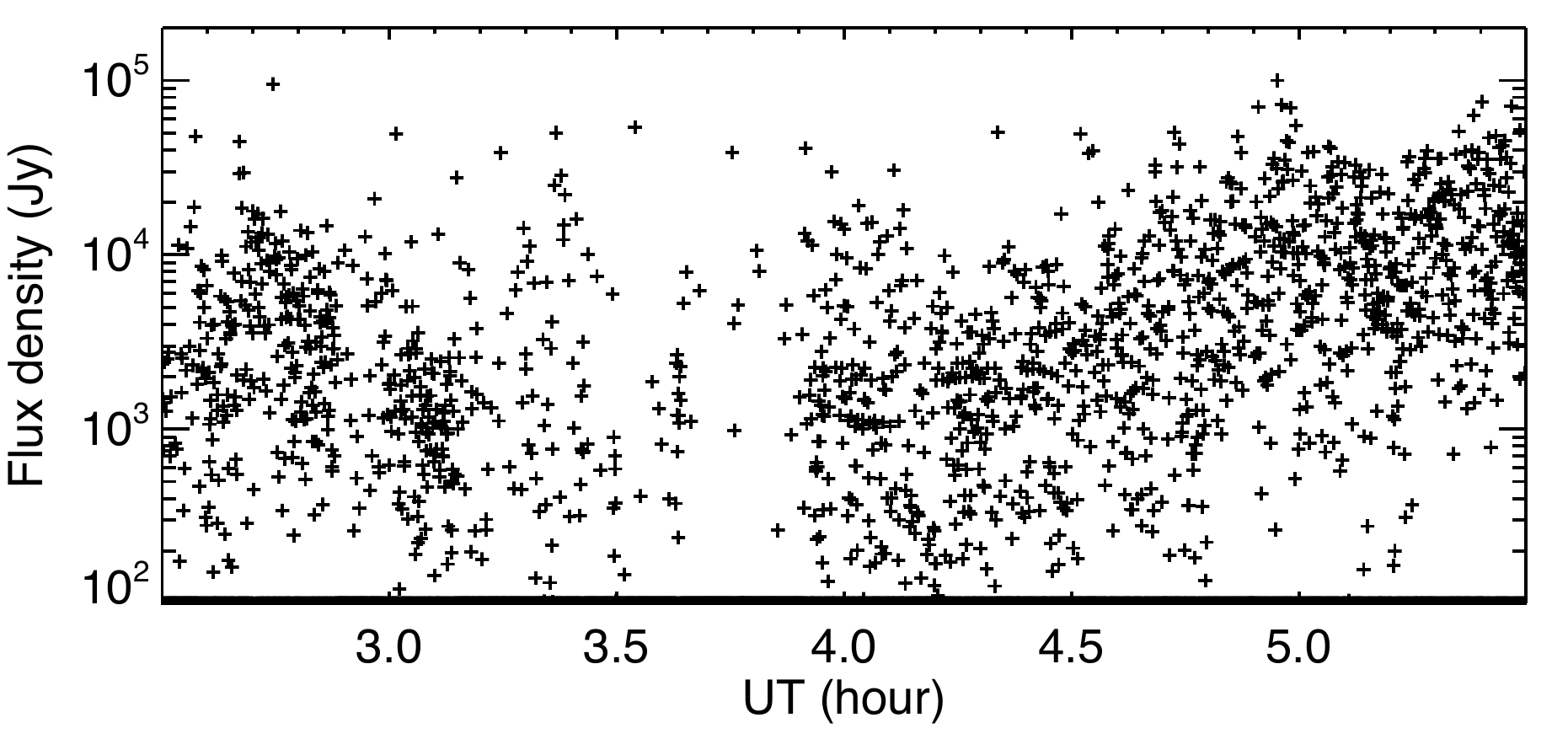}
        \label{}
    \end{subfigure}%
      \vspace{-0.5em}
        \begin{subfigure}[c]{0.47\textwidth}
        \centering
        \caption{}
        \includegraphics[width=\textwidth]{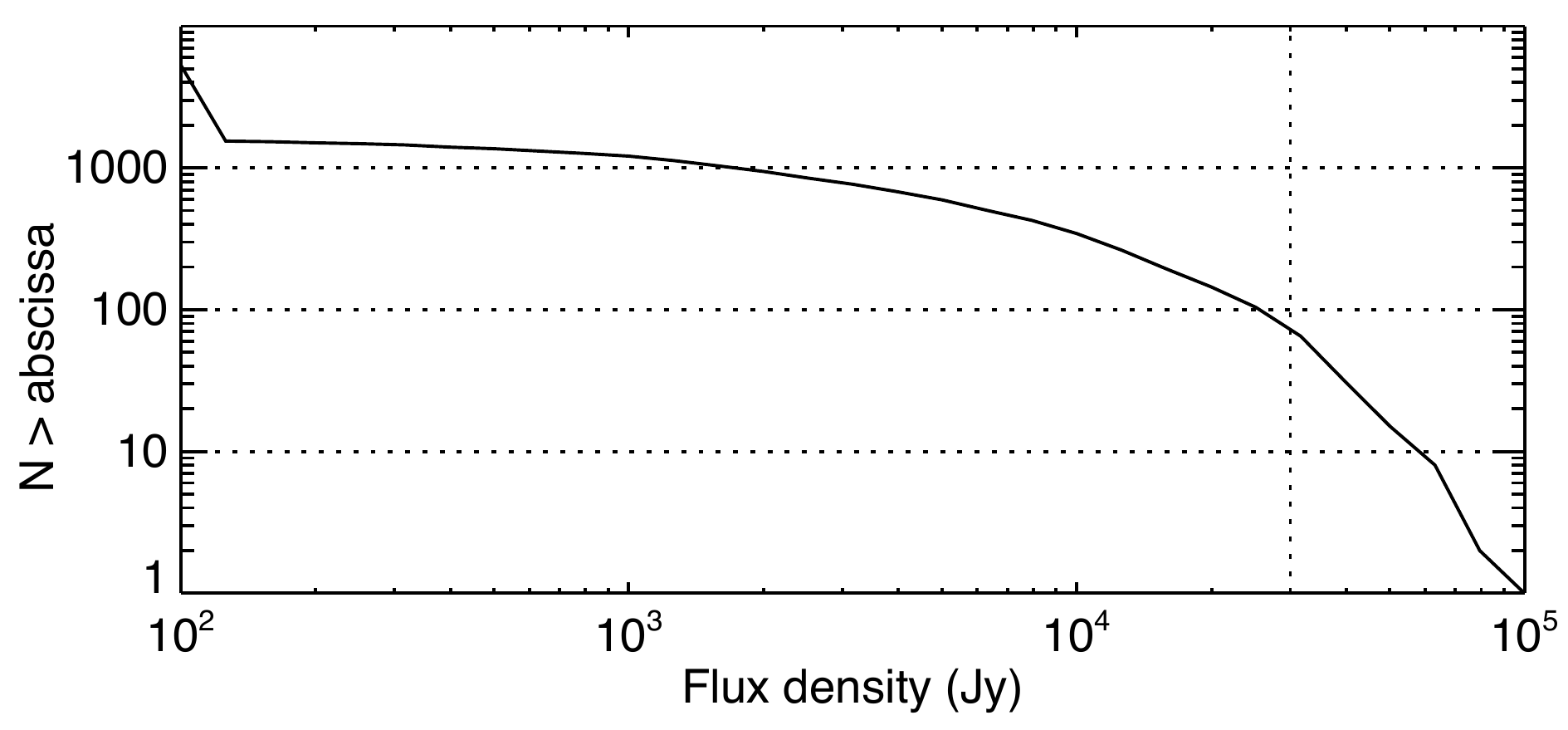}
        \label{}
    \end{subfigure}%
   \vspace{-0.5em}
 
\caption{\textbf{(a)} Calibrated flux density of the Jupiter emission detected on 2017/02/11 between 02:30 and 05:30 UT with the NDA, averaged over the range of 15--25 MHz after background subtraction. Black '+' symbols are the measurements at 1 sec time resolution, whereas the red line is a running average over 10 sec. \textbf{(b)} High-pass filtered flux densities obtained by subtracting the 10 sec average from 1 sec measurements. Only values $\ge$100 Jy are displayed. \textbf{(c)} Cumulative distribution function of the values of panel \textbf{(b)}. 
} 
\label{FigJ20170211}
\end{figure}

For step (ii) we use an observation of Jupiter simultaneously taken to the LOFAR one, carried out at the NDA. For this observation, the NDA observes simultaneously in right-hand (RH) and left-hand (LH) circular polarizations from 10 to 40 MHz in 400 spectral channels at a time resolution of 1 second. Hourly calibration sequences on noise sources of known flux density are embedded in the data and allow us to calibrate the observations in absolute flux density (Jy), with an accuracy $\sim20$\%. From NDA data, we know that the first Jupiter burst at about 02:45 UT is RH elliptically polarized, whereas the drifting emission bands starting around 04:00 UT are LH elliptically polarized. For Stokes-I, we summed the RH and LH signals to obtain the total intensity. We removed the main fixed-frequency RFI and the main broadband spikes (recognized as non-Jupiter signal by integration over the 26-40 MHz range). After subtraction of a background (computed in each frequency channel) the cleaned calibrated dynamic spectrum was averaged over the 15--25 MHz range to obtain the time series displayed in Fig. \ref{FigJ20170211}a (black '+' symbols) together with a running average over 10 seconds (red line). Fig. \ref{FigJ20170211}b displays the high-pass filtered flux densities obtained by subtracting the 10 second average from 1 second measurements. The bursty spikes in this high-pass filtered time-series will be used for comparison to the results of our processing below (Sect. \ref{sec:discussion}). The cumulative distribution function of the values of Fig. \ref{FigJ20170211}b is displayed in Fig. \ref{FigJ20170211}c. We obtain similar results within a factor $\le$2 performing the same analysis on Stokes-V. From that figure, we see for example that $\sim$100 high-pass filtered flux density measurements exceed $3\times10^4$ Jy. By comparing this curve to the actual number of data points of emission detected in the LOFAR data we can determine the sensitivity of our observations and processing (Sect. \ref{sec:discussion}).

\subsection{Extraction of Jupiter's signal} \label{sec:JupSig}  


To obtain both the Jupiter signal ($I_{J1}$) and the sky background in the Jupiter observation $I_{S1}$ we first need a RFI mask. Since Jupiter is as bright as the RFI, we used a modified version of the RFI mitigation pipeline presented in \citet{Turner2017pre8}. The following steps are performed: (1) find RFI on the ON-beam above 30 MHz (where no Jupiter emission is present) using the algorithm PATROL (\citealt{Zakharenko2013}, \citealt{Vasylieva2015}) to flag entire time steps, (2) find RFI in the OFF-beam (beam 2) using only PATROL to flag entire time steps and frequency channels, and (3) combine the RFI masks from step (1) and (2) together to obtain a final RFI mask. This mask is then applied to Obs~$\#$1 and this dataset is used as the Jupiter signal ($I_{J1}$). 


Next, we find $I_{S1}$ for Obs~$\#$1 using the least Jupiter-contaminated OFF-beam (beam 2) and during a time interval (3740 - 3830 seconds after the start of the observation) where Jupiter's emission was minimal. To find the SEFD we apply the RFI mask from step (3) to the raw data. Then at each frequency we compute the 10$\%$ quantile of the distribution of intensities (using this quantile allows for minimal influence from any Jupiter emission or remaining RFI). The level of the 10$\%$ quantile is lower than the mean, therefore, $I_{S1}$ has to be corrected. Quantitatively, the 10$\%$ quantile ($\mu_{10}$) for a Gaussian distribution with moments~($\mu$,~$\sigma_{g}$~) is
\begin{align}
\mu_{10}             \sim& \mu - 1.3 \sigma_{g}    ,       \label{eq:mu10}\\
\frac{\mu_{10}}{\mu} \sim& 1 - \frac{1.3}{\sqrt{n_{pol} \ b \ \tau_{r} }}, \label{eq:mu10final}
\end{align} 
where $n_{pol}$ is the number of polarizations (2), $b$ is the frequency resolution ($b = 3.05$ kHz), and $\tau_{r}$ is the time resolution ($\tau_{r} = 10.5$ msec). The factor of 1.3 in equation \eqref{eq:mu10} and \eqref{eq:mu10final} was determined using a standard Gaussian distribution. Therefore, the term $I_{S1}$ used in the analysis is obtained from the measured value ($\mu_{10}$) using
\begin{align}
    I_{S1} = \mu  =   \mu_{10} \left( 1 - \frac{1.3}{\sqrt{n_{pol} \ b \ \tau_{r} }}\right)^{-1}. 
\end{align}

\section{Signal processing and observables} \label{sec:pipeline}

At low radio frequencies, any observed field not containing a A-team source (especially Cas-A, Cyg-A, Vir-A and Tau-A) is dominated by the Galactic background, that is an intense diffuse radio emission. We use this emission as a calibrator. Before doing so, however, the data have to be cleaned of RFI.

\subsection{Processing pipeline for Stokes-I}\label{sec:process}
The data of Observation~$\#$2 with ($I_{sim}$) and without the added Jupiter signal were run through the Stokes-I data reduction pipeline described in \citet{Turner2017pre8}. This pipeline performs RFI mitigation, finds the time-frequency (t-f) response function of the telescope and normalizes the data by this function, and rebins the data in broader t-f bins. For RFI mitigation we use four different techniques (\citealt{Offringa2010}; \citealt{Offringa2012PhD}; \citealt{Offringa2012AA}; \citealt{Zakharenko2013}; \citealt{Vasylieva2015}, and references therein) that are combined together for optimal efficiency and processing time. The result of the RFI mititation is an array (mask, $m_{I}$) of the same resolution as the data with a value of either 0 (polluted pixels) or 1 (clean pixels). Subsequently, the data is rebinned to a time and frequency resolution of 1 second and 45 kHz. This rebinned data is the input into the post-processing pipeline (Sect. \ref{sec:Qs}). The original method used in \citet{Turner2017pre8} to find the time-frequency response function of the telescope (hereafter, method 1) is biased if some astrophysical emission or left-over RFI is present in the raw dynamic spectrum since the mean of the data is used to create the function. The nominal raw sensitivity of the LOFAR observations is 208 Jy (Table \ref{tb:setup}) where the expected flux from most exoplanets is less than 100 mJy (\citealt{Gr2007}; \citealt{Griessmeier17PREVIII}). Therefore, for exoplanets we do not expect that the emission will be bright enough to be seen in the raw dynamic spectrum. However, when we test large Jupiter scaling factors (e.g. $\alpha = 10^{-2}$) this is no longer the case.   

Therefore, we introduce a new method (hereafter, method 2) to find the time-frequency response function that is less biased towards bright emission in the raw dynamic spectrum. In the pipeline (1) we divide the raw data into sections of 4000 spectra (42 seconds), (2) we apply the RFI mask to the raw data, (3) we create an integrated spectrum from the 10$\%$ quantile of the distribution of intensities at each frequency, (4) we correct the average of the 10$\%$ quantile such that it is equivalent to the mean using equation \eqref{eq:mu10final}, then (5) we find a second order polynomial fit at each frequency over all time sections, and (6) we create and save the 2-d time-frequency response surface made from the polynomial fits. As expected, method 1 and method 2 obtain the same results when $\alpha$ is very small (e.g. below $\alpha=10^{-5}$). When $\alpha$ is large, method 2 is more robust. In addition, method 2 is computationally faster than method 1; therefore method 2 is the preferred method for finding the time-frequency function and will be used in the analysis of this paper.


\subsection{Processing pipeline for Stokes-V}\label{sec:Vprocess}
For Stokes-V, the processing pipeline has to be partially adapted. For example, the raw Stokes-V data includes both negative and positive values, so we cannot use the same exact approach as done in the Stokes-I pipeline. Just like in the Stokes-I pipeline, we want to perform RFI mitigation, normalize the data by a time-frequency (t-f) response function of the telescope, and rebin the data into larger t-f bins.  Additionally, the output of the Stokes-I pipeline is used in the Stokes-V processing pipeline. Therefore, Stokes-V is always ran after processing Stokes-I.

\subsubsection{Pipeline for normal operation}

In this section, we will describe the normal operation (i.e. no Jupiter signal added) of the Stokes-V pipeline. The raw Stokes-V data does not have an average of 0 as a function of frequency as we have for calibrated Stokes-I data. We perform 3 steps to center the data around 0. 

(1) We divide the raw Stokes-V ($V$) by the raw Stokes-I ($I$) data to get rid of any large-scale instrumental systematics
\begin{align}
    v = \frac{V}{I}. \label{Eq:VdivI}
\end{align} 

(2) We subtract $v$ at each frequency by its time average ($<v(f)>_{t}$) to center the dynamic spectrum around 0 
\begin{align}
    v(f)^{'} = v(f) - <v(f)>_{t}.  \label{eq:v_minus_time}
\end{align}
The time average $<v(f)>_{t}$ contains instrumental systematics that are not common between the Stokes-I and Stokes-V polarizations. Additionally, $<v(f)>_{t}$ is taken over 42 seconds in the default pipeline. 

(3) We multiply $v^{'}$ by the RFI-mitigated and normalized Stokes-I data ($I_{cor}$)  
\begin{align}
    V^{'} = v^{'}I_{cor}. \label{eq:V_cor}
\end{align}
$I_{cor}$ is calculated in the Stokes-I pipeline as
\begin{align}
    I_{cor} = \left(\frac{I}{<I(f)>} \right)m_{I}, \label{eq:I_cor}
\end{align}
where $<I(f)>$ is the frequency response function of the telescope (i.e. obtained with method 1 and method 2 described in \citealt{Turner2017pre8} or in Section \ref{sec:process}, respectively) and $m_{I}$ is the Stokes-I RFI mask. Step (3) ensures that the units of $V^{'}$ are the same as in $I_{cor}$. As described below all or a subset of these steps are used in each part of the Stokes-V pipeline.

To find the RFI mask for Stokes-V ($m_{V}$), we use the following procedure. First, we divide the raw Stokes-V data into slices of 42 seconds (4000 spectra). Then we perform steps (1), (2), and (3) to find $V^{'}$. The time-average in step (2) is done separately on each slice of data. In step (3), we use the frequency response function of the Stokes-I data derived using the 10$\%$ quantile of the distribution of intensities at each frequency (\citealt{Turner2017pre8}). The resulting dynamic spectrum $V^{'}$ is then ran through the RFI mitigation code described in \citet{Turner2017pre8} which produces $m_{V}^{'}$. The final mask ($m_{V}$) used in the analysis is the $V^{'}$ mask multiplied by the $I$ mask ($m_{V} = m_{V}^{'}m_{I}$). 

In the construction of the pipeline, we discovered that the systematic difference between the frequency responses of the Stokes-I and Stokes-V signals changed throughout the observation. Therefore, a Stokes-V t-f response function is also needed for the analysis. To find the Stokes-V t-f response function of the telescope we first divide the raw dynamic spectrum $V$ into sections of 4000 spectra. We then perform step (1) and apply the RFI mask $m_{V}$ to $v$. Next, we perform step (2) but only save the average of each frequency in $v$ over each time section. Then we fit the averages with a second order polynomial at each frequency over all time sections. Finally, we create and save the 2-d t-f response surface made from the polynomial fits. 

To obtain the final RFI-mitigated and normalized $V^{'}$ dynamic spectrum we again perform steps (1), (2), and (3) with some variations. Instead of using the time-average in step (2) we subtract by the Stokes-V t-f response surface. In Step (3), to find $I_{cor}$ we normalize the raw data by the Stokes-I t-f response surface using method 2 (Section \ref{sec:process}). Just like the Stokes-I normalized data, $V'$ is in units of the SEFD. After step (3), we multiple the normalized data by the Stokes-V mask $m_{V}$. After normalization and RFI-mitigation, $V'$ is then rebinned to the same resolution as the Stokes-I data (1 second and 45 kHz) and this is dynamic spectrum that will be input into the post-processing code.

\subsubsection{Pipeline for Jupiter analysis}

For the Jupiter analysis, the data of Obs $\#2$ with ($V_{sim}'$) and without the added Jupiter signal were processed through the new Stokes-V pipeline. The $V_{sim}'$ used in the pipeline follows Equation \eqref{eq:transfer_V} but $V$ is substituted with $V'$  
\begin{align}
      V^{'}_{sim} =& V^{'}_{S2} + \alpha V^{'}_{J1} \left(\frac{I_{S2}}{I_{S1}}\right) \left(\frac{S_{S1}}{S_{S2}}\right). \label{eq:transfer_V2}
\end{align}
The normalized Stokes-V Jupiter data $V^{'}_{J1}$ is found using a slight variation of the 3 steps in the pipeline
\begin{align}
    V^{'}_{J1} =& v^{'}_{J1}I_{corJ} \\
               =& \left(v_{J1} - <v_{S1}>_{t}\right) \left(I_{J1} - I_{S1}\right).
\end{align}
We have to subtract $v_{J1}$ by the average of the OFF-beam $<v_{S1}>_{t}$ because the average of the Jupiter beam is skewed due to the immense Jupiter emission. We determined $<v_{S1}>_{t}$ with the same beam (Beam 2) and time interval (3740-3830 seconds after the start of the observation) as in the calculation of $I_{S1}$.


\subsection{Post-processing pipeline: Observables of the exoplanet signal}\label{sec:Qs} 

In the following section, we present the post-processing pipeline. After processing the data we compute several observable quantities that we named Q1 to Q4 for the ON- and OFF-beam and examine their behavior over time or frequency. The input dynamic spectrum for the observables is the RFI-mitigated, normalized, and rebinned data (Sects. \ref{sec:process} and \ref{sec:Vprocess}; Fig. \ref{fig:dynspec_Q1}a). For the Stokes-V data, the analysis is performed on 3 different variants of the data: $|V^{'}|$, $V^{'}+$, and $V^{'}-$, where
\begin{align}
    V'+ =& 
    \begin{cases}
    V^{'}, & \text{if } V^{'} > 0\\
    0,              & \text{otherwise}
    \end{cases},\\
    V^{'}- =& \begin{cases}
    (-V^{'}), & \text{if } V^{'} < 0\\
    0,              & \text{otherwise}
    \end{cases}.
\end{align}
This will allow us to determine whether right-hand or left-hand polarization can be seen in the analysis. 

The observable quantities fall into two general categories: extended emission (Q1) or burst emission (Q2 - Q4). Below is the list of observables we defined (inspired by the methods used by \citealt{Zarka1997pre4} and further developed by \citealt{Vasylieva2015}):
\begin{itemize}
    \item Q1: Extended emission observables
     \begin{itemize}
         \item Q1a (Time-series): Total power of the dynamic spectrum integrated over all frequencies and rebinned in time to a specified time interval ($TI$; 2 minutes for the default pipeline) (Fig. \ref{fig:dynspec_Q1}b)
         \item Q1b (Integrated spectrum): Total power of the dynamic spectrum integrated over all time and rebinned in frequency to a specified frequency interval ($FI$; 0.5 MHz for the default pipeline) (Fig. \ref{fig:dynspec_Q1}c)
     \end{itemize}  
    \item Q2 (Normalized high-pass filtered time-series): The normalized high-pass filtered time-series ($y$) 
      \begin{align}
       y          &= \frac{(x - x_{s}) - <(x - x_{s})>}{\sigma_{(x - x_{s})}},  \label{eq:y}
      \end{align}
where $x$ is the time-series of the dynamic spectrum integrated over all frequencies but not rebinned in time and $x_{s}$ is the low-pass filtered data (low-frequency component) created by running a sliding window of $w$ seconds over $x$ (in the default pipeline $w = 10$ time bins). We subtract by the mean $<(x - x_{s})>$ to center $y$ around 0. Finally, the time-series is normalized by its standard deviation in order to unify the thresholds. An example of $y$ can be found in Fig. \ref{fig:AllQ}a.    

\begin{figure}[!th]
\begin{center}
  \vspace{-0.5em}
   \begin{subfigure}[I]{0.46\textwidth}
        \centering
        \caption{}
        \includegraphics[width=\textwidth]{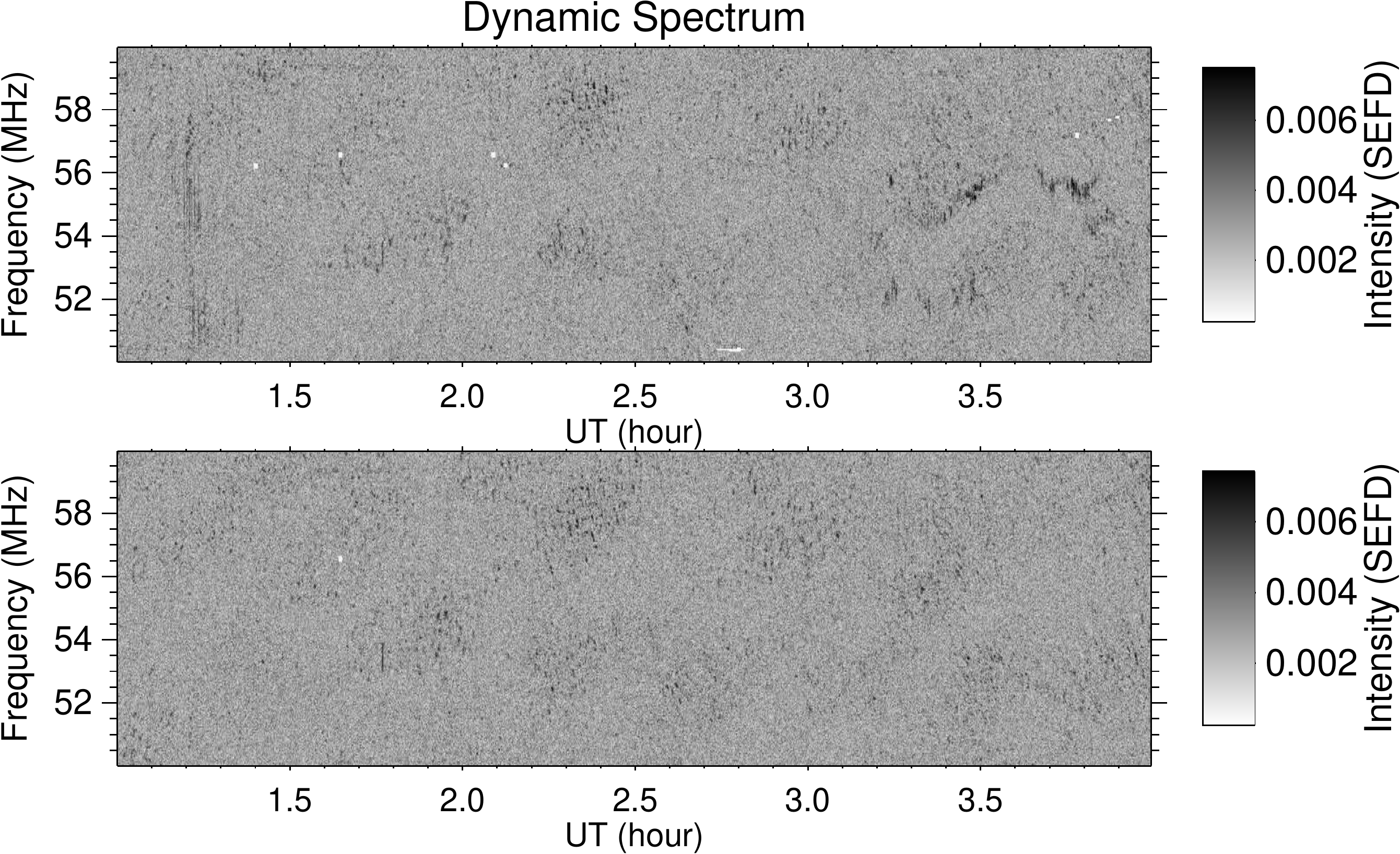}
        \label{}
    \end{subfigure}%
      \vspace{-0.5em}
      \begin{subfigure}[I]{0.46\textwidth}
        \centering
        \caption{}
        \includegraphics[width=\textwidth]{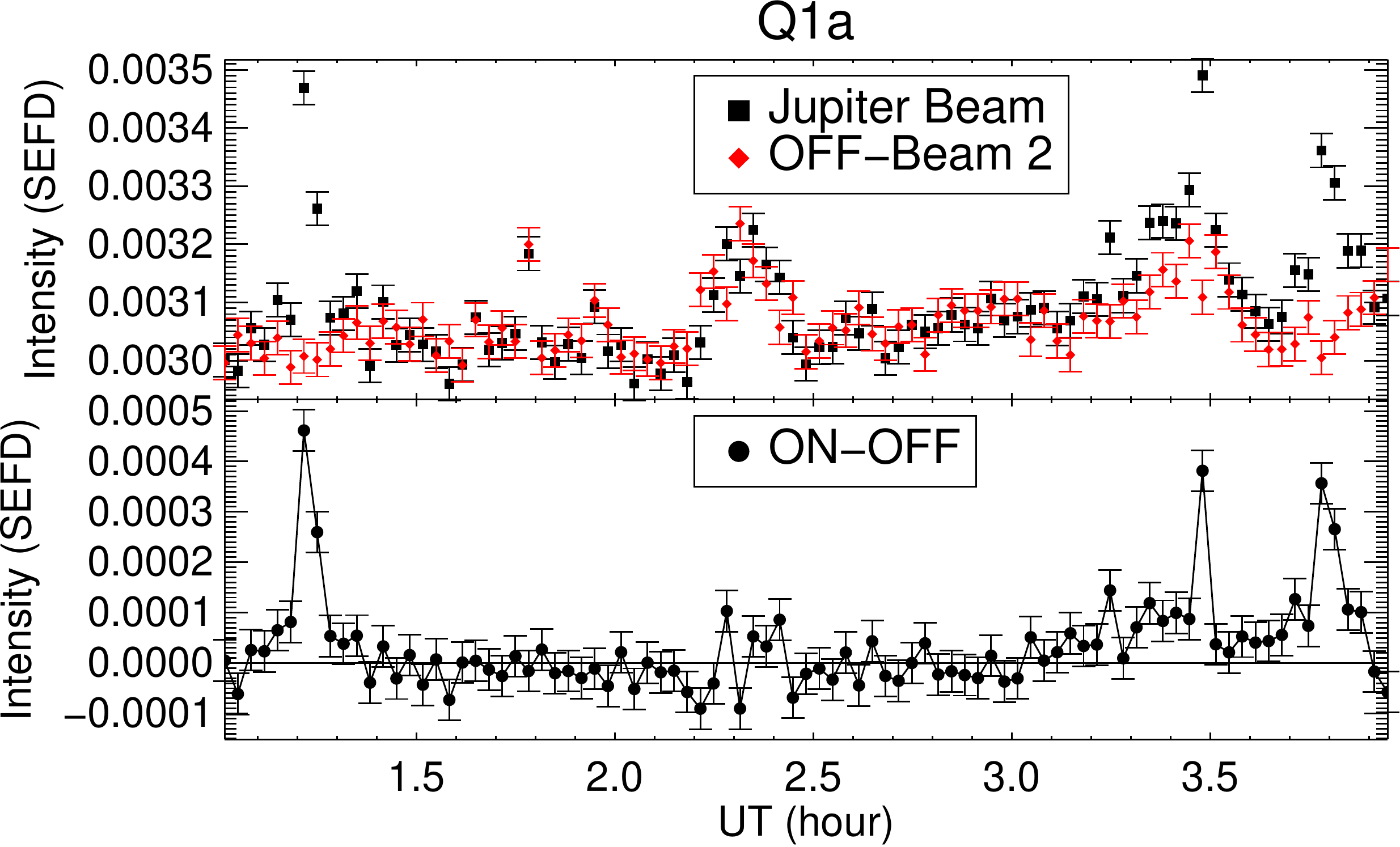}
        \label{}
    \end{subfigure}%
  \vspace{-0.5em}
   \begin{subfigure}[I]{0.46\textwidth}
        \centering
        \caption{}
        \includegraphics[width=\textwidth]{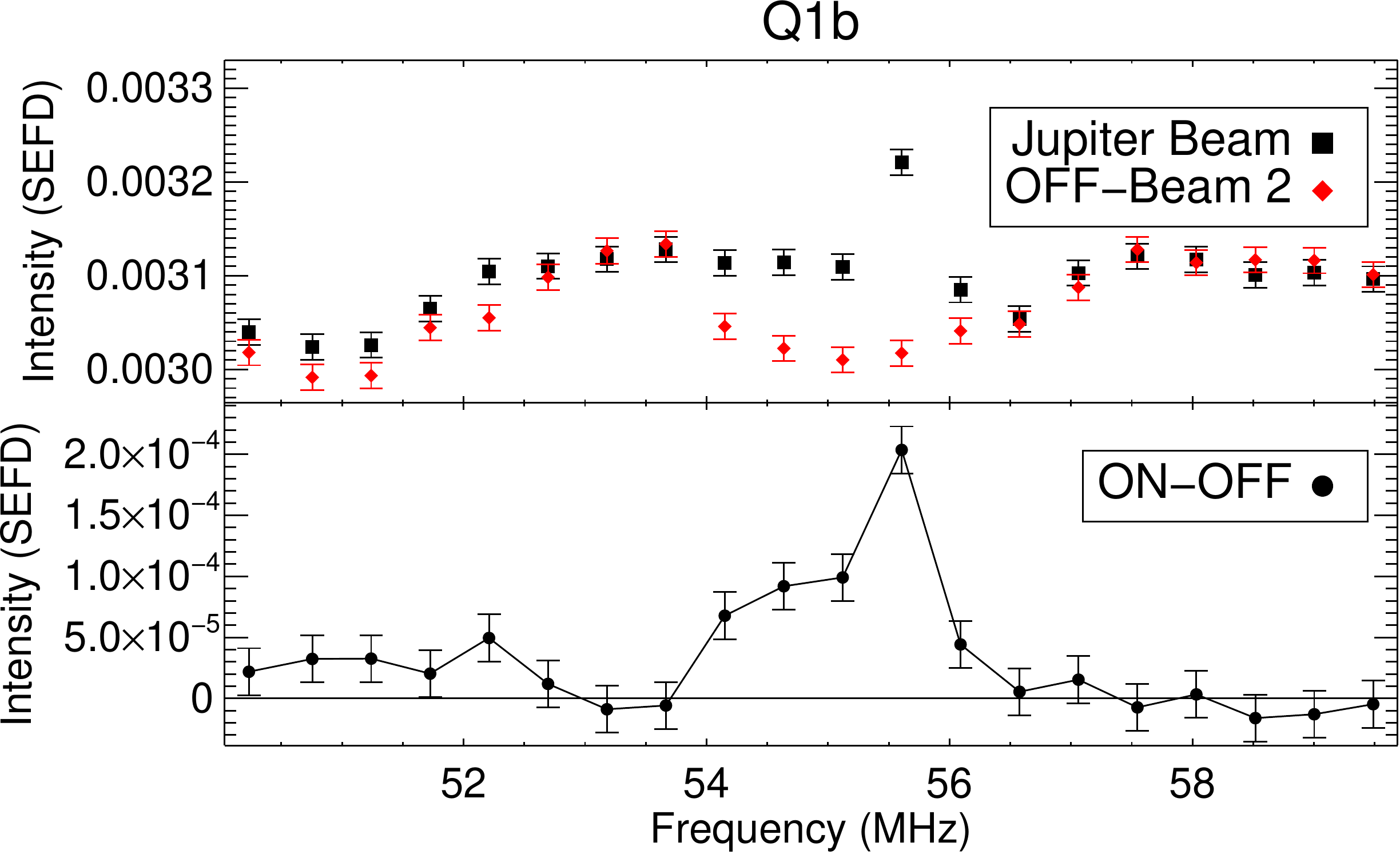}
        \label{}
    \end{subfigure}%
  \vspace{-0.5em}


  \end{center}
\caption{Dynamic spectra and extended emission observable Q1 in Stokes-V (|$V^{'}$|) for a scaling parameter of $\alpha = 10^{-4}$. \textbf{(a)} Dynamic spectra for the ON-beam (top) and the OFF-beam (bottom). \textbf{(b)} Q1a (time-series integrated over all frequencies). \textbf{(c)} Q1b (integrated spectrum summed over all times). See Sect. \ref{sec:Qs} for a detailed description of each observable. For all plots the black squares are the ON-beam, red diamonds are the OFF-beam, and black circles are the difference between beams. The error bars in panels \textbf{(b)} and \textbf{(c)} were calculated assuming pure Gaussian noise ($\sigma = 1/\sqrt{b\tau}$). }  
\label{fig:dynspec_Q1}
\end{figure}

\hspace{4ex} We further examine Q2 by creating a scatter plot of the ON-beam values versus the corresponding OFF-beam values (Fig. \ref{fig:AllQ}b). In this plot, peaks that appear only in the ON-beam would be visible on the right edge of the cloud of points. An example for Q2 of simulated data is given in Fig. \ref{fig:Scatter_demo}a. Due to residual low-level RFI or ionospheric fluctuations, high values of Q2 frequently occur simultaneously in the ON- and OFF-beam (points close to the main diagonal in Fig. \ref{fig:Scatter_demo}a). For this reason, we implemented an elliptical correction, as described in Appendix \ref{app:elliptical-correction}. After the elliptical correction is applied, the Q2 distribution of the sky noise data points is much closer to circular, which makes the signal data points more easily detectable. This is demonstrated in Fig. \ref{fig:Scatter_demo}b. The analysis of real data (Sect. \ref{sec:anaysis}) shows that this elliptical correction does indeed facilitate the detection of astrophysical signals in the target beam and gives a better sensitivity (i.e. allows the detection of fainter signals). The sensitivity is increased by half an order of magnitude using the elliptical correction. This is demonstrated in Figure \ref{fig:Q4Detectionplot}, where the comparison of panels a (without) and e (with elliptical correction), b (without) and f (with elliptical correction), c (without) and g (with elliptical correction), or panels d (without) and h (with elliptical correction) shows a marked difference.

\begin{figure*}[!th]
\begin{center}
\begin{tabular}{ccc}
   \begin{subfigure}[c]{0.32\textwidth}
     \vspace{-0.5em}
        \centering
        \caption{}
        \includegraphics[width=\textwidth]{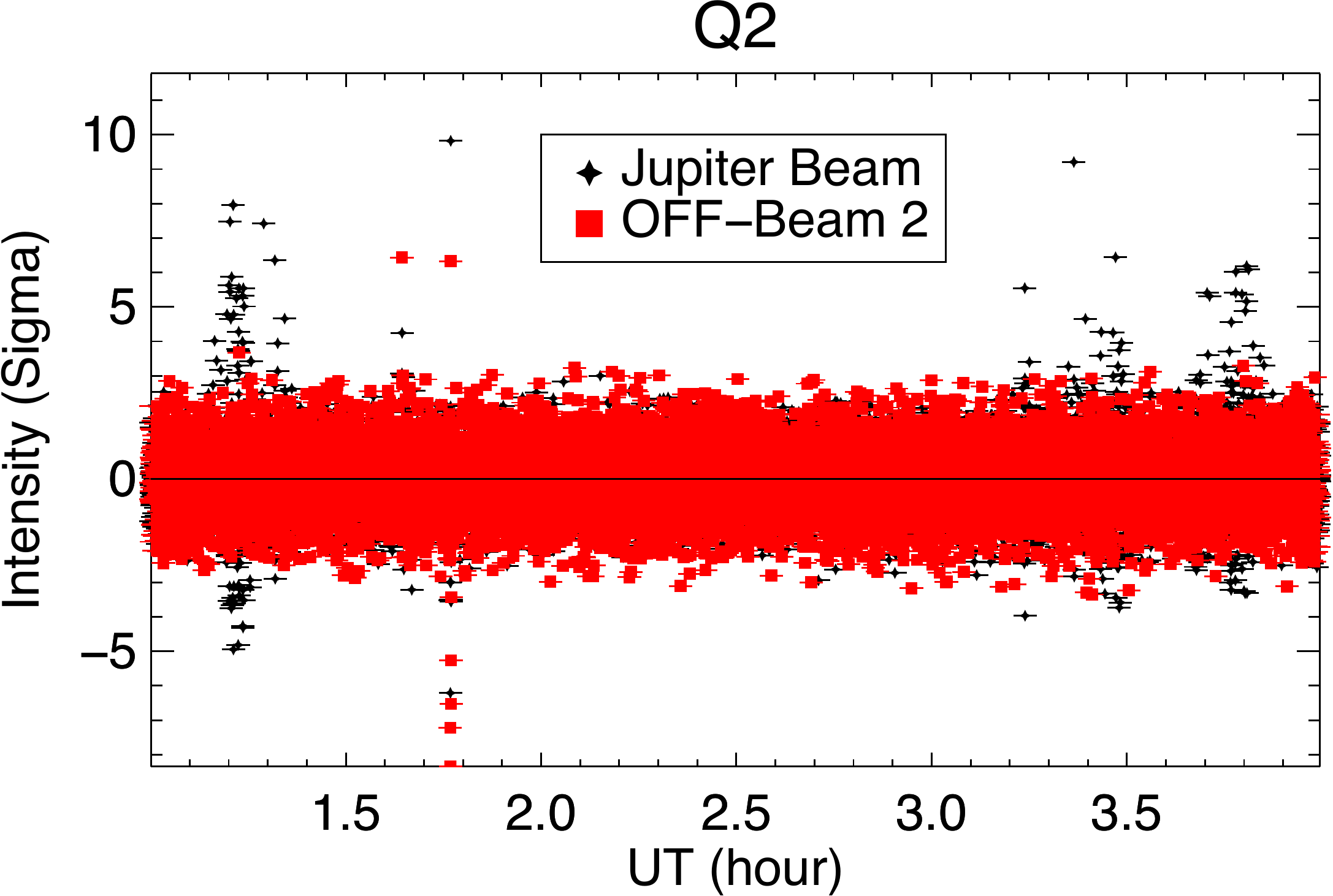}
        \label{}
    \end{subfigure}%
    
    &  \begin{subfigure}[c]{0.23\textwidth}
        \centering
        \caption{}
        \includegraphics[width=\textwidth]{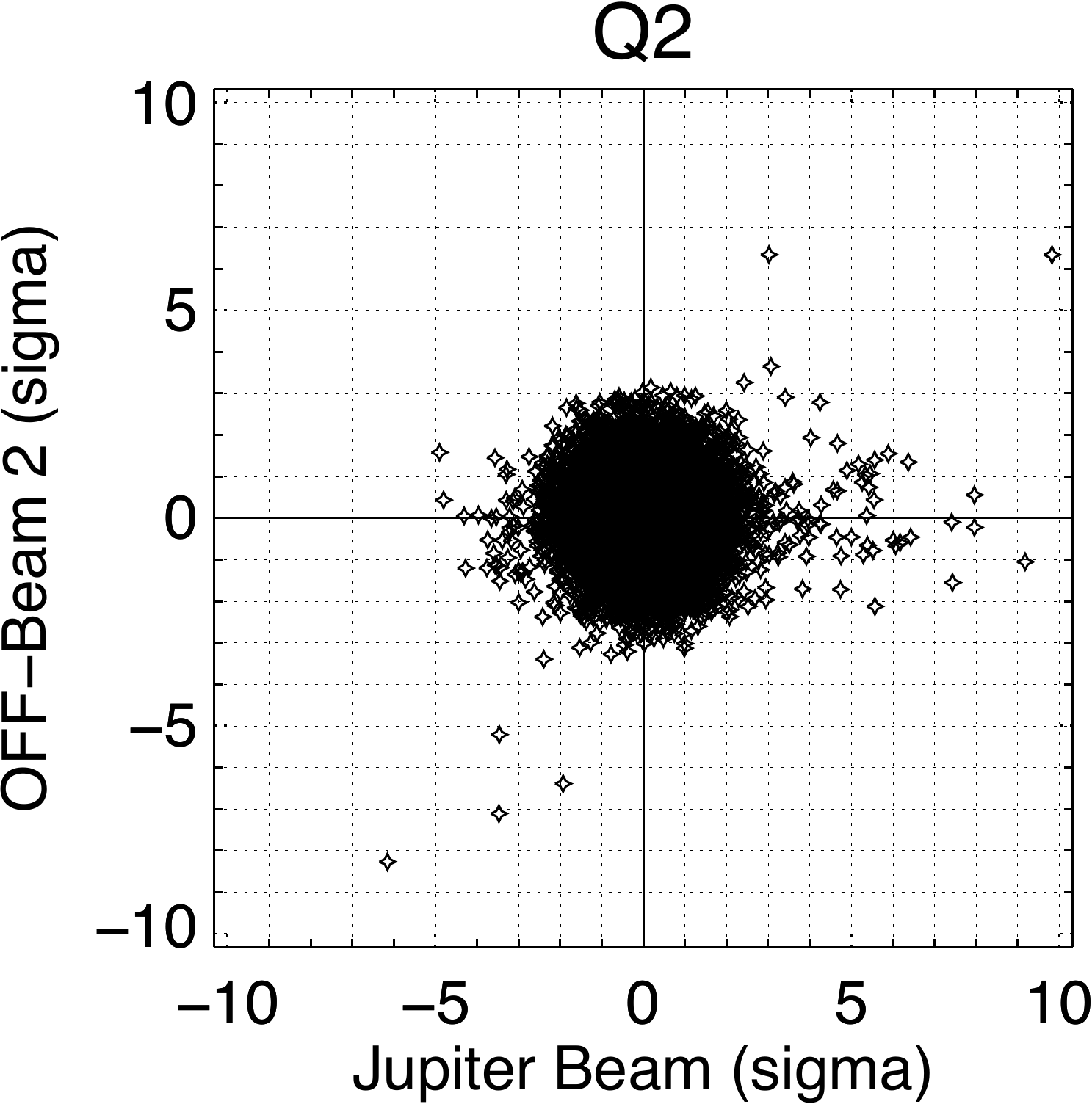} 
        \label{}
    \end{subfigure}%
    &
      \begin{subfigure}[c]{0.32\textwidth}
        \centering
        \caption{}
        \includegraphics[width=\textwidth]{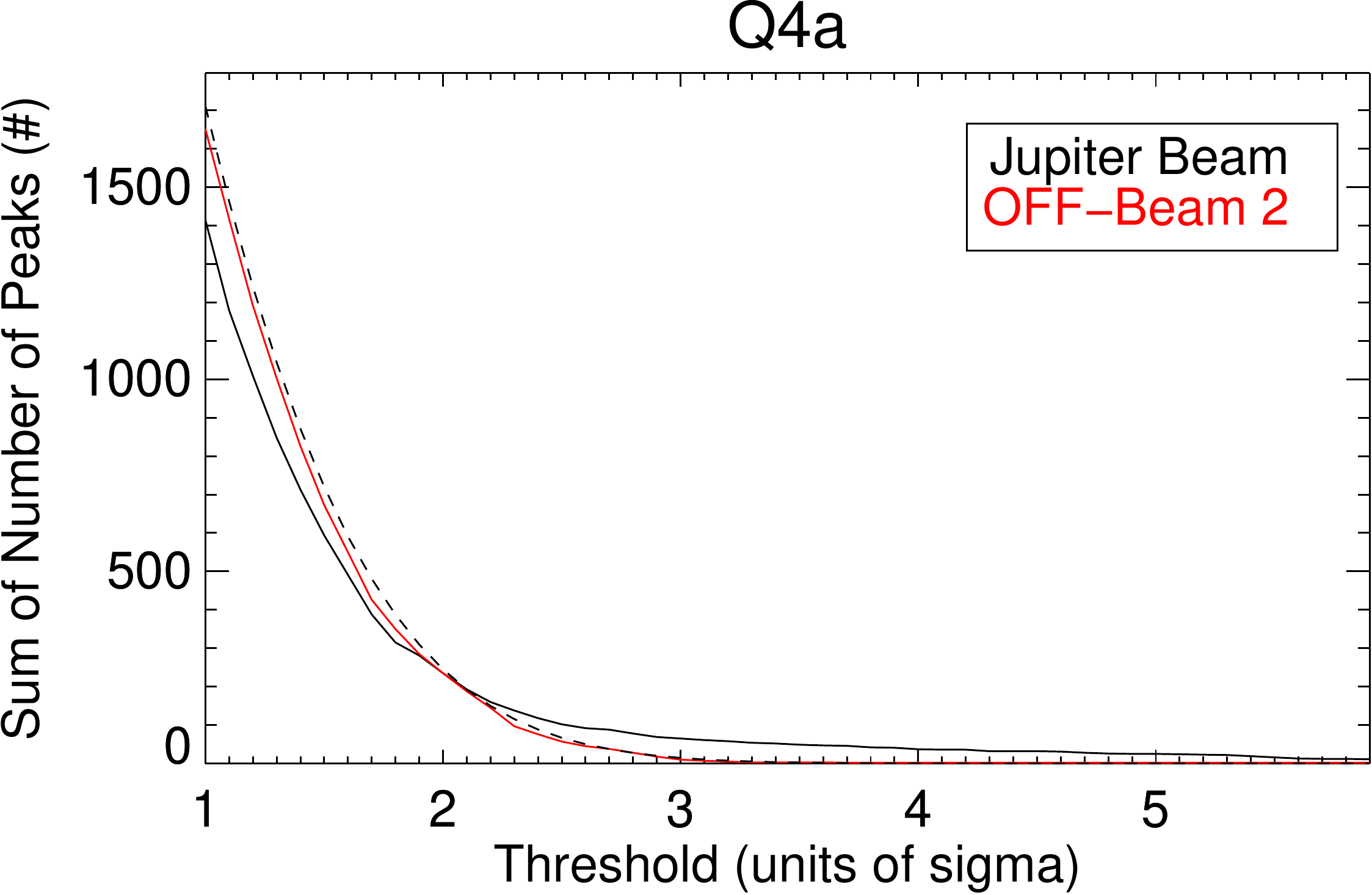}
        \label{}
    \end{subfigure}%
    \\
      \vspace{-0.5em}
       \begin{subfigure}[c]{0.32\textwidth}
        \centering
        \caption{}
        \includegraphics[width=\textwidth]{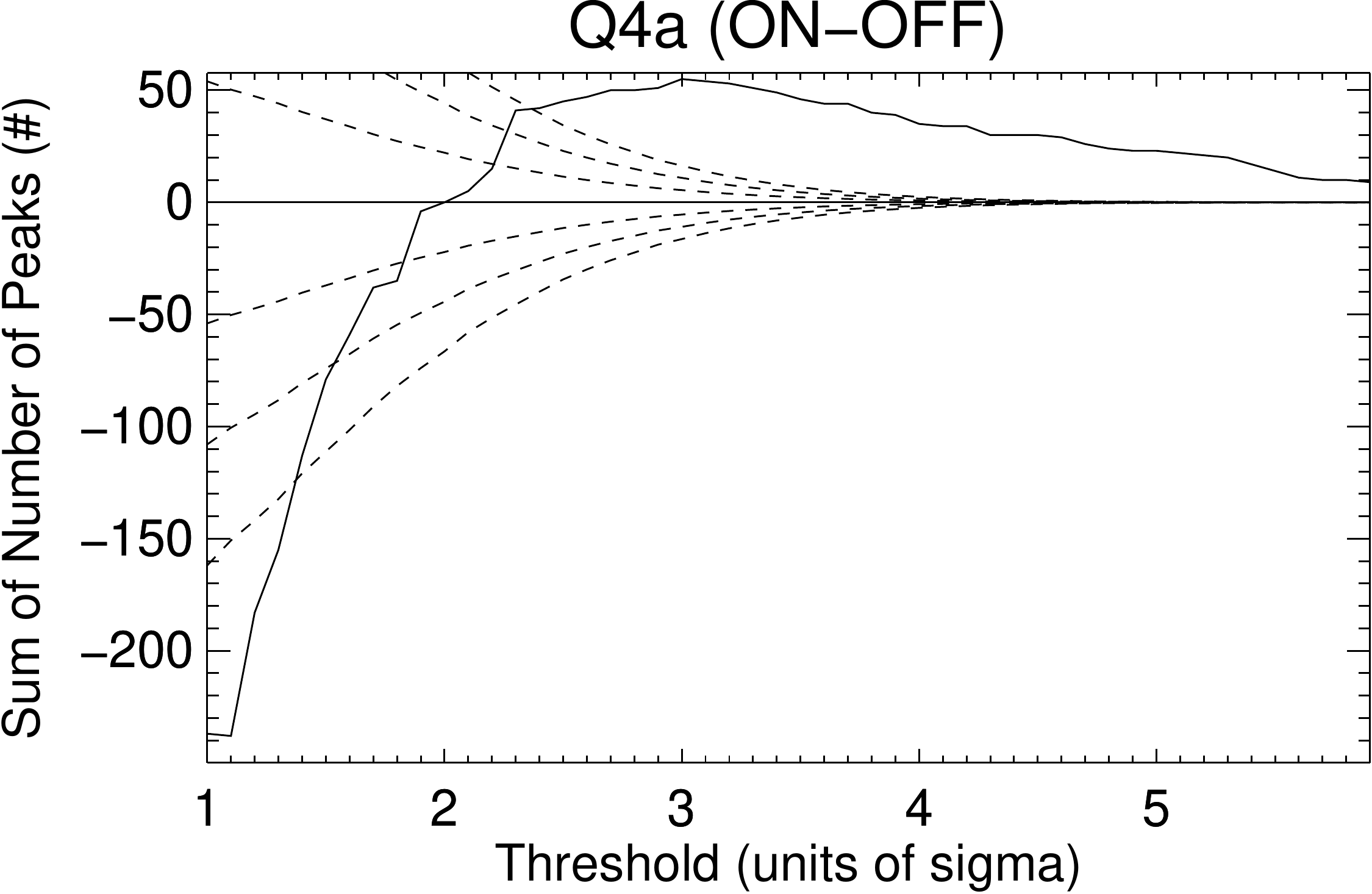}
        \label{}
    \end{subfigure}%
    &
  \begin{subfigure}[c]{0.32\textwidth}
        \centering
        \caption{}
        \includegraphics[width=\textwidth]{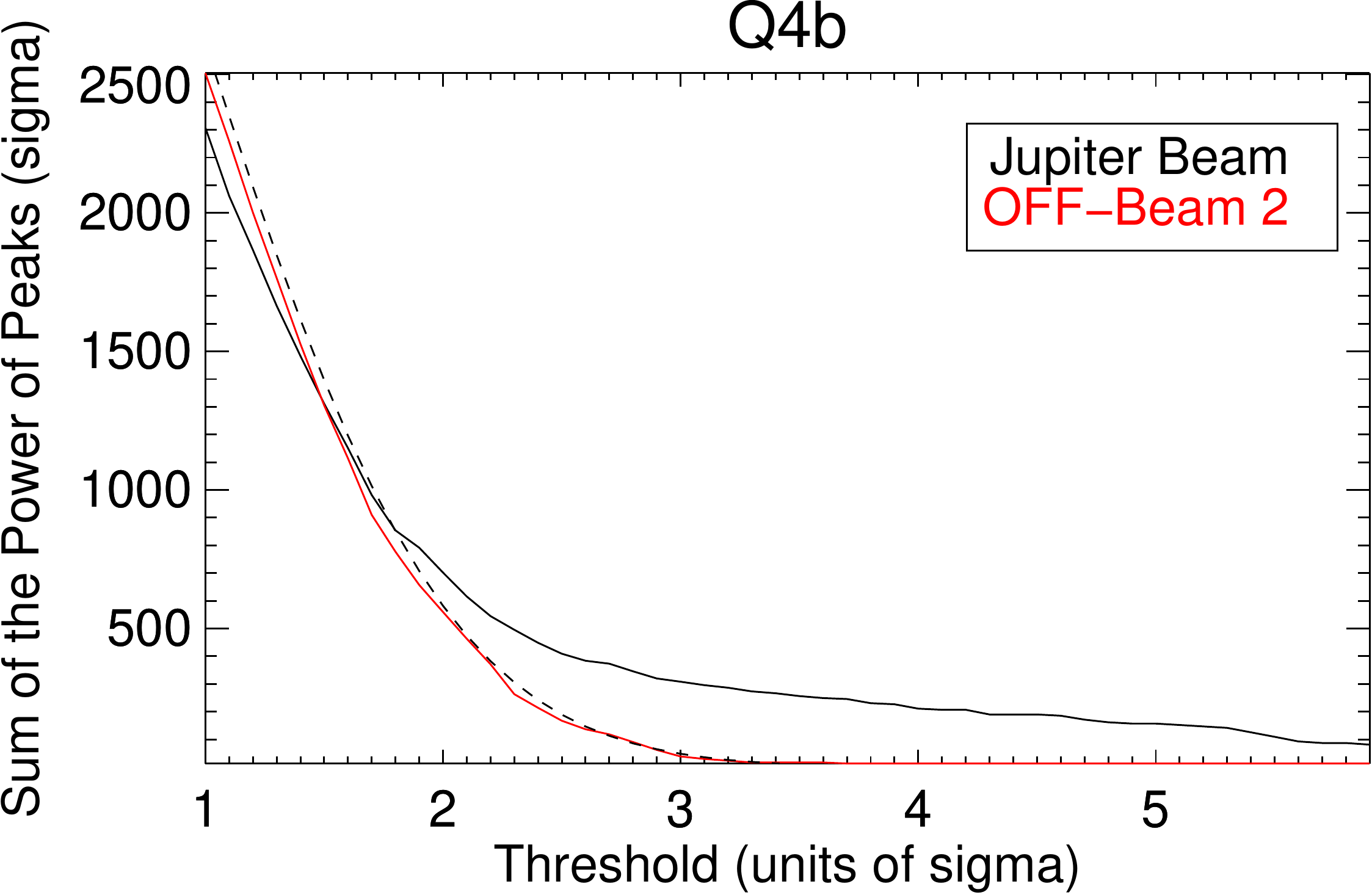} 
        \label{}
    \end{subfigure}%
    &
      \begin{subfigure}[c]{0.32\textwidth}
        \centering
        \caption{}
        \includegraphics[width=\textwidth]{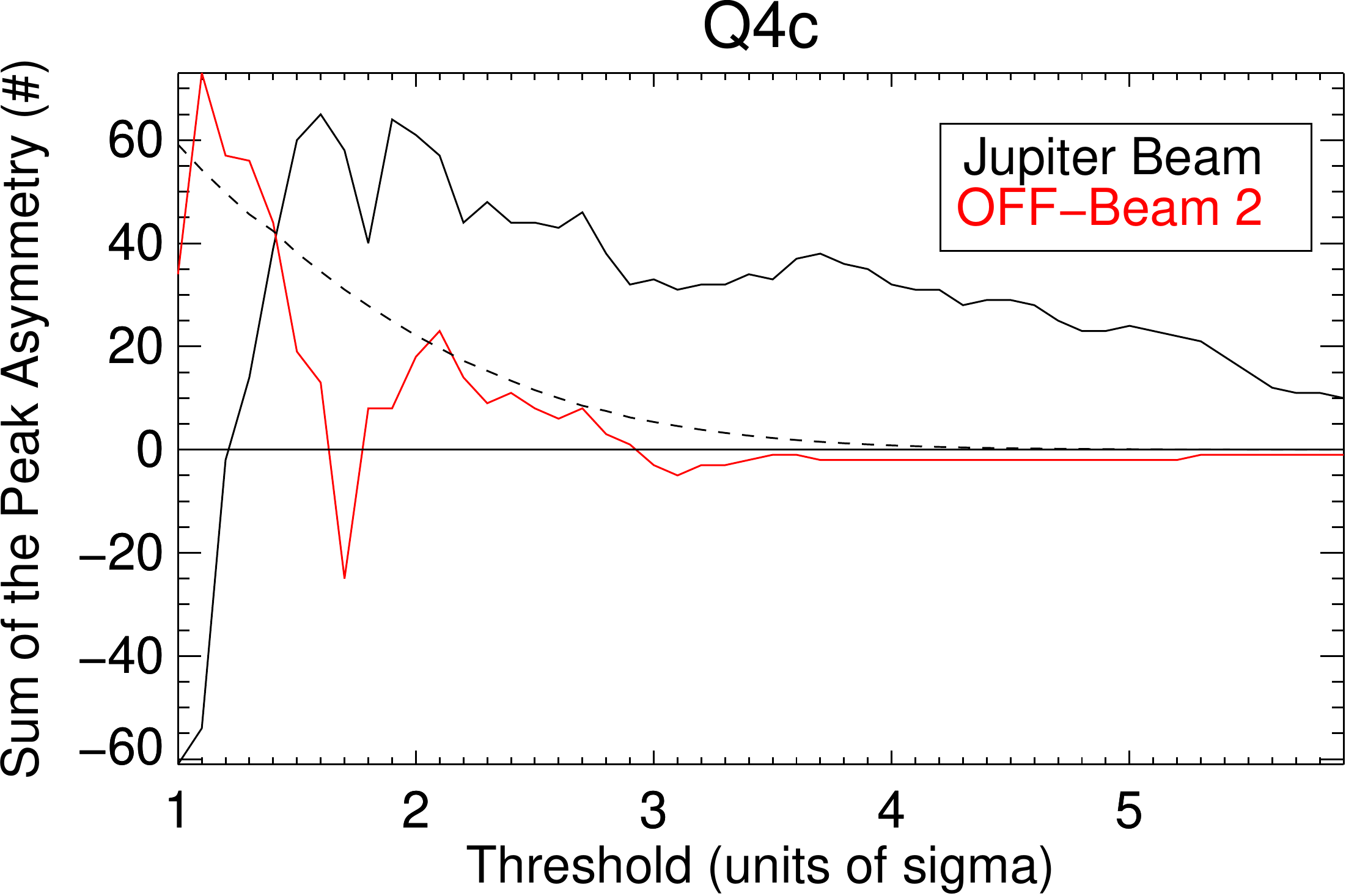}
        \label{}
    \end{subfigure}%
    \\
      \vspace{-0.5em}
       \begin{subfigure}[c]{0.32\textwidth}
        \centering
        \caption{}
        \includegraphics[width=\textwidth]{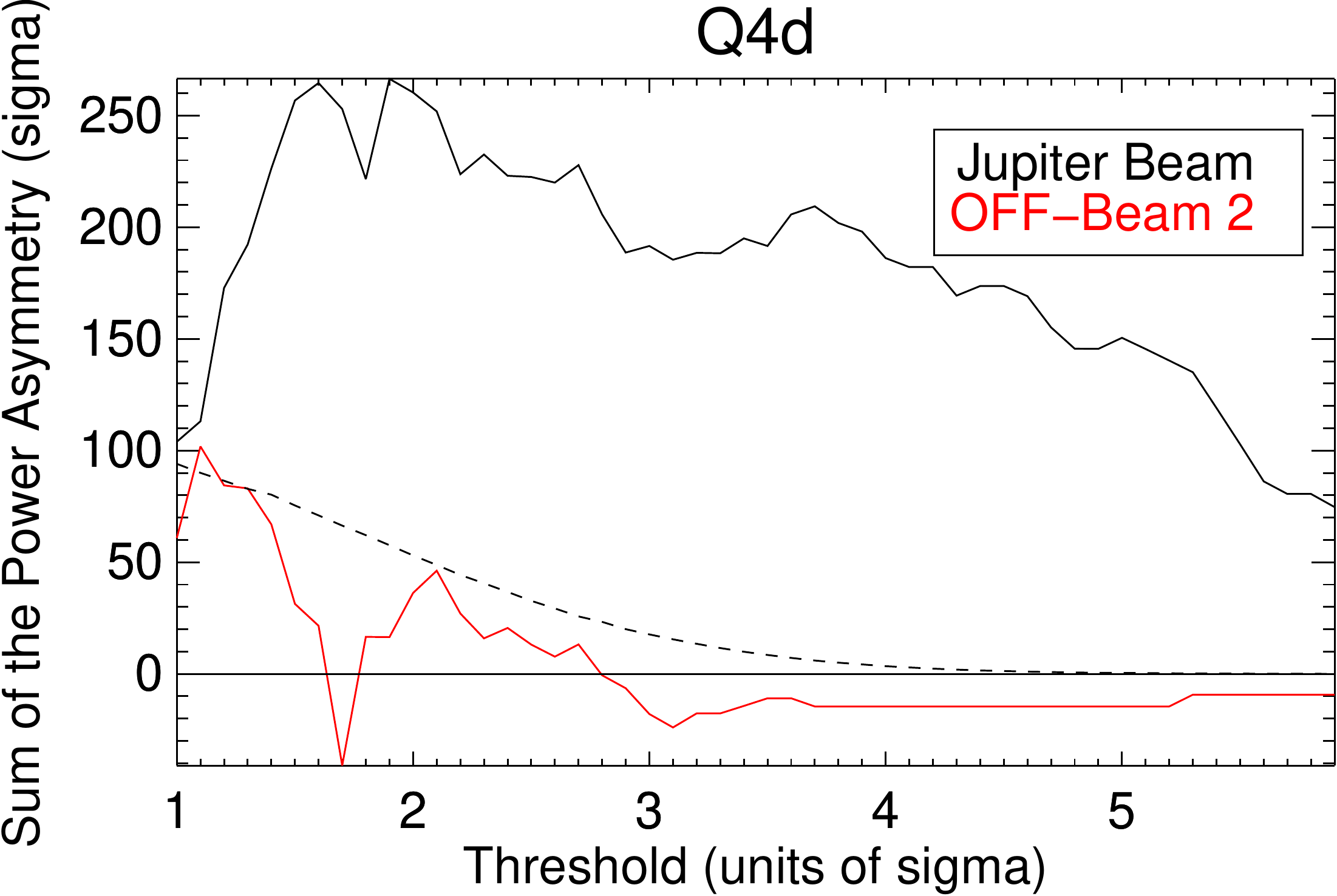}
        \label{}
    \end{subfigure}%
    &
  \begin{subfigure}[c]{0.32\textwidth}
        \centering
        \caption{}
        \includegraphics[width=\textwidth]{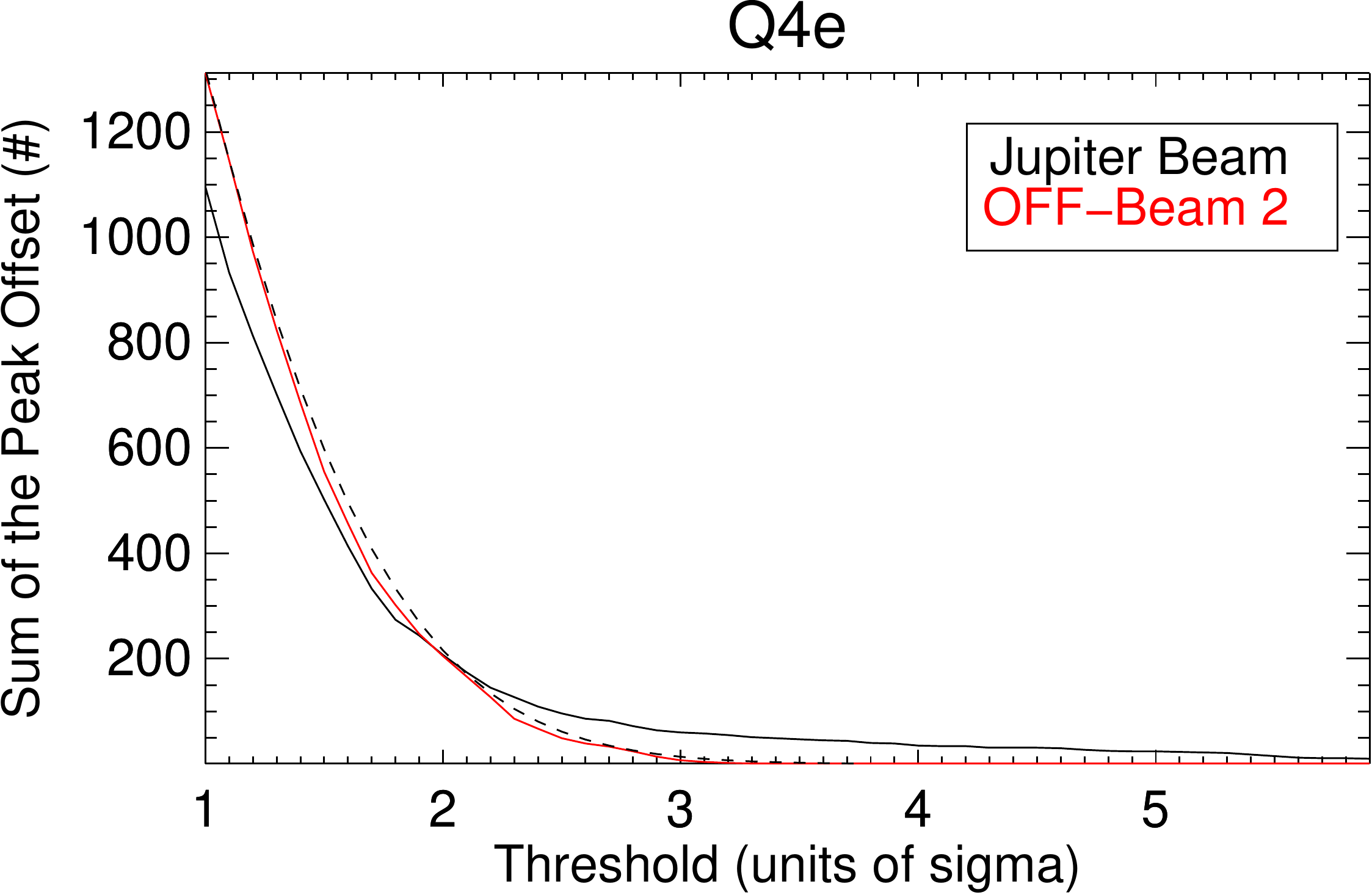} 
        \label{}
    \end{subfigure}%
    &
      \begin{subfigure}[c]{0.32\textwidth}
        \centering
        \caption{}
        \includegraphics[width=\textwidth]{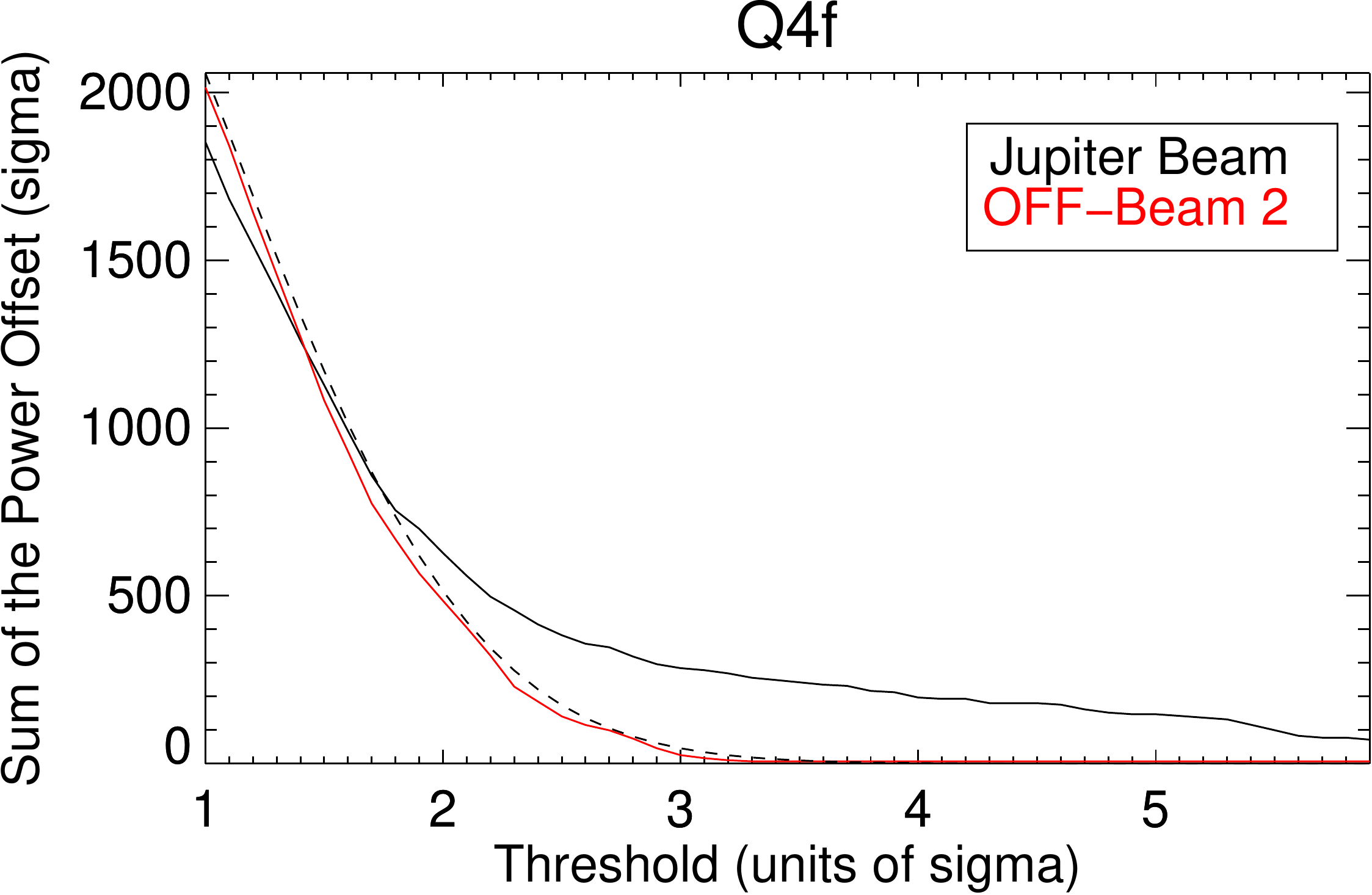}
        \label{}
    \end{subfigure}%
      \vspace{-0.5em}
\end{tabular}    
\end{center}
\caption{Observable quantities (Q2 and Q4) in Stokes-V (|$V^{'}$|) for a scaling value of $\alpha = 10^{-4}$. \textbf{(a)} Q2 (high-passed filtered intensities) vs time. \textbf{(b)} Q2 scatter plot for the ON- and OFF-beam. \textbf{(c)} Q4a (number of peaks). \textbf{(d)} Difference of ON - OFF for Q4a. \textbf{(e)} Q4b (power of the peaks). \textbf{(f)} Q4c (peak asymmetry). \textbf{(g)} Q4d (power asymmetry). \textbf{(h)} Q4e (peak offset). \textbf{(i)} Q4f (power offset). See Sect. \ref{sec:Qs} for a detailed description of each observable. For all plots the black lines are the ON-beam and the red lines are the OFF-beam. The dashed line for panels \textbf{(c)}, \textbf{(e)}, \textbf{(f)}, \textbf{(g)}, \textbf{(h)}, and \textbf{(i)} is the mean of the derived Q values from 10000 different Gaussian distributions with the same length as Q2. The dashed lines for panel \textbf{(d)} are the 1, 2, 3$\sigma$ statistical limits of the difference between all the Q values derived from two different Gaussian distributions (each run 10000 times).
}
\label{fig:AllQ}
\end{figure*}

\begin{figure*}[tbh!]
\begin{tabular}{cc}
      \begin{subfigure}[c]{0.47\textwidth}
        \centering
        \caption{}
        \includegraphics[width=\textwidth]{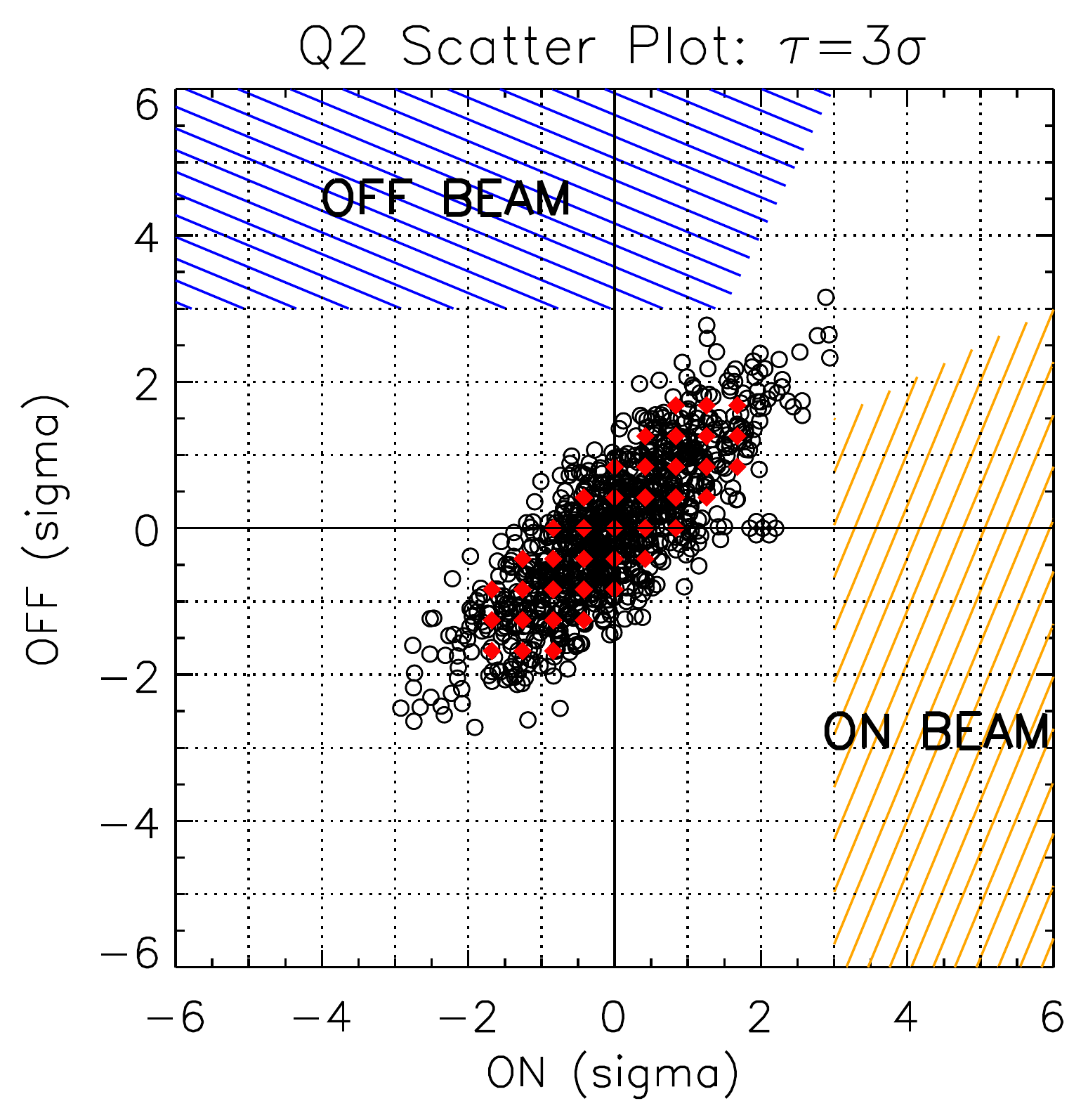}
        \label{}
    \end{subfigure}%
    &
          \begin{subfigure}[c]{0.47\textwidth}
        \centering
        \caption{}
        \includegraphics[width=\textwidth]{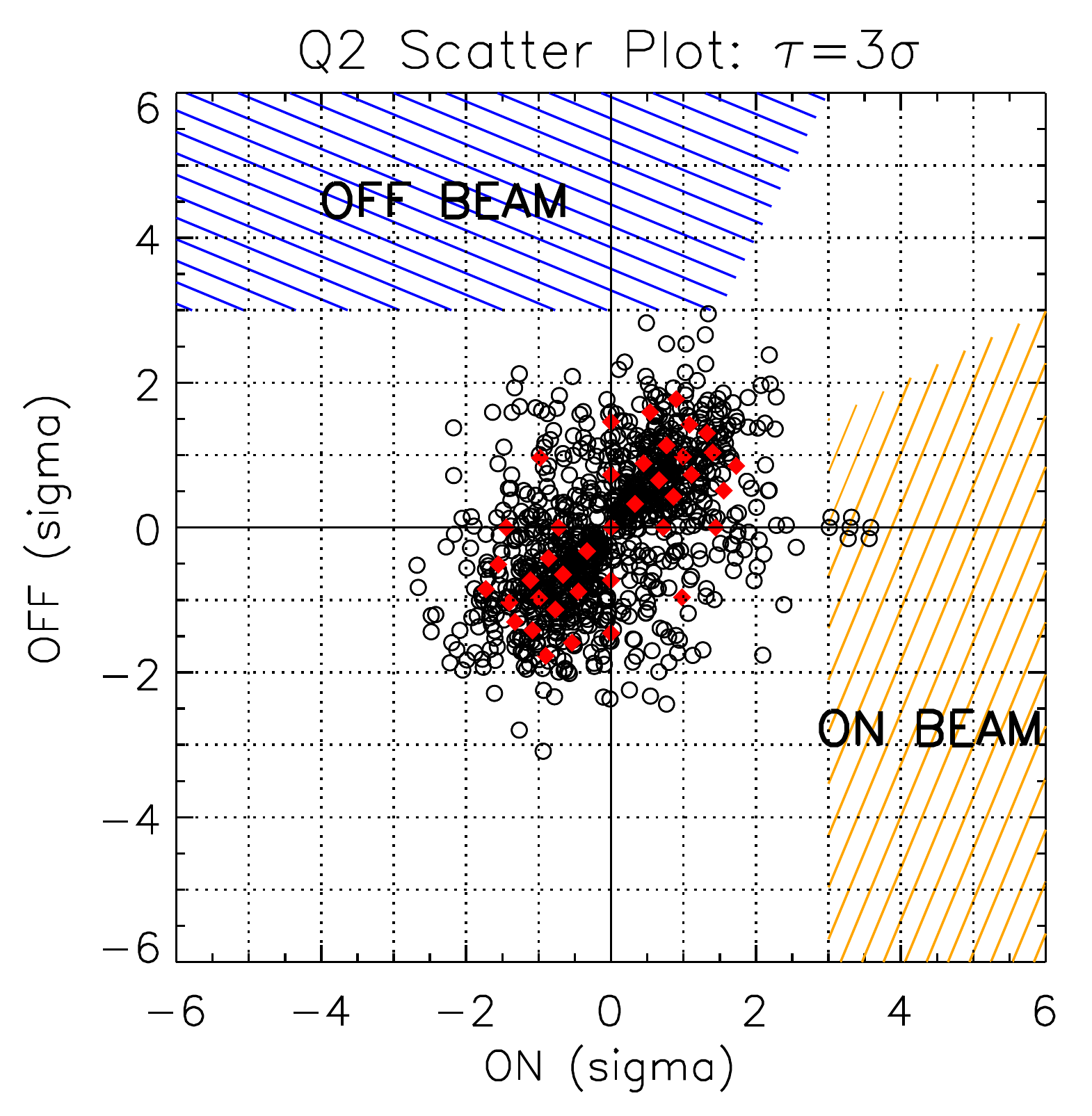}
        \label{}
    \end{subfigure}%
    
\end{tabular}
\caption{Simulated data-points (black) and test-points (red) to demonstrate the observable quantity Q2 and the effect of the elliptical correction. \textbf{(a)} Q2 before the elliptical correction. \textbf{(b)} Q2 after the elliptical correction. X-axis: Q2 (normalized high-pass filtered intensities) for the ON-beam. Y-axis: same for the OFF-beam. Points with high values in the ON- and OFF-beam (i.e. close to the main diagonal) are due to either residual RFI or ionospheric fluctuations. One of the main detection criteria is based on the number of points with high values only in the ON-beam or only in the OFF-beam. For this, the regions used in Q3e and Q4e are hatched (orange for the ON-beam, and blue for the OFF-beam; see text for the precise definition) for the case of $\tau = 3\sigma$ (i.e. a threshold of 3$\sigma$). This figure also illustrates the effect of the elliptical correction described in Appendix \ref{app:elliptical-correction}. The red test data-points allow for the visualization of the displacement of individual points that leads to the circularization of the cloud. Using these red data-points, it can be seen that data-points on the x- and y-axis are the least affected by this procedure; data-points close to the main diagonal are most strongly affected. The black points represent what we expect from an observation, namely sky noise plus a few signal datapoints (visible in panel a at ON$\sim$2.0 and OFF$\sim$0.0 in this example). Before the elliptical correction, there are no points in the orange and blue hatched regions. After the elliptical correction, the injected data-points are indeed in the orange hatched region.}
\label{fig:Scatter_demo}
\end{figure*} 

\begin{figure*}[!phtb]
\begin{center}
\begin{tabular}{cc}
  \vspace{-0.5em}
       \begin{subfigure}[c]{0.28\textwidth}
        \centering
        \caption{}
        \includegraphics[width=\textwidth]{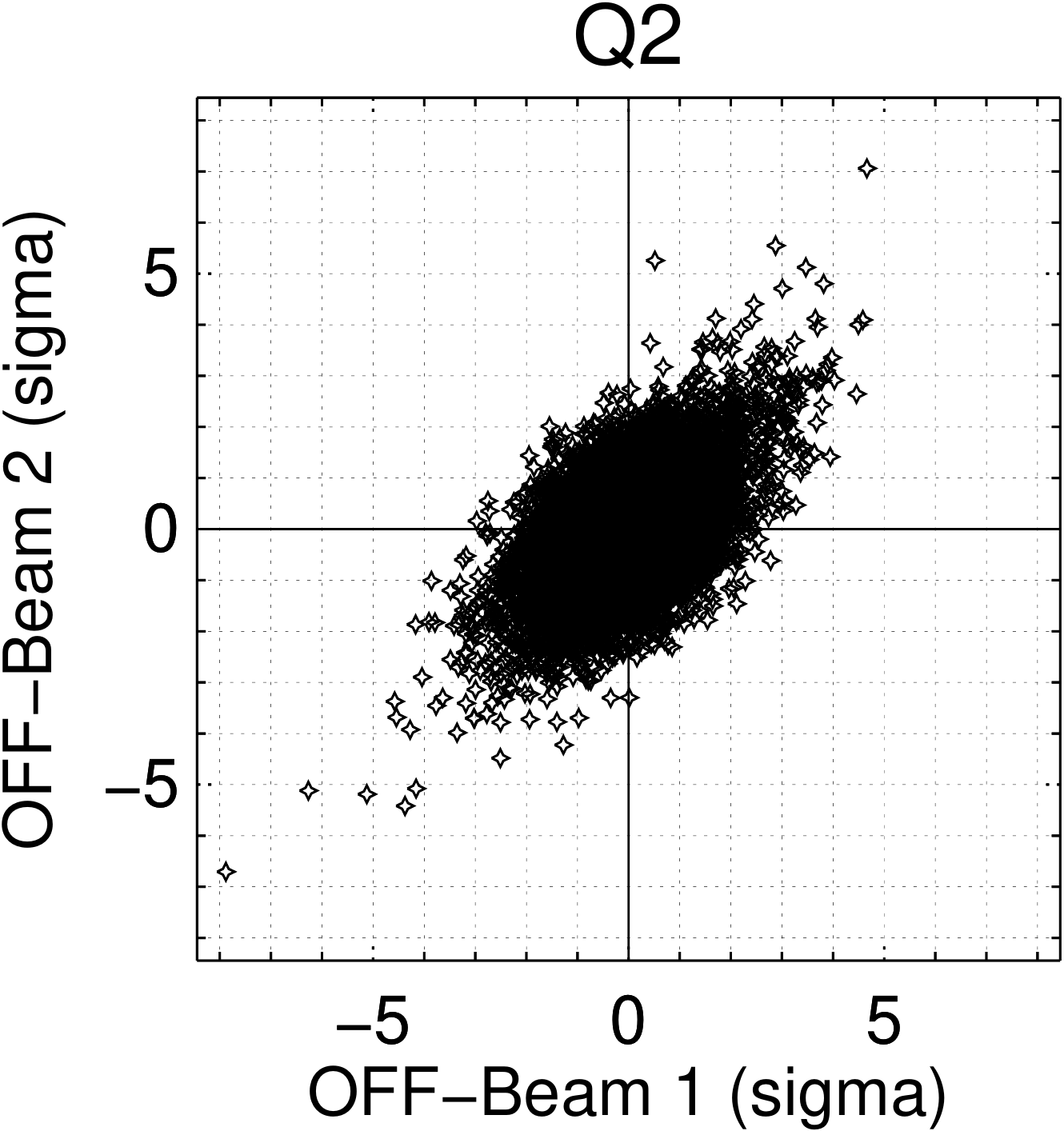}
        \label{}
    \end{subfigure}%
    &
           \begin{subfigure}[c]{0.28\textwidth}
        \centering
        \caption{}
        \includegraphics[width=\textwidth]{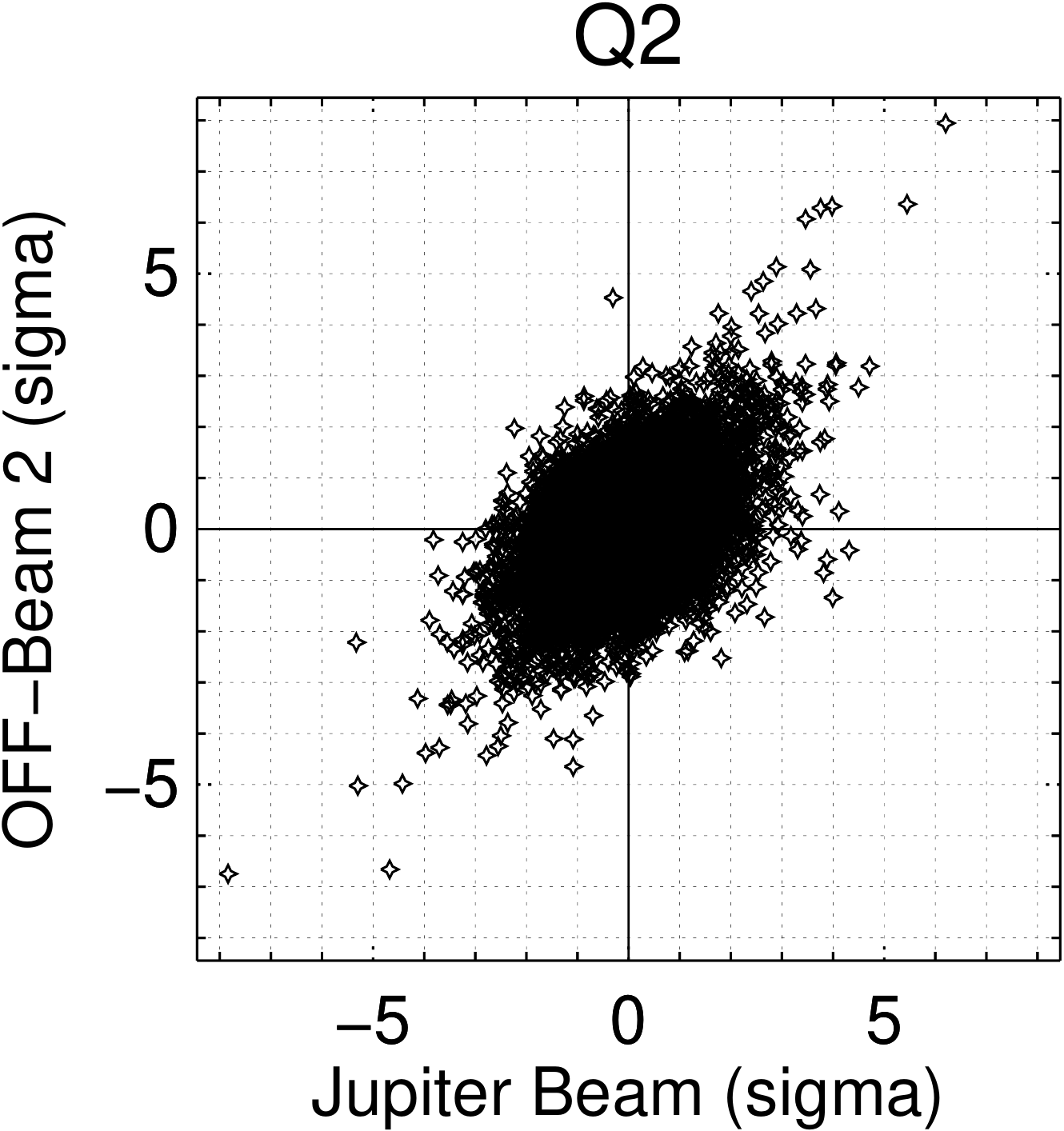}
        \label{}
    \end{subfigure}%
    \\
      \vspace{-0.5em}
           \begin{subfigure}[c]{0.33\textwidth}
        \centering
        \caption{}
        \includegraphics[width=\textwidth]{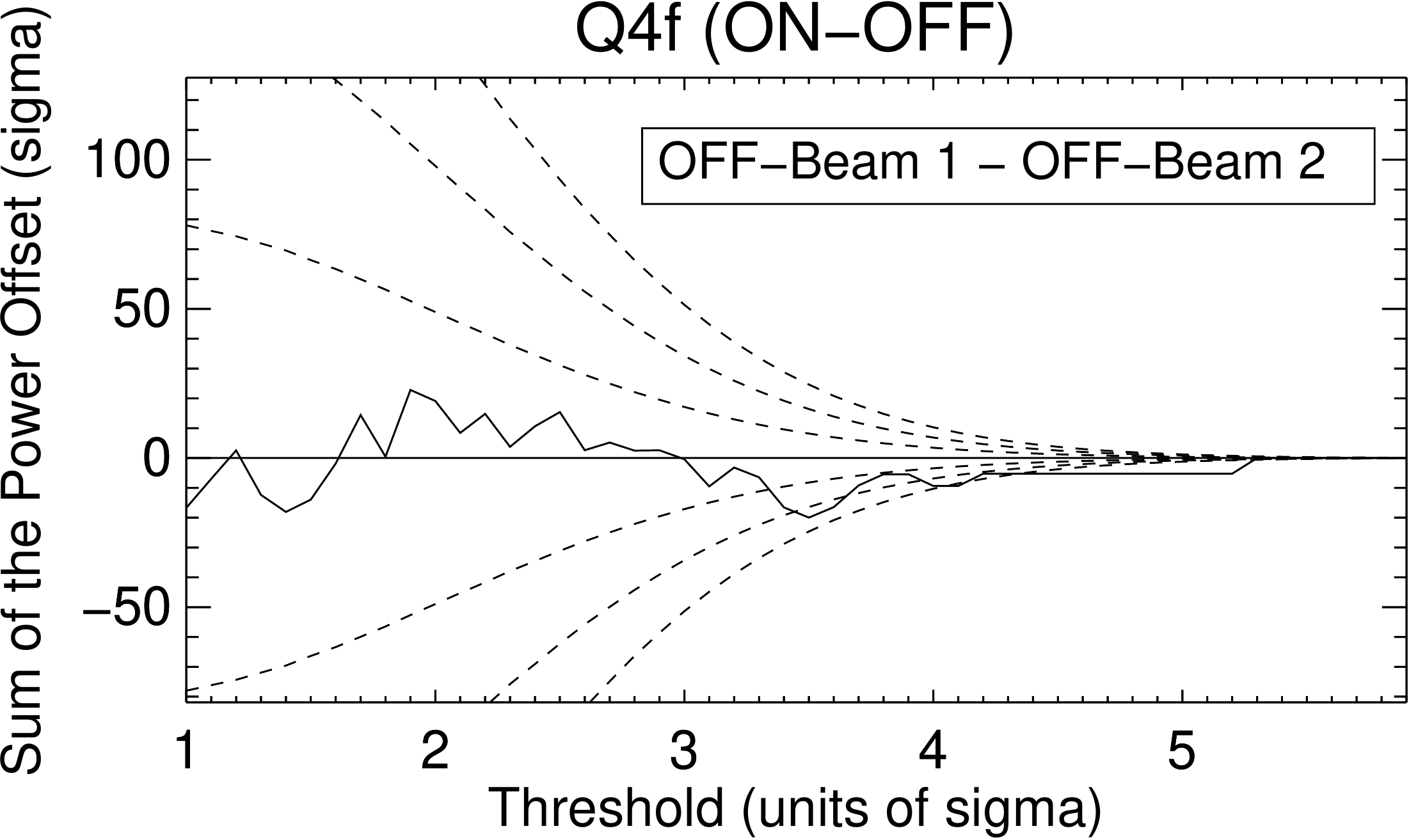}
        \label{}
    \end{subfigure}%
    &
           \begin{subfigure}[c]{0.33\textwidth}
        \centering
        \caption{}
        \includegraphics[width=\textwidth]{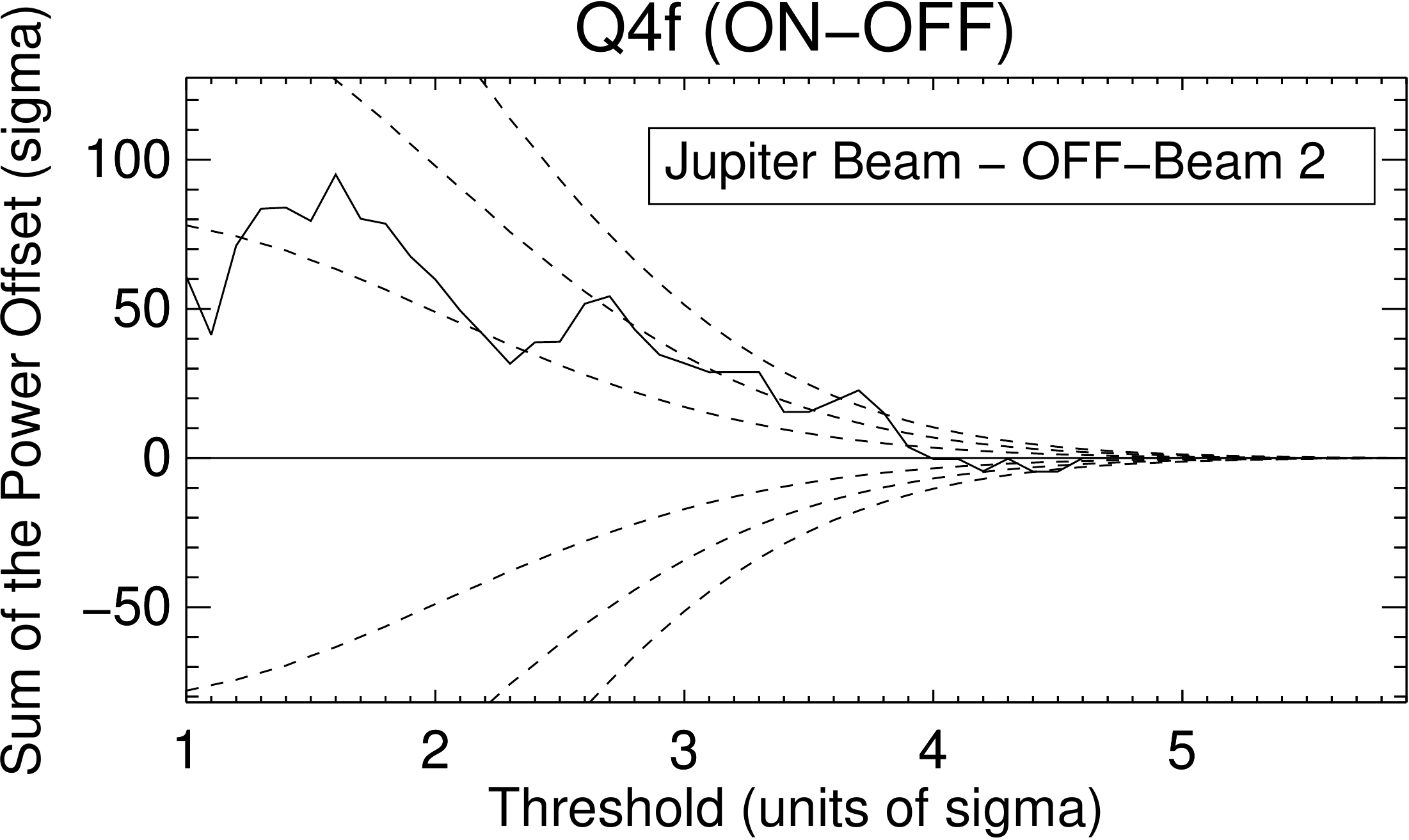}
        \label{}
    \end{subfigure}%
    \\
      \vspace{-0.5em}
           \begin{subfigure}[c]{0.28\textwidth}
        \centering
        \caption{}
        \includegraphics[width=\textwidth]{{Sky_50-60_EllC_Q2}.pdf}
        \label{}
    \end{subfigure}%
    &
           \begin{subfigure}[c]{0.28\textwidth}
        \centering
        \caption{}
        \includegraphics[width=\textwidth]{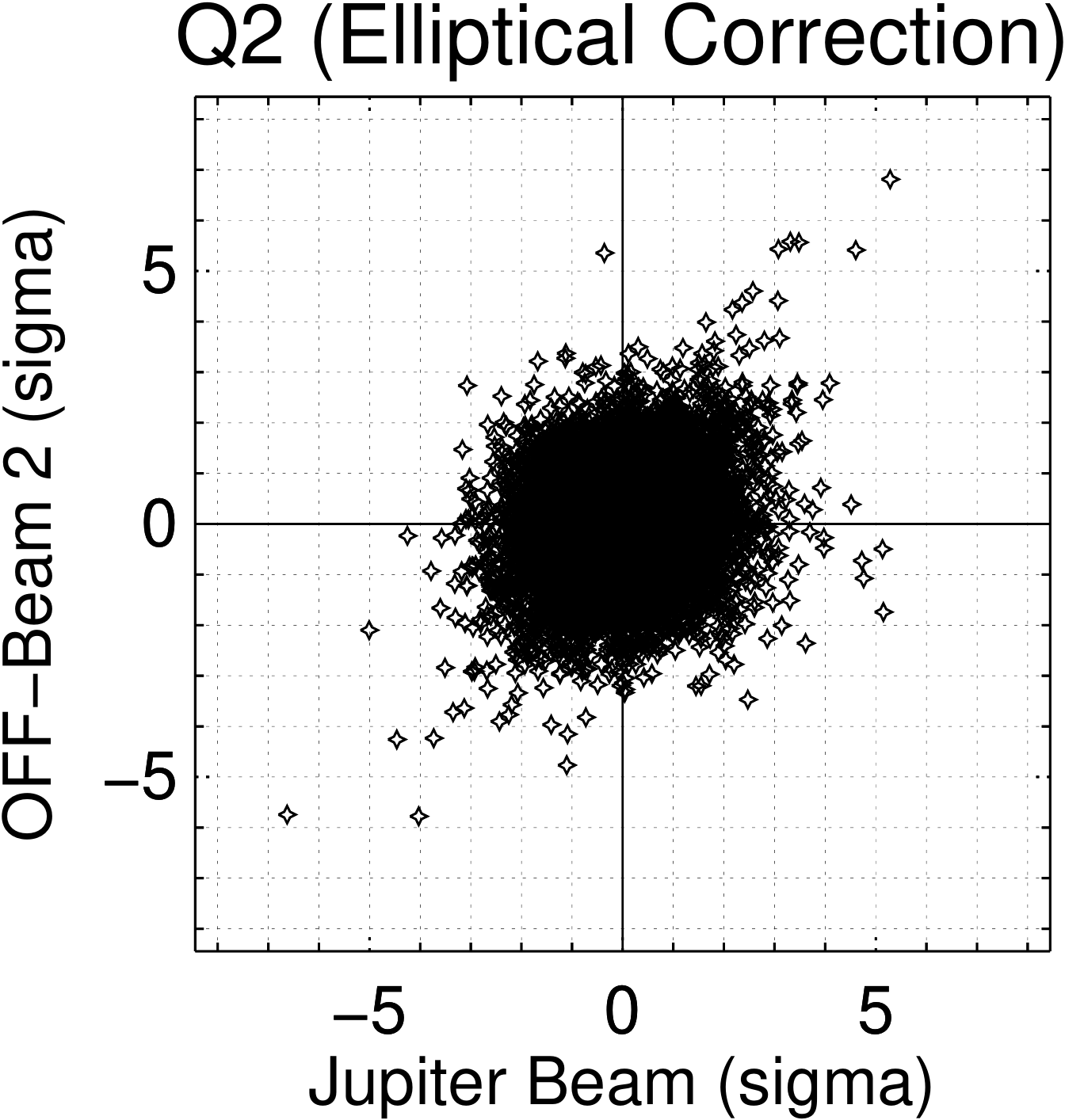}
        \label{}
    \end{subfigure}%
    \\
      \vspace{-0.5em}
           \begin{subfigure}[c]{0.33\textwidth}
        \centering
        \caption{}
        \includegraphics[width=\textwidth]{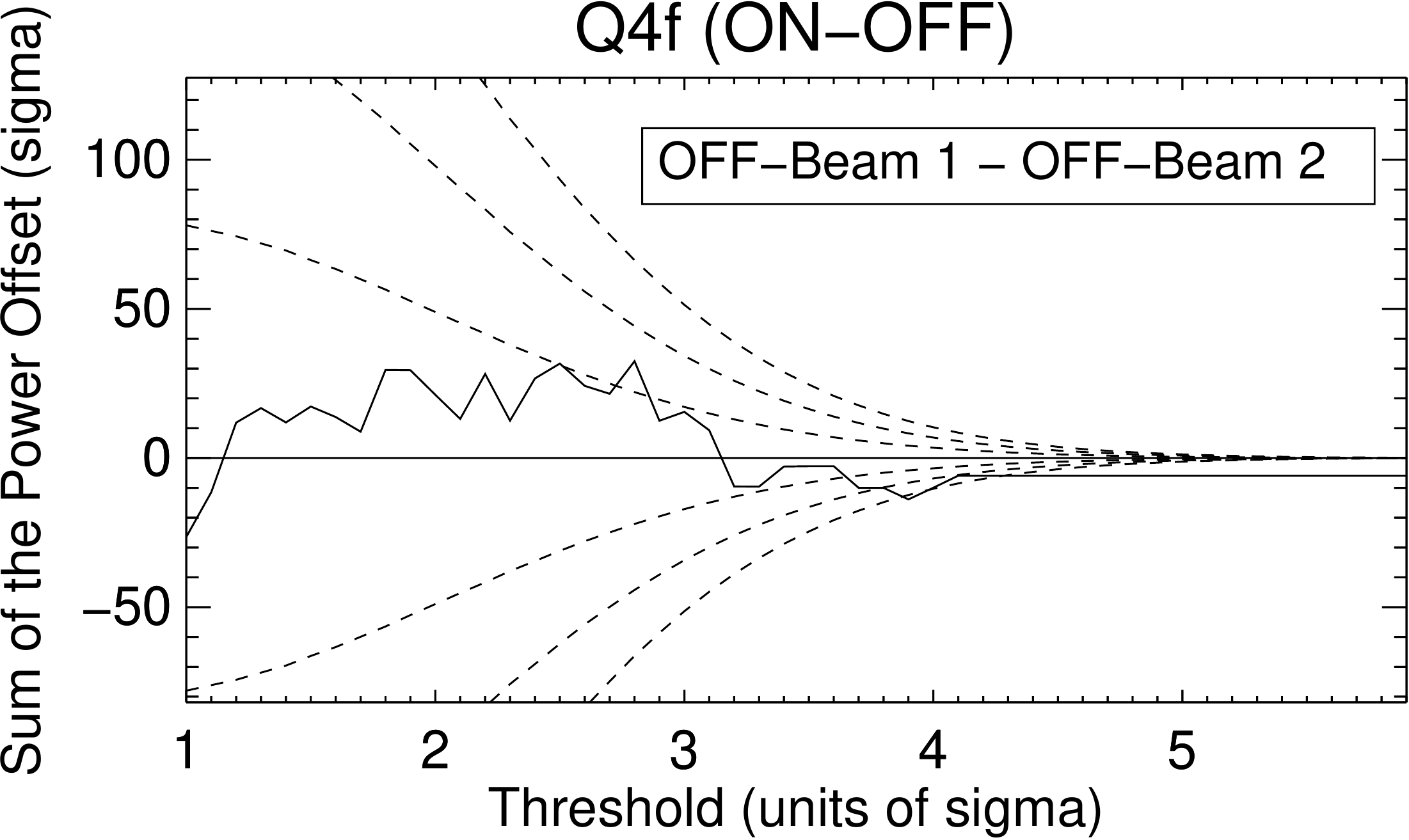}
        \label{}
    \end{subfigure}%
    &
           \begin{subfigure}[c]{0.33\textwidth}
        \centering
        \caption{}
        \includegraphics[width=\textwidth]{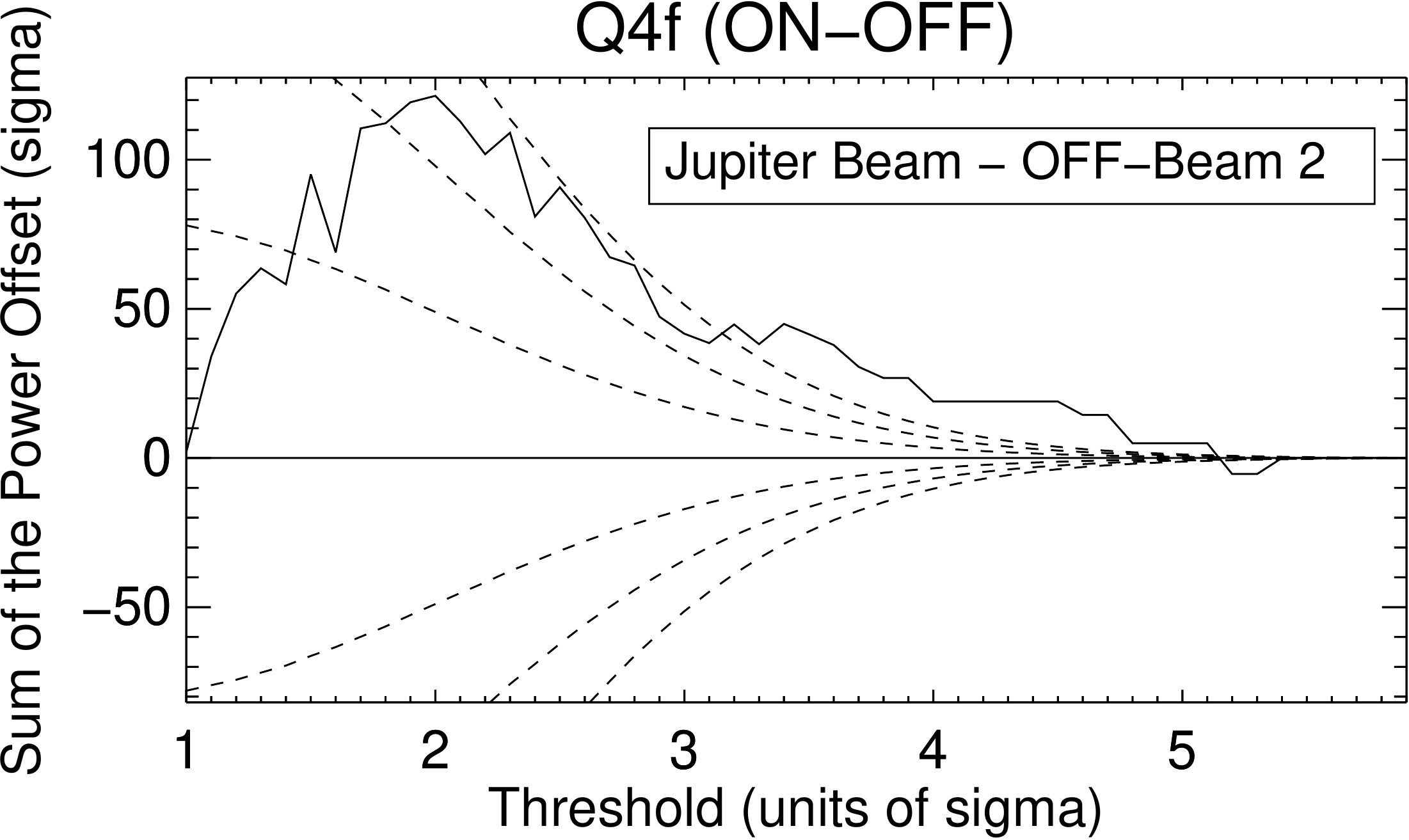}
        \label{}
    \end{subfigure}%
      \vspace{-0.5em}

\end{tabular}
\end{center}
\caption{Plots of Q2 and Q4f (Power Offset) showing the detection limit for Stokes-I ($\alpha = 10^{-3.5}$) for the frequency range 50--60 MHz. \textbf{(a)} and \textbf{(b)} Q2 before elliptical correction. \textbf{(c)} and \textbf{(d)} Q4f difference of the two beams before elliptical correction. \textbf{(e)} and \textbf{(f)} Q2 after elliptical correction. \textbf{(g)} and \textbf{(h)} Q4f difference of the two beams after elliptical correction. The comparison of the two OFF-beams from Obs~$\#$2 can be found in the left column (panels \textbf{a}, \textbf{c}, \textbf{e}, \textbf{g}) and the comparison of ON-beam (Jupiter) vs OFF-beam 2 can be found in the right column (panels \textbf{b}, \textbf{d}, \textbf{f}, \textbf{h}). The dashed lines for panels \textbf{(c)}, \textbf{(d)}, \textbf{(g)}, and \textbf{(h)} are the 1, 2, 3 $\sigma$ statistical limits derived from two different Gaussian distributions (each run 10000 times). Panel \textbf{(h)} shows an excess of ON vs OFF points at $\ge~2~\sigma$ statistical significance for signals up to a threshold of 4$\sigma$. For comparison, in panel \textbf{(g)} all the excess points are mostly below the $1\sigma$ statistical significance level. We find by performing Gaussian simulations that the probability to obtain a curve like panel \textbf{(g)} is $\sim$99$\%$, whereas the probability to reproduce a curve like panel \textbf{(h)} is $\sim$10$^{-4}$. 
}
\label{fig:Q4Detectionplot}
\end{figure*}

\hspace{4ex} Next, the observables Q3 and Q4 are defined to systematically and statistically explore the parameter space of Q2 (e.g. $y$).     
    \item Q3: Time-series of broadband burst emission from Q2 for one threshold $\tau$ (in units of sigma)
      \begin{itemize}
         \item Q3a (Number of Peaks): Number of peaks per $TI$ where $y \ge \tau$ (Fig. \ref{fig:Q3})
         \item Q3b (Power of Peaks): Sum of the power of peaks per $TI$ where $y \ge \tau$
         \item Q3c (Peak Asymmetry): Number of peaks per ~$TI$~where $y \ge \tau$ subtracted by number of peaks where $y \le - \tau$
         \item Q3d (Power Asymmetry): Sum of the power of peaks per $TI$ where $y \ge \tau$ subtracted by the sum of $|\text{power}|$ of peaks where~$y~\le~-\tau~$
         \item Q3e (Peak Offset): Number of peaks per $TI$ where $y \ge \tau$ for the ON (OFF) beam and exceeding the corresponding OFF (ON) values by a factor $\ge 2$
         \item Q3f (Power Offset): Sum of the power of peaks per $TI$ where $y \ge \tau$ for the ON (OFF) beam and exceeding the corresponding OFF (ON) values by a factor $\ge 2$
      \end{itemize}
    \item Q4a to Q4f: Each observable in Q3 is summed over all times and plotted versus the threshold value $\tau$ (Figs. \ref{fig:AllQ}c--i)
\end{itemize}

When examining Q3 and Q4, the ON- and OFF-beam are always compared to each other and plotted against a reference curve with the same number of elements. This reference curve is created by taking the mean of the derived Q values from 10000 different Gaussian distributions of random values. When we subtract the ON- and OFF-beam Q value, then the reference curve is the standard deviation of the difference between all the Q values derived from two different Gaussian distributions (each run 10000 times). By default, Q4 is calculated from $\tau = 1...6\sigma$ with a step size of 0.1$\sigma$. Q4 is more effective at finding excess faint emission than Q3 since it is summed over all times. Once a detection is found in Q4, then Q3 can be used to localize the emission in time (e.g. Fig \ref{fig:Q3}a). The reason for evaluating Q3a and Q4a are to determine if the ON-beam has more positive peaks than the OFF-beam thus indicative of burst emission. The power of the peaks (Q3b and Q4b) highlights more clearly any potential excess. The peak (Q3c and Q4c) and power asymmetry (Q3d and Q3d) are useful at determining whether there is an asymmetry in the signal distribution. These observables are similar to the skewness but are more adapted to a small numbers of outliers. An excess of positive peaks over negative ones could be evidence of bursts. Finally, the peak (Q3e and Q4e) and power offset (Q3f and Q4f) are the best discrimination of real burst emission because they directly correlate any detection against the other beam. Additionally, ionospheric effects and any remaining low-level RFI will be concentrated on the diagonal; the peak and power offset mitigate these effects especially after elliptical correction. See Fig. \ref{fig:Scatter_demo} for an illustration of where these observables lie in the parameter space of the scatter plot of Q2.           

\begin{figure*}[tbh!]
\begin{center}
\begin{tabular}{cc}

   \begin{subfigure}[c]{0.45\textwidth}
        \centering
        \caption{}
        \includegraphics[width=\textwidth]{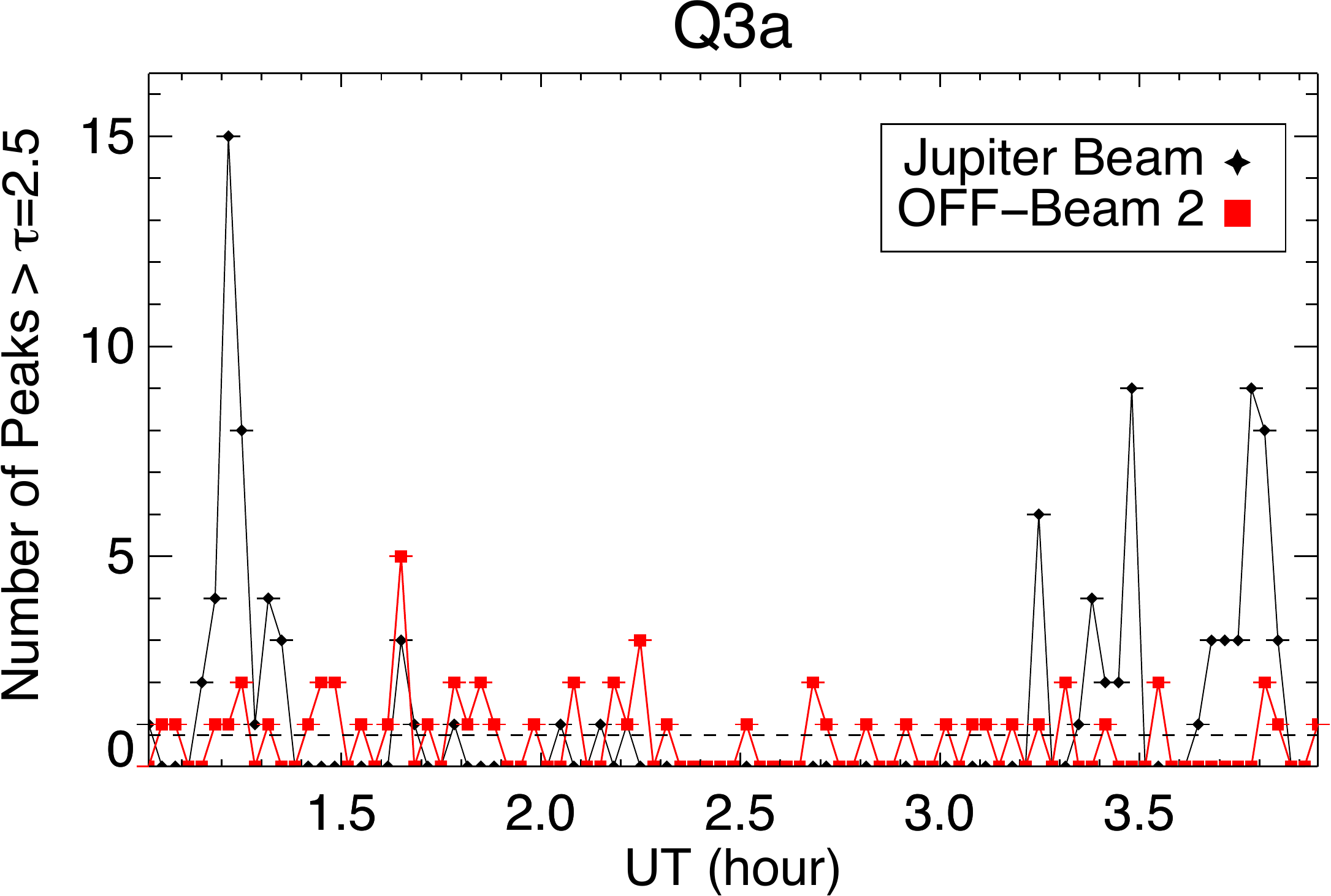}
        \label{}
    \end{subfigure}%
    &
       \begin{subfigure}[c]{0.45\textwidth}
        \centering
        \caption{}
        \includegraphics[width=\textwidth]{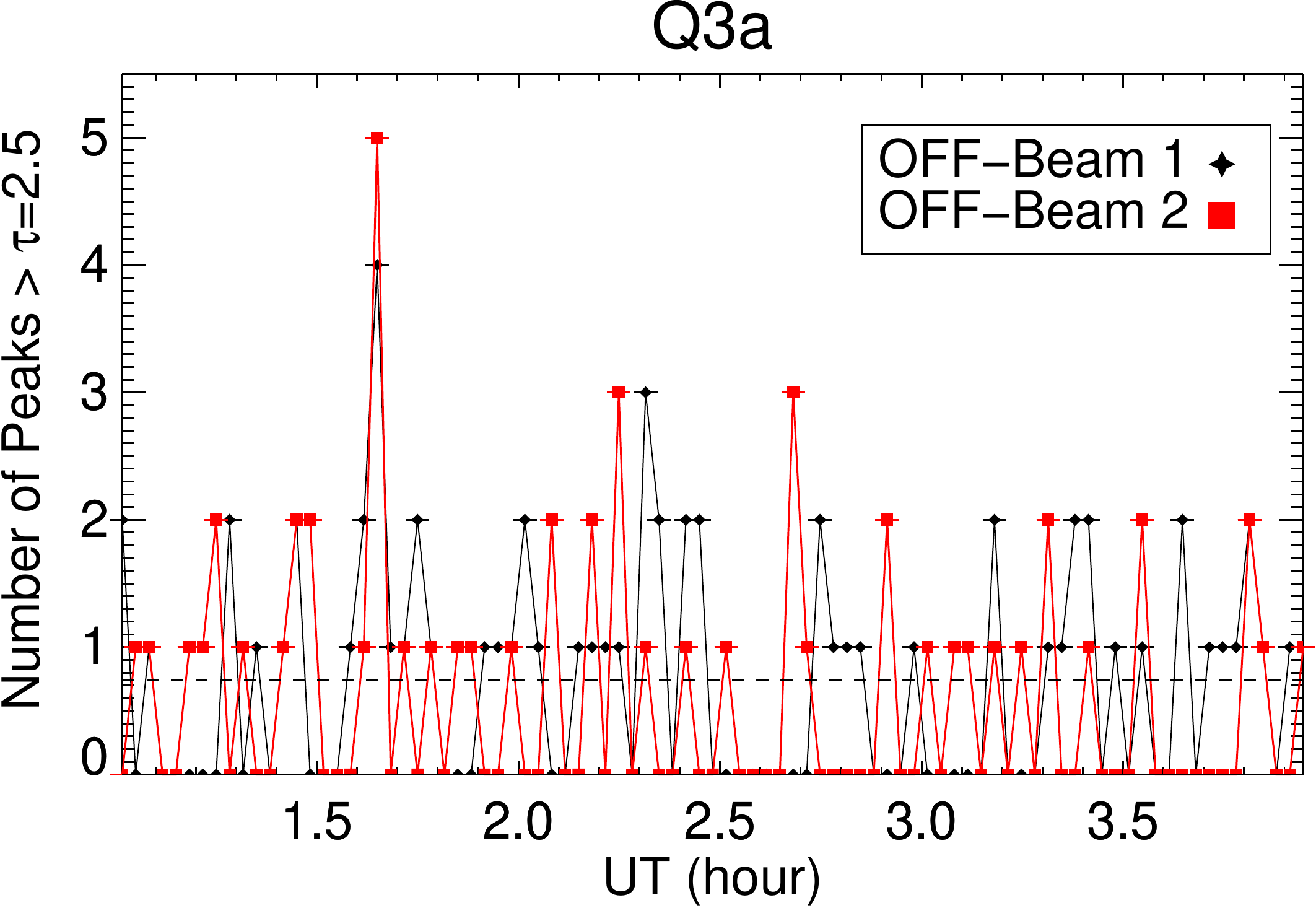}
        \label{}
    \end{subfigure}%

\end{tabular}   
\end{center}
\vspace{-1em}
\caption{Comparison of the observable quantity Q3a between the ON-beam (Jupiter) and OFF-beam 2 \textbf{(a)} and the 2 OFF-beams \textbf{(b)} in Stokes-V (|$V^{'}$|) for a scaling value $\alpha = 10^{-4}$ and threshold $\tau = 2.5\sigma$. See Sect. \ref{sec:Qs} for a detailed description of Q3a. For all plots the black lines and the red lines correspond to two different beams. The dashed line is the mean of the derived Q values from 10000 different Gaussian distributions with the same length as the time interval ($TI$). Jupiter's emission is mainly localized between 1.2--1.4 UT and 3.2--3.9 UT, whereas the bright emission at 1.7 UT can be seen in both OFF beams. }
\label{fig:Q3}
\end{figure*} 

\begin{figure*}[!pht]
\begin{center}
\begin{tabular}{cc}
  \begin{subfigure}[c]{0.35\textwidth}
        \centering
        \caption{}
        \includegraphics[width=\textwidth]{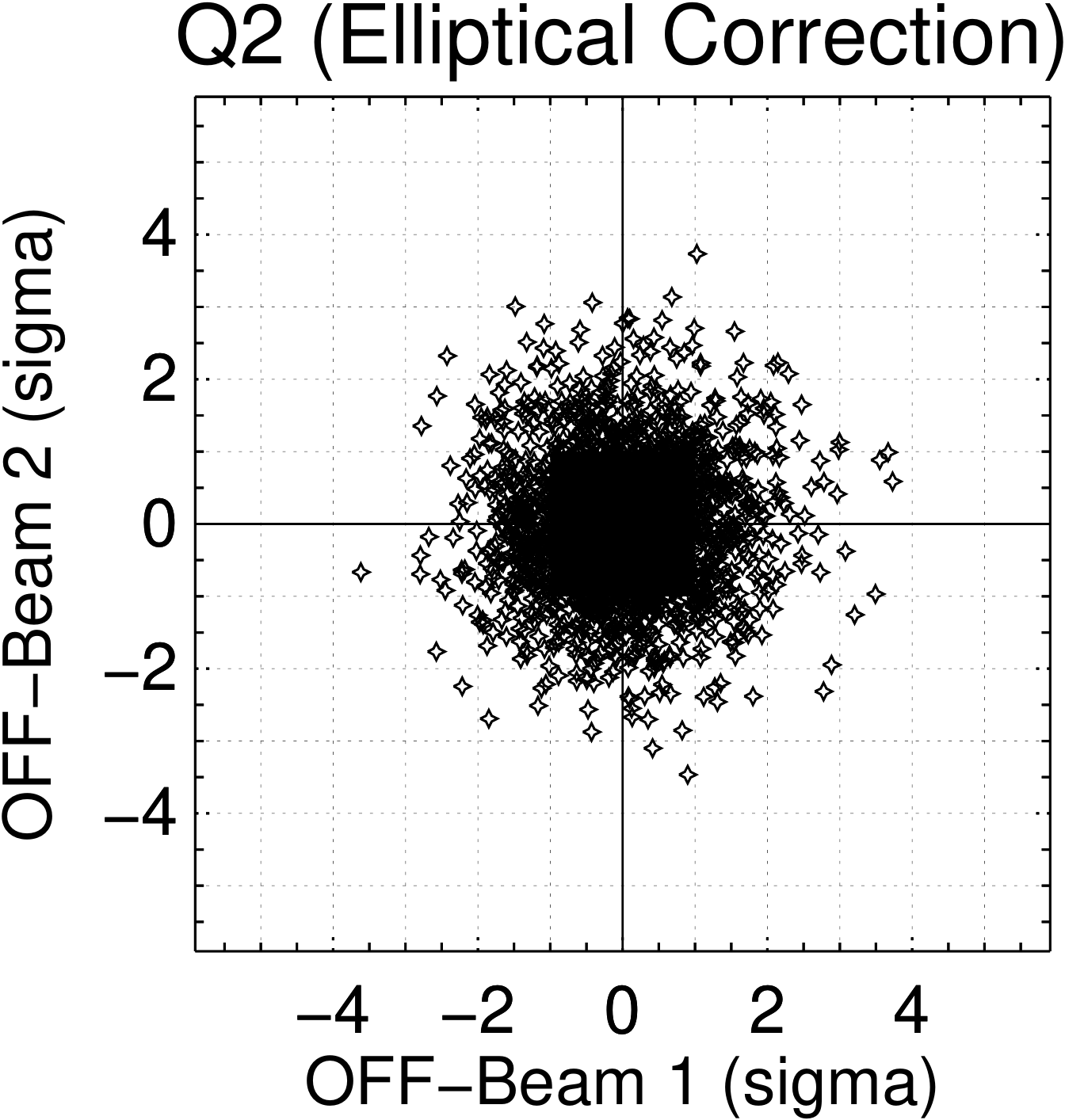}
        \label{}
    \end{subfigure}%
&
  \begin{subfigure}[c]{0.35\textwidth}
        \centering
        \caption{}
        \includegraphics[width=\textwidth]{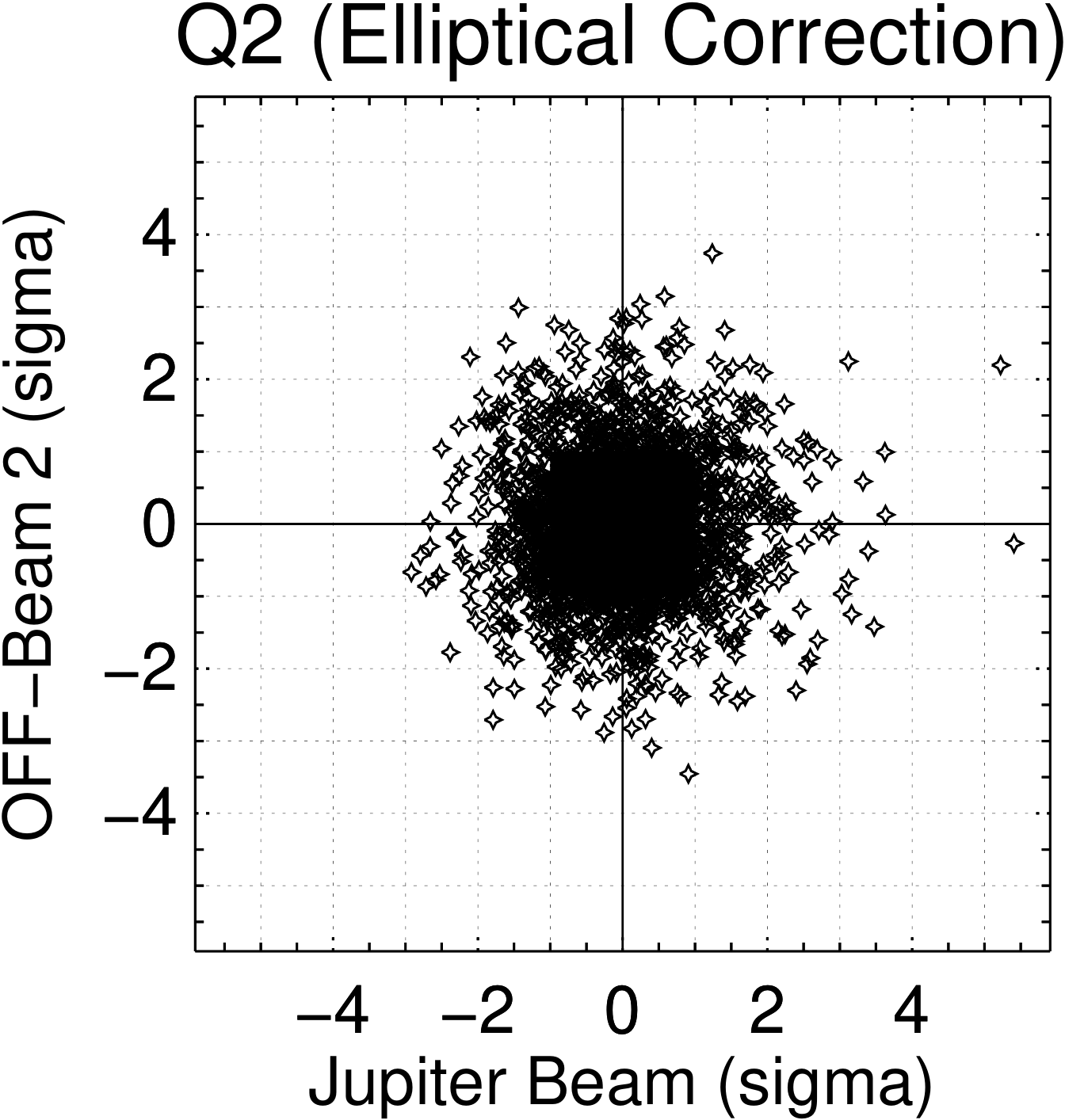}
        \label{}
    \end{subfigure}%
\\
  \begin{subfigure}[c]{0.45\textwidth}
        \centering
        \caption{}
        \includegraphics[width=\textwidth]{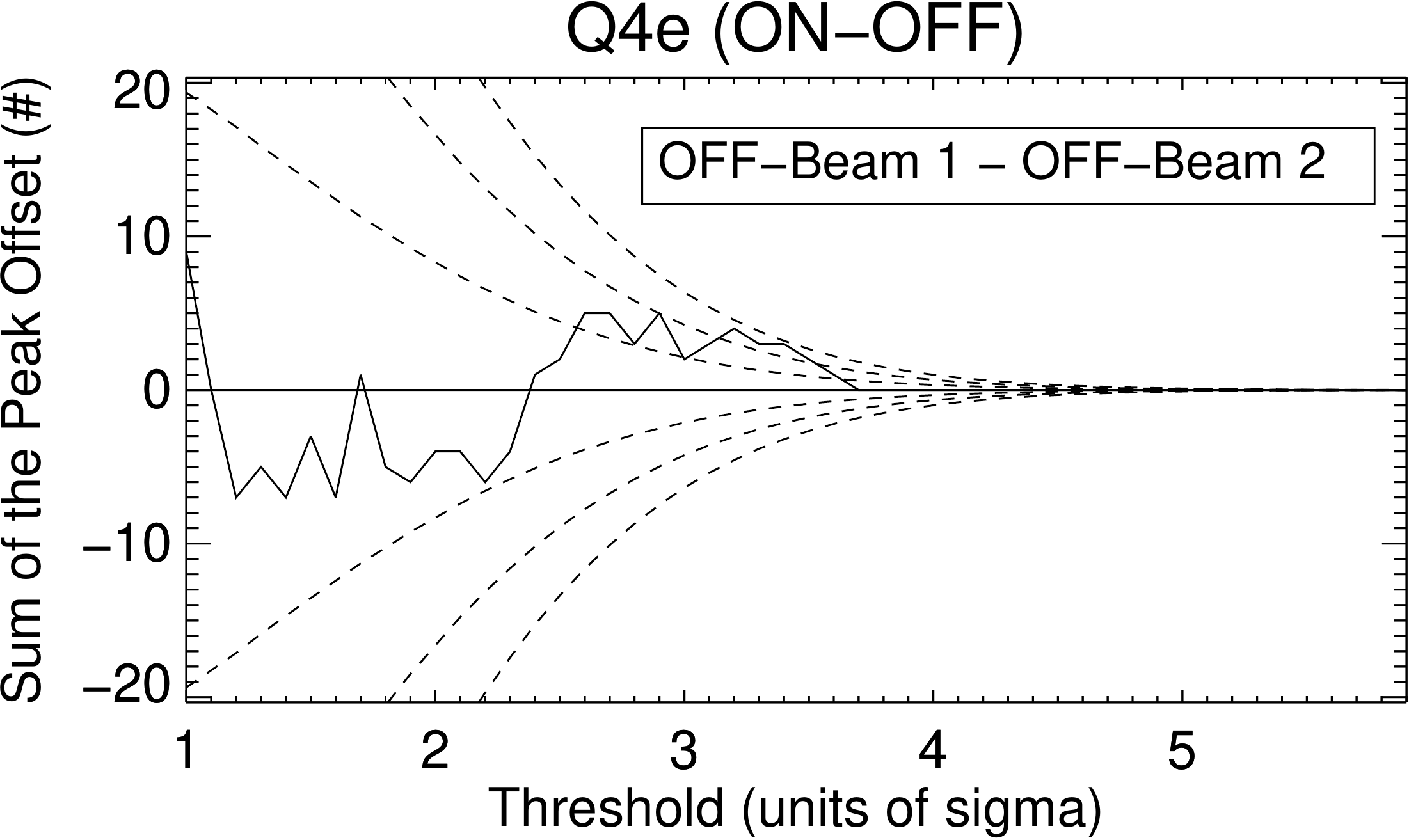}
        \label{}
    \end{subfigure}%
&
  \begin{subfigure}[c]{0.45\textwidth}
        \centering
        \caption{}
        \includegraphics[width=\textwidth]{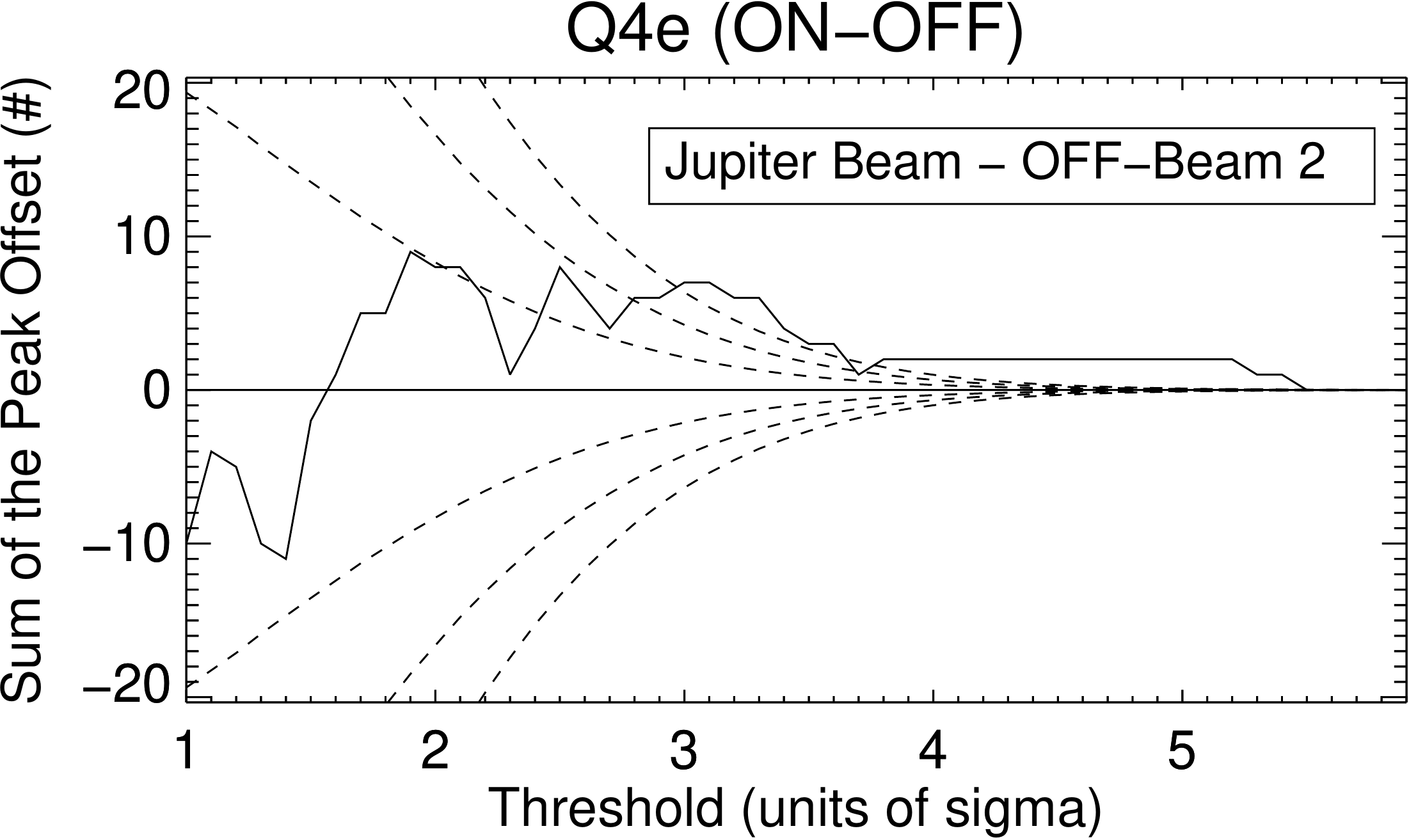}
        \label{}
    \end{subfigure}%
\\
  \begin{subfigure}[c]{0.45\textwidth}
        \centering
        \caption{}
        \includegraphics[width=\textwidth]{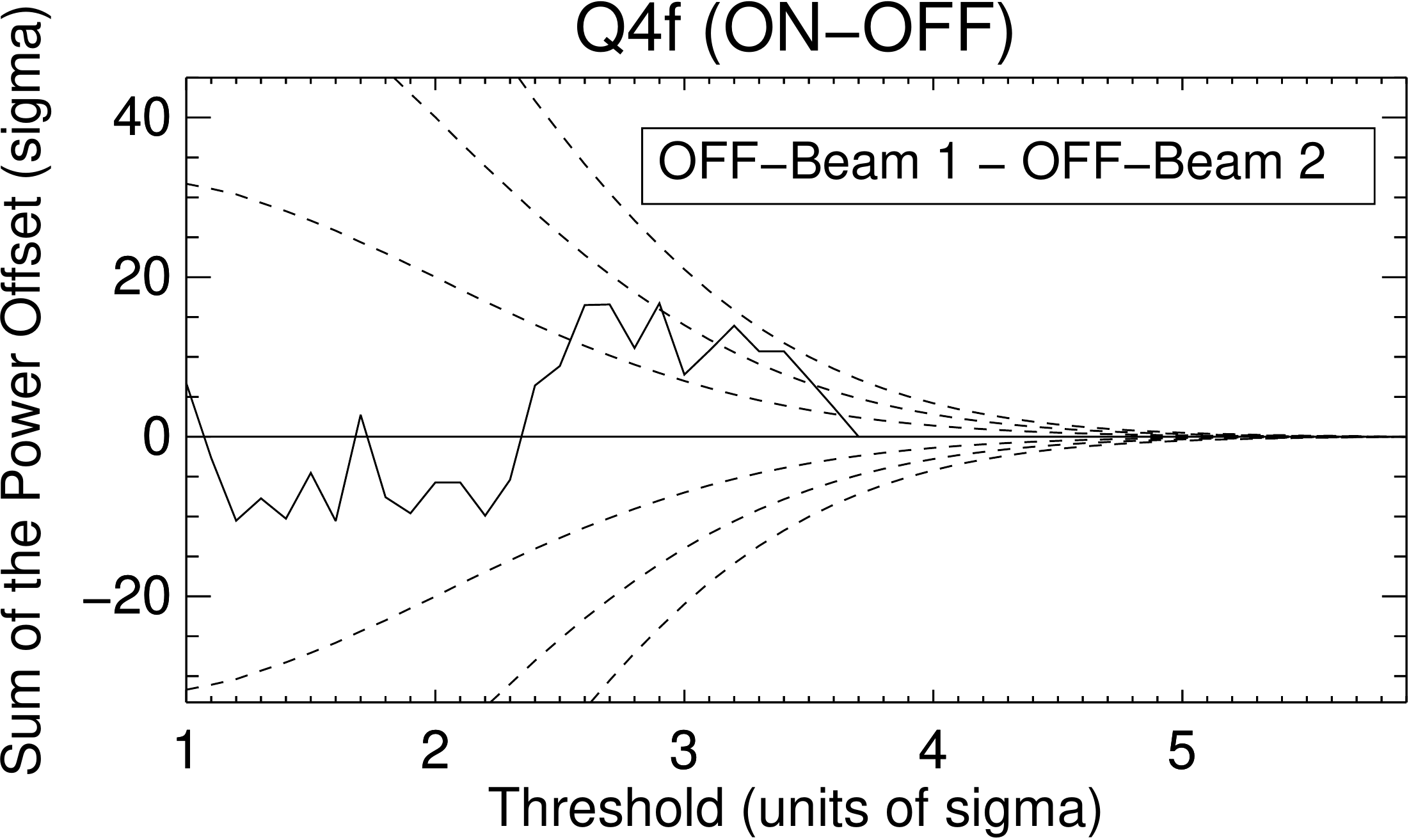}
        \label{}
    \end{subfigure}%
&
  \begin{subfigure}[c]{0.45\textwidth}
        \centering
        \caption{}
        \includegraphics[width=\textwidth]{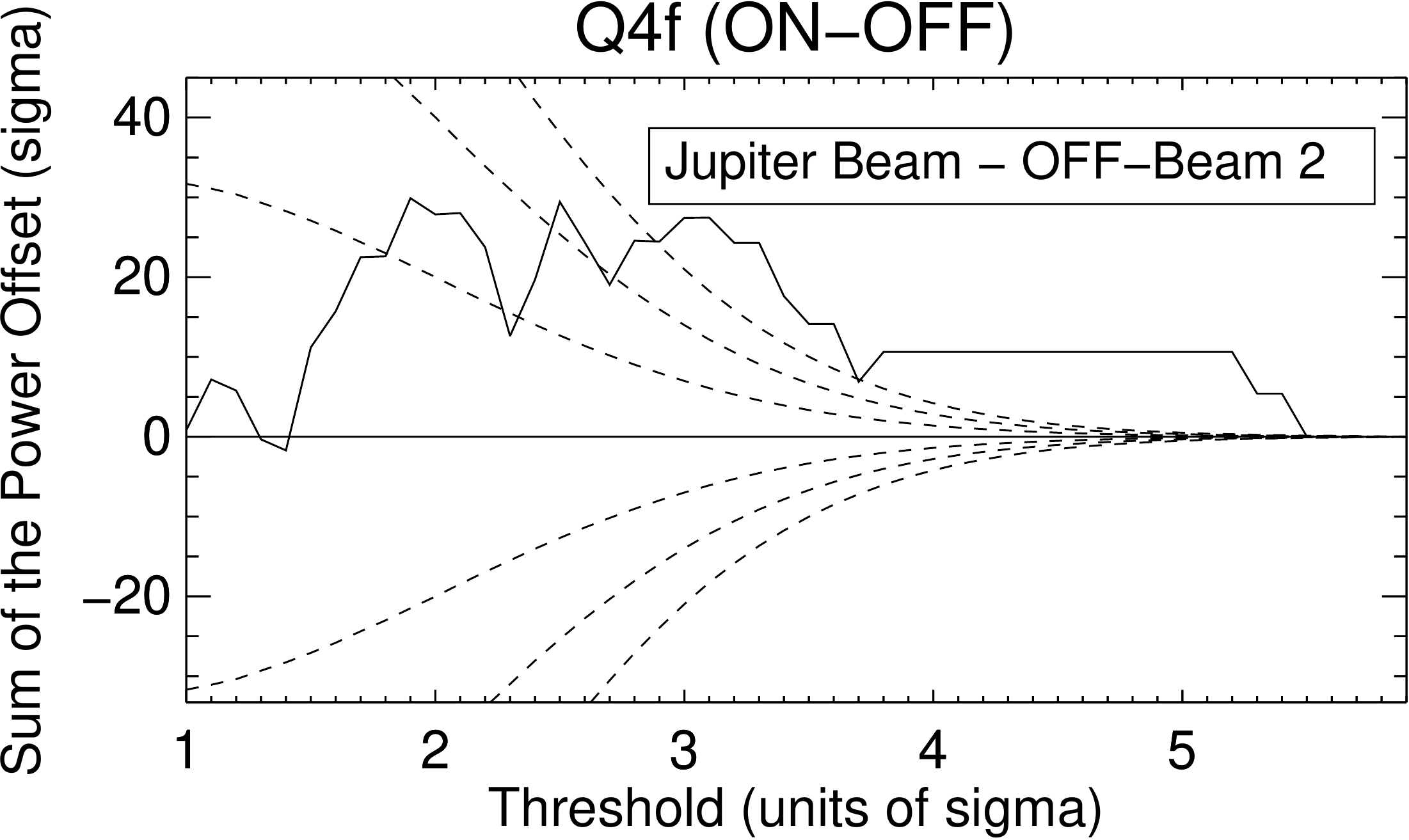}
        \label{}
    \end{subfigure}%

\end{tabular}
\end{center}
\caption{ Plots of Q2, Q4e (Peak Offset), and Q4f (Power Offset) showing the detection limit in Stokes-V (V$^{'}$+) with an $\alpha = 10^{-4.5}$ for the frequency range 53.5--56.5 MHz. \textbf{(a)} and \textbf{(b)} Q2 after elliptical correction. \textbf{(c)} and \textbf{(d)} Q4b difference of the two beams. \textbf{(e)} and \textbf{(f)} Q4f difference of the two beams. The comparison of the two OFF-beams from Obs~$\#$2 can be found in the left column (panels \textbf{a}, \textbf{c}, \textbf{e}) and the comparison of ON-beam (Jupiter) vs OFF-beam 2 can be found in the right column (panels \textbf{b}, \textbf{d}, \textbf{f}). The dashed lines for panels \textbf{(c-f)} are the 1, 2, 3 $\sigma$ statistical limits derived from two different Gaussian distributions (each run 10000 times). Panels \textbf{(d)} and \textbf{(f)} show an excess of ON vs OFF points at $\ge~3~\sigma$ statistical significance for signals up to a threshold of 6$\sigma$. For comparison, in panels \textbf{(c)} and \textbf{(e)} most the excess points are below the $2\sigma$ statistical significance level. We find by performing Gaussian simulations that the probability to get by chance a curve like panel \textbf{(e)} is $\sim$20$\%$, whereas the probability to obtain a curve like panel \textbf{(f)} is $\sim$10$^{-5}$. 
\\
}
\label{fig:Q4Detectionplot2}
\end{figure*}
\section{Data Analysis and Results} 
\label{sec:anaysis}

In this study, the analysis is performed using 11 different scaling factors ($\alpha$) between $10^{-2}$ to $10^{-7}$ in steps of $10^{+0.5}$. We use Jupiter emission from 15--25 MHz added to Obs~$\#$2 in 4 frequency ranges (20--30, 30--40, 40--50, 50--60 MHz). The comparison of the two OFF-beams in Obs~$\#$2 with each other is used as a benchmark for what could be considered a detection. This test proved to be highly important for Stokes-I as the OFF-beams contain non-Gaussian noise and there is unknown systematic noise (e.g. low-level RFI, non-corrected instrumental effects, ionospheric differences) in the data (e.g., see \citealt[Figure 4]{Turner2017pre8}).    

A summary of the parameters used in the post-processing can be found in Table \ref{tb:observable_setup}. The rebin time of the processed data ($\delta t$) is a very important parameter because this defines the timescale over which we search for excess peaks in Q2. The frequency ($\Delta \nu$) and time range ($\Delta T$) over which we calculate these observables is 10 MHz and 3 hours, respectively. Additionally, we include a threshold cut on the rebinned RFI mask. The rebinned mask no longer consists only of values of 0 (polluted pixels) and 1 (clean pixels) since it was rebinned and clean pixels were mixed with polluted pixels. The mask threshold we use in our analysis is 90$\%$, meaning a pixel will not be used in the analysis if $\ge 10\%$ of the original pixels were contaminated. We use a time interval ($TI$) of 2 minutes and a frequency interval ($FI$) of 0.5 MHz for Q1b.

\begin{table}[!ht]
\centering
\caption{Nominal parameters for the post-processing setup}
\begin{tabular}{ccc}
\hline 
\hline
Parameter & Value   & Units  \\ 
\hline
\hline 
Width of frequency range ($\Delta \nu$)                    & 10       & MHz \\
Time Range ($\Delta T$)                                 & 3        & hours\\
Rebin time of processed data ($\delta \tau$)    & 1        & secs\\
Mask threshold                               & 90       & $\%$\\
Time interval ($TI$)                         & 2        & minutes \\
Frequency interval ($FI$)                    & 0.5     & MHz \\
Low-pass filter smoothing window ($w$)       & 10       & secs\\
Threshold ($\tau$) range                     & 1 - 6    & sigma \\
\hline
\end{tabular}
\label{tb:observable_setup}
\end{table}  

Figs. \ref{fig:dynspec_Q1}, \ref{fig:AllQ}, and \ref{fig:Q3} show the observable quantities Q1, Q2, Q3a, and Q4 in Stokes-V for a value of $\alpha = 10^{-4}$. This test case is very useful to demonstrate how each observable behaves. In this case the ON-beam can be seen to have additional flux when compared to the OFF-beam in all the Q values (Figs. \ref{fig:dynspec_Q1}, \ref{fig:AllQ}, \ref{fig:Q3}). For Q3a, it can be seen that Jupiter's emission is mainly localized between 1.2--1.4 UT and 3.2--3.9 UT (Fig. \ref{fig:Q3}a) where emission around 1.7 UT can be seen in both OFF beams (Fig.~\ref{fig:Q3}b). This is a good example demonstrating that two OFF beams are required to confirm a detection.

The extended emission observables Q1a and Q1b are very useful for Stokes-V. For Stokes-I they are useful only when the simulated exoplanet emission is very bright ($\alpha = 10^{-2} - 10^{-3}$) and can be seen by eye in the processed dynamic spectrum. Additionally, when we run the analysis of $V^{'}+$ and V$^{'}-$ the right-hand and left-hand polarizations are easily separated. The dominant source of variations in Q1a and Q1b are changes in the ionosphere. Ionospheric variations are the limiting factor in distinguishing real emission from background variations.    

The observables Q2 - Q4 are more effective at detecting fainter burst emission. The best observables to detect the faintest emission are Q4e and Q4f (Peak/Power Offset). We can reliably detect emission from Jupiter down to a value of $\alpha = 10^{-3.5}$ for Stokes-I and $\alpha = 10^{-4.0}$ for Stokes-V with the elliptical correction when adding Jupiter to the range 50 - 60 MHz. For the Stokes-V detection limit ($\alpha = 10^{-4.0}$), additional flux in the ON-beam can be seen in all Q values (Figs. \ref{fig:dynspec_Q1}, \ref{fig:AllQ}, \ref{fig:Q3}). Fig. \ref{fig:Q4Detectionplot} shows the Stokes-I analysis for $\alpha = 10^{-3.5}$ for both the ON- vs. OFF-beam and OFF-beam 1 vs OFF-beam 2. The main criteria we use to confirm a detection are (1) Q4f is distinctly different than the OFF-beam 1 vs. OFF-beam 2 comparison plot (Fig. \ref{fig:Q4Detectionplot}g), (2) Q4f shows an excess $\ge 2\sigma$ statistical significance (dashed lines in Fig. \ref{fig:Q4Detectionplot}), and (3) the detection curve is always positive for thresholds between $1.5-4.5\sigma$. \\
\indent The detection limits for each frequency range are summarized in Table \ref{tb:scaling}. We get the most constraining detection limit for the frequency range 50 - 60 MHz. Our detection limits for 40 - 50 MHz and 30 - 40 MHz are half an order of magnitude less sensitive, where the detection limit for 20 - 30 MHz is an order of magnitude less sensitive. This is expected since the frequency-response curve of LOFAR sharply peaks at 58 MHz and has only limited sensitivity at the lowest frequency range (e.g. Figure 1 in \citealt{Turner2017pre8}).\\
\indent Next, we test the robustness of the detection limits by varying the parameters of the post-processing from those in Table \ref{tb:observable_setup}. We vary the rebin time of processed data ($\delta \tau$), the smoothing window ($w$), the value of the slope for the Peak/Power Offset (Q4e and Q4f), the frequency range, and the time range. Our detection limit for Stokes-I did not significantly improve when we varied these parameters. For Stokes-V, we improved our detection limit to $\alpha = 10^{-4.5}$ (half an order of magnitude) when examining the $V^{'}+$ data from 3--4 UT and over the frequency range 53.5--56.5 MHz (Figure \ref{fig:Q4Detectionplot2}). Therefore, our detection limit is fairly robust against the exact parameters used in the analysis but can be improved with a thorough frequency and time search. The signal from Jupiter is detected with Q4f until the data is binned to a $\delta \tau$=30 seconds for Stokes-I and $\delta \tau$=4 minutes for Stokes-V. For Q1a (the time-series), we can detect signal for Stokes-V up to a $\delta \tau$=30 minutes but with lower significance ($\sim 2\sigma$). Therefore, assuming that radio emission from an exoplanet is similar to Jupiter's, searching over a variety of timescales with Q1a and Q4f will be helpful for a detection. This result also indicates that our method of analysis for beam-formed data can be applied to various setups of beam-formed observations and dynamic spectra extracted from rephased calibrated visibilities of imaging pipelines (Loh et al. in prep). \\
\indent Furthermore, we tested whether the date of observation or the position on the sky has a noticeable effect in our detection limits. For Obs~$\#$3, we find detection limits for Stokes-I that are half an order of magnitude less sensitive from those found using Obs~$\#$2. However, for Stokes-V we obtain similar detection limits to Obs~$\#$2. Next, performing the analysis on Obs~$\#$4 we find detection limits that are similar to Obs~$\#$2. Therefore, our detection limits (Table \ref{tb:scaling}) are also insensitive to where in the sky we are pointed at, provided that the observations were taken under good conditions.\\
\begin{table}[!ht]
\centering
\caption{Summary of the scaling factor upper limits found in the analysis}
\begin{tabular}{ccc}
\hline 
\hline
Frequency Range (MHz) & Stokes-I $\alpha$  & Stokes-V $\alpha$   \\
\hline
\hline
\multicolumn{3}{c}{Obs~$\#$2} \\
\hline
50 - 60                 & 10$^{-3.5}$       & 10$^{-4.5}$\\
40 - 50                 & 10$^{-3}$         & 10$^{-4}$ \\
30 - 40                 & 10$^{-3}$         & 10$^{-4}$ \\
20 - 30                 & 10$^{-2.5}$         & 10$^{-3.5}$\\
\hline
\multicolumn{3}{c}{Obs~$\#$3} \\
\hline
50 - 60                 & 10$^{-3}$         & 10$^{-4.5}$ \\
40 - 50                 & 10$^{-2.5}$       & 10$^{-4}$\\
30 - 40                 & 10$^{-2.5}$       &10$^{-4}$\\
20 - 30                 & 10$^{-2.0}$       &10$^{-3.5}$ \\
\hline
\multicolumn{3}{c}{Obs~$\#$4} \\
\hline
50 - 60                 & 10$^{-3.5}$  & 10$^{-4.5}$\\
40 - 50                 & 10$^{-3}$    & 10$^{-4}$\\
30 - 40                 & 10$^{-3}$   & 10$^{-4}$\\
20 - 30                 & 10$^{-2.5}$  & 10$^{-3.5}$\\
\hline
\end{tabular}
\label{tb:scaling}
\end{table}

\indent Finally, we quantify the statistical significance of our detection limits. First, we normalize the observable Q4f (ON-OFF, i.e. the solid line in Fig.~\ref{fig:Q4Detectionplot}h) by the statistical limit (1 $\sigma$, i.e. the first dashed line in Fig.~\ref{fig:Q4Detectionplot}h) and calculate the average value of Q4f for threshold values between 1.5 and 4.5. With this, we obtain an average value $<$Q4f$>$ of 2.23 for Stokes-I and 4.34 for Stokes-V. Next, we compare these values to those obtained in the case when both the ON- and the OFF-beam only contain random noise. We generate random distribution of points for the ON- and OFF-beam (generating an artificial equivalent of Q2), and calculate Q4f and $<$Q4f$>$. We find that the probability of a false positive for obtaining a signal like Jupiter is 1.4$\times10^{-5}$ for Stokes-V for $\alpha=10^{-4.5}$ at 53.5-56.5 MHz and 3.2$\times10^{-4}$ for Stokes-I for $\alpha = 10^{-3.5}$ at 50-60 MHz. This corresponds to a statistically significant detection of 3.6$\sigma$ for Stokes-I and 4.3$\sigma$ for Stokes-V. As a final step, we compare these values to those obtained in observations without any astrophysical signal (i.e. when comparing the two OFF beams, Figs.~\ref{fig:Q4Detectionplot}g and \ref{fig:Q4Detectionplot2}e). In that case, we find that the false positive rate is 99$\%$ for Stokes-I and 20$\%$ for Stokes-V. Therefore, these observations are thus classified as non-detections.

\section{Discussion}\label{sec:discussion} 
In the following, we will compare the detection limits for two of our observables: Q1a and Q4f. We can estimate an upper limit from 50--60 MHz using the Stokes-V Q1a (time-series) observable (e.g. Fig. \ref{fig:dynspec_Q1}b). The standard deviation of the difference of the two sky beams is $3.7\times10^{-5}$ of the theoretical SEFD. Therefore, the 1-sigma sensitivity from these observations would be 62 mJy using a SEFD for 24 stations of 1.7 kJy (\citealt{vanHaarlem2013}). This flux density is $\sim$1.3 times higher than the sensitivity expected for LOFAR beam-formed observations:
\begin{align}
      \sigma_{LOFAR}  &=  \frac{S_{S} }{N\sqrt{ b \tau }}, \label{eq:sigmat}\\
      \sigma_{LOFAR}  &= \frac{ S_{S} }{ 24\sqrt{2 mins \times 10 MHz }} =  48 \ \text{mJy},
\end{align}
where $S_{S}$ is the SEFD with a value of 40 kJy (\citealt{vanHaarlem2013}). \citet{Turner2017pre8} found that the Stokes-I sensitivity using this same method was 850 mJy due to fluctuations in the ionosphere ($\sim$50 times the theoretical sensitivity). If we rebin Q1a to longer timescales the standard deviation decreases with $\sim$~$t^{-1/2}$ white noise dependence. Additionally, we find no evidence of red noise in the Stokes-V time-series using the time-averaging method (\citealt{Pont2006}; \citealt{Turner2016}). Performing the same procedure in Stokes-I, \citet[Fig. 4]{Turner2017pre8} found that there was a great deal of red noise in the time series (RMS of red noise $\sim$ 0.5 RMS of white noise) due to non-Gaussian ionospheric variations between the two beams.  

We demonstrated that we can detect the Jupiter signal down-scaled by a factor $\alpha=10^{-4.5}$ for Stokes-V in all Q values but with the largest significance in Q4f (Fig \ref{fig:Q4Detectionplot2}f). Our detection in Q4f consists of $\sim$30 data-points in the NDA calibration data exceeding $3\times10^{4}$ Jy with a threshold $\ge 2\sigma$ (Fig \ref{fig:Q4Detectionplot2}d). Therefore, this limit corresponds to a flux density of $\sim\alpha\times4 \times 10^4$ Jy = 1.3 Jy using the value of Jupiter's absolute flux density corresponding to 30 data-points from Fig. \ref{FigJ20170211}c. We also find that this flux density is $\sim$1.3 times the theoretical sensitivity ($\sigma_{LOFAR}$ = 1 Jy) expected for LOFAR beam-formed observations using equation \eqref{eq:sigmat} when $\tau$ = 1 sec and $b$ = 3 MHz. This factor of 1.3 is mostly due to ionospheric variations that were not mitigated during the post-processing and partly due to the fact that our criteria for a burst detection is a statistical significance $\ge 2\sigma$ (Section \ref{sec:anaysis}; Fig. \ref{fig:Q4Detectionplot}h). 

Our comparison shows that for both Q1a and Q4f we are at 1.3$\times$ the theoretical sensitivity of LOFAR. Both observables are complementary to each other and should be used in parallel since they probe different timescales and emission structures. 

\begin{table*}[!htb]
\centering
\caption{Detection limit of LOFAR LBA beam-formed observations found by observing ``Jupiter as an exoplanet''}
\begin{tabular}{cccc}
\hline 
\hline
 $S_{J}(\text{ref})$ [Jy at 5 AU] & Distance [pc] & Stokes-I $\alpha_{J}$  & Stokes-V  $\alpha_{J}$\\
\hline
\hline
4$\times 10^{4}$\ (a)                    &  5             & 1$\times10^{7} $  & 1$\times10^{6}$ \\
''                                    &  10             & 4$\times10^{7} $    & 4$\times10^{6}$\\
''                                    &  20             & 2$\times10^{8}$     & 2$\times10^{7}$\\ 
\hline
4$\times 10^{5}$\ (b)                       &  5           & 1$\times10^{6}$ & 1$\times10^{5}$\\  
''                                    &  10             & 4$\times10^{6}$    & 4$\times10^{5}$\\
''                                    &  20             & 2$\times10^{7}$    & 2$\times10^{6}$\\
\hline
6 $\times 10^{6}$\ (c)                    &  5           &  6$\times10^{4}$   & 6$\times10^{3}$\\
''                                   &  10              & 3$\times10^{5}$    &  3$\times10^{4}$\\
''                                   &  20              & 1$\times10^{6}$   & 1$\times10^{5}$\\
\hline
\end{tabular}
\tablefoot{All calculations were done with equation \eqref{eq:alpha} where the scaling factor $\alpha = 10^{-3.5}$ for Stokes-I and $\alpha = 10^{-4.5}$ for Stokes-V and $S_{J}(obs) = 3\times10^{4}$ Jy (Sect. \ref{sec:scaleJup}, Figure \ref{FigJ20170211}). \textbf{(a)} The level of Jupiter's burst emission exceeded in $\ge$50$\%$ of Jupiter bursts (\citealt[Figure 7]{Zarka2004}). \textbf{(b)} The mean level of Jupiter's burst emission exceeded in $\sim1\%$ of Jupiter bursts. \textbf{(c)} Maximum peak of Jupiter's S-burst emission (\citealt{Queinnec2001}). }
\label{tb:detectionlimits}
\end{table*}

One may wonder why bothering with the complex observables to achieve the sensitivity expected for beam-formed observations. The answer is that they allow us to detect confidently a signal and distinguish it from false positives at a 1.5--2$\sigma$ level, whereas simple detection of a spike in beam-formed data requires generally a $\sim10\sigma$ level to be considered as reliable. Thus we actually gain a factor $>5$ in effective sensitivity (detection capability) with our method. Also, our method allows for the detection of relatively sparse and short bursts that would be washed out by averaging over long integrations.

The $\alpha$ value found in our analysis can be decomposed into three separate physical factors (strength of emission compared to Jupiter, relative Jupiter flux levels, and distance): 
\begin{align}
   \alpha &= \alpha_{J} \left(\frac{S_{J}[\text{ref}]}{S_{J}[\text{obs}]}\right) \left( \frac{5 \ AU}{d}\right)^2, \notag \\
          &= \alpha_{J} \left(\frac{S_{J}[\text{ref}]}{S_{J}[\text{obs}]}\right) \left( \frac{2.4\times10^{-5} \ pc}{d}\right)^2,  \label{eq:alpha}
\end{align}
where $\alpha_{J}$ is the scaling factor of the emission compared to Jupiter, $d$ is the distance, $S_{J}(\text{obs})$ is the flux density of the observed Jupiter signal in Obs~$\#$1 calibrated using NDA, and $S_{J}(\text{ref})$ is a reference flux density value of Jupiter to which the putative exoplanet signal is compared. The Jupiter signal ($S_{J}[\text{obs}]$) of $\sim3 \times 10^4$ Jy is more than a factor 100 below the peak value reached by Jupiter's decametric emission (up to $6 \times 10^6$ Jy; \citealt{Queinnec2001}) observed from the Earth, at 5 AU range. For $S_{J}(\text{ref})$, we use Jupiter's radio emission levels and occurrence rates given in \citet[Figure 7]{Zarka2004}. Jupiter does emit decameter emission quasi-continuously but the most energetic emission can be found in bursts. During a fairly active emission event, the median flux density of Jupiter's decametric bursts at 5 AU is $\sim4\times10^{5}$ Jy. This flux density is exceeded by $\sim1\%$ of all detected Jupiter bursts, whereas the level $\sim4\times10^{4}$ Jy is exceeded by $\ge$ 50$\%$ of Jupiter bursts.

We find that we can detect an exoplanetary polarized signal intrinsically $10^{5}$ times stronger than Jupiter's emission strength from a distance of 5 pc using equation \eqref{eq:alpha} and taking the mean level of Jupiter's decametric bursts as the reference flux ($S_{J}[\text{ref}]=4\times10^{5}$ Jy) that would occur for a few minutes within an observation of a few hours. A stronger signal may be detected more often, a weaker one more rarely. In Table \ref{tb:detectionlimits}, we show the values of $\alpha_{J}$ that would be required for the emission to be detectable in Stokes-I and Stokes-V (circular polarization), respectively. 




Such signals are indeed expected to exist. According to most models, the strongest emission up to $10^{6-7}$ times Jupiter's radio emission is expected for close-in planets, especially massive hot Jupiters \citep{Zarka2001,Zarka2007,Gr2007,Griessmeier11RS}. However, rapidly rotating planets with strong internal plasma sources have also been suggested to produce radio emission at detectable levels at orbital distances of several AU from their host star \citep{Nichols2011,Nichols2012}. Furthermore, the expected radio flux is a function of the age of the exoplanetary host star, with stronger radio signals being expected for planets around young stars \citep{Stevens2005,Griessmeier05AA,Griessmeier07PSS,Gr2007}, and for planets around stars with frequent and powerful coronal mass ejections \citep{GriessmeierPREVI,Griessmeier07PSS,Gr2007}.

Sources beyond 20 pc would need to be extremely intense ($\ge 10^{7}\times$ Jupiter's mean level of burst emission; Table \ref{tb:detectionlimits}), and may be beyond the reach of LOFAR. If the structure of the emission is different from that of Jupiter bursts (e.g. longer bursts of several minutes), the above sensitivity may be improved by an order of magnitude or more.

Finally, let us mention that detection of a radio signal from an exoplanetary system will only constitute the first step. Even though the planetary emission is expected to be much stronger than the stellar emission \citep[see e.g.][]{Griessmeier05AA}, one would have to confirm that the signal is indeed produced by the exoplanet rather than its host star. 
The most direct indication would be the detection of radio emission from a transiting planet, with the planetary emission disappearing during secondary eclipses.
Secondly, stellar and planetary radio emission have different polarization properties \citep{Zarka1998}, making polarization a very powerful tool even beyond signal detection.
Thirdly, one would have to search for a periodicity in the detected signal, and compare its period to the stellar rotation period \citep[or, more precisely, the beat period between the stellar rotation and the planetary orbit, see e.g.][]{Fares10}, and (if known) the planetary rotation period.

Ancillary data which would help with the interpretation of a radio signal include: stellar lightcurves (correlation with stellar flares), stellar magnetic field maps (e.g. obtained by Zeeman-Doppler-Imaging), the stellar rotation rate, data on the stellar wind (e.g. obtained by astrospheric absorption) or at least a good estimation of the stellar age, the exoplanet's orbital inclination \citep[see][]{Hess2011} and the planetary rotation rate.

\section{Conclusions and perspectives} 

Our analysis shows that our pipeline for beam-formed LOFAR Stokes-V data can detect signals of $10^{-4.5}$ times the intensity of Jupiter's polarized emission. This corresponds to either a Jupiter-like planet at a distance of 13000 AU, or an exoplanet with $10^{5}$ times Jupiter's mean radio flux for strong burst emission ($4\times 10^5$ Jy; \citealt{Zarka2004}) at a distance of 5 pc (Table \ref{tb:detectionlimits}). According to frequently employed scaling laws (e.g. \citealt{Zarka2001,Zarka2007,Gr2007}; Zarka et al. 2018, submitted), one can expect exoplanetary radio emission up to $10^{6-7}$ times Jupiter's flux. This also means our pipeline could potentially detect radio emission from the exoplanets 55 Cnc (12 pc), Tau Bo\"{o}tis (16 pc), and Upsilon Andromedae (13 pc) if their emission can reach $10^{5}$ times the peak flux value reached by Jupiter's decametric burst emission ($\sim6 \times 10^6$ Jy; \citealt{Queinnec2001}). We have observed all these planets using LOFAR; the analysis using this pipeline is currently on-going, and will be the subject of a follow-up article.\\
\indent In this study, we present the Stokes-V processing and post-processing extension of our beam-formed reduction pipeline (\citealt{Turner2017pre8}). We show that the Stokes-V pipeline can reach $\sim$1.3 times the theoretical sensitivity of LOFAR. With the post-processing improvement our pipeline can now be applied to various setups of beam-formed data from different telescopes (e.g. LOFAR, UTR-2) and dynamic spectra extracted from radio imaging observations (Loh et al. in prep).\\
\indent On a slightly longer timescale, NenuFAR \citep{Zarka2012,nenufar} will allow more sensitive observations, with an improvement in sensitivity by a significant factor compared to LOFAR's core below 35 MHz. This is precisely the frequency range where we believe most exoplanetary systems will emit.\\
\indent The Square Kilometer Array (SKA) will be even more sensitive (with an improvement in sensitivity by a factor $\sim$30 compared to LOFAR; \citealt{Zarka2015SKA}). It will only observe at frequencies above 50 MHz, but there are cases where exoplanetary radio emission is expected to extend to frequencies of a few 100 MHz. This is the case for young and massive planets \citep{Griessmeier18Handbook} as well as in the case of a unipolar induction mechanism between a hot Jupiter and its parent star \citep{Zarka2007}, making the SKA a promising instrument for exoplanet radio studies \citep{Zarka2015SKA,Griessmeier18Handbook}.\\
\indent Besides improvements in telescope sensitivity, many more nearby exoplanets with short orbital periods are likely to be discovered by the upcoming Transiting Exoplanet Survey Satellite mission (TESS; \citealt{Ricker2015}) and ground-based transit surveys such as the Next-Generation Transit Survey (NGTS; \citealt{Wheatley2017}) and the Kilodegree Extremely Little Telescope (KELT; \citealt{Pepper2007}). For example, TESS is predicted to find hundreds of planets within 50 pc and a dozen exoplanets within 10 pc (\citealt{Sullivan2015}). These new exoplanets may be good candidates for the exoplanetary radio emission search because our detection capability is strongly dependent on distance (Equation \ref{eq:alpha}; Table \ref{tb:detectionlimits}).

\section*{Acknowledgements} 
J. Turner was funded by the National Science Foundation Graduate Research Fellowship under Grant No. DGE-1315231. We also thank PNP and PNST for their support. 

This research has made use of the Exoplanet Orbit Database (\citealt{Wright2011exo}), Exoplanet Data Explorer at exoplanets.org (\citealt{Han2014}), Extrasolar Planet Encyclopaedia (exoplanet.eu) maintained by J. Schneider (\citealt{Schneider2011}), and NASA's Astrophysics Data System Bibliographic Services. 

This paper is based (mostly) on data obtained with the International LOFAR Telescope (ILT) under project codes LC2$\_$018, LC5$\_$DDT$\_$002, LC6$\_$010, and LC7$\_$013. LOFAR (\citealt{vanHaarlem2013}) is the Low Frequency Array designed and constructed by ASTRON. It has observing, data processing, and data storage facilities in several countries, that are owned by various parties (each with their own funding sources), and that are collectively operated by the ILT foundation under a joint scientific policy. The ILT resources have benefitted from the following recent major funding sources: CNRS-INSU, Observatoire de Paris and Universit\'{e} d'Orl\'{e}ans, France; BMBF, MIWF-NRW, MPG, Germany; Science Foundation Ireland (SFI), Department of Business, Enterprise and Innovation (DBEI), Ireland; NWO, The Netherlands; The Science and Technology Facilities Council, UK. We use the TGSS survey (\citealt{Intema2017}) in our study when determining the locations of the OFF-beams and we thank the staff of the GMRT that made these this survey possible. GMRT is run by the National Centre for Radio Astrophysics of the Tata Institute of Fundamental Research. NDA Jupiter observations were used for the flux calibration in our analysis. The NDA is hosted by the Nan\c{c}ay Radio Observatory/ Unit\'{e} Scientique de Nan\c{c}ay of the Observatoire de Paris (USR 704-CNRS, supported by Universit\'{e} d'Orl\'{e}ans, OSUC, and Region Centre in France).

We also thank the anonymous referee for their helpful comments during the publication process.

\bibliographystyle{aa} 
\bibliography{reference.bib} 


\begin{appendix}
\section{Jupiter Scaling Derivation} \label{app:Jupiter}

\subsection{Scaling Jupiter's Signal in Total Intensity}

When observing the sky with a radio telescope, we measure, for a signal of antenna temperature $T_{A}$, a specific intensity $I$ proportional to the received power
\begin{align}
    I  =& \frac{2k}{\lambda^2} T_{A},\label{eq:Inu}
\end{align}
where $\lambda$ is the wavelength of interest. The \textit{unpolarized} flux density $S$ for an unresolved source is
\begin{align}
    S = \frac{2k T_{A}}{A_{e}}, \label{eq:Snu}
\end{align}
where $A_{e}$ is the effective area of the telescope, i.e. $A_{e} = \lambda^2/\Omega$, and $\Omega$ is the solid angle of the telescope beam in the approximation where the main beam largely dominates. The flux density measured is independent of the radio telescope performing the measurement.

When observing the Galaxy (sky background) with a radio telescope the intensity $I_{S}$ would be 
\begin{align}
    I_{S} =  \frac{2k}{\lambda^2} T_{SG}, 
\end{align}
where $T_{SG}$ is system noise temperature for an observation of the Galaxy. $T_{SG}$ is the sum of the noise contributions in the beam
\begin{align}
    T_{SG} =  T_{G} + T_{i},
\end{align}
where $T_{G}$ is the antenna temperature measured for the Galaxy, i.e. $T_{G} = 60 K \lambda^{2.55}$, and $T_{i}$ is the instrumental noise. By definition the System Equivalent Flux Density ($S_{S}$) would be 
\begin{align}
    S_{S} = \frac{2 k T_{SG}}{A_{e}} = \frac{2 k \left(T_{G} + T_{i}\right)}{A_{e}}.  \label{eq:SEFD}
\end{align}
The units of $S_{S}$ are in Jy. Then, the background intensity of the sky $I_{S}$ measured in the data would be 
\begin{align}
    I_{S} = \frac{2k}{\lambda^2}  \left(T_{G} + T_{i}\right). \label{eq:IS}
\end{align}
If we compare two different sky observations (i.e. $I_{S1}$ and $I_{S2}$) with different instruments and at different wavelengths we have 
\begin{align}
    \frac{I_{S1}}{I_{S2}} = \left(\frac{\lambda_{2}}{\lambda_{1}}\right)^2  \left( \frac{T_{G1} + T_{i1}}{T_{G2} + T_{i2}}\right).  \label{eq:IScompare}
\end{align}

When observing Jupiter with a radio telescope, the flux density $S_{J}$ would be 
\begin{align}
    S_{J} = \frac{2k T_{SJ}}{A_{e}},
\end{align}
where $T_{SJ}$ is the system noise temperature of an observation of Jupiter. The system noise temperature $T_{SJ}$ now includes contributions from Jupiter, the Galaxy, and the instrument
\begin{align}
    T_{SJ} = T_{AJ} + T_{G} + T_{i},
\end{align}
where $T_{AJ}$ is the observed antenna temperature for Jupiter. $T_{AJ}$ is $>>$ than both $T_{i}$ and $T_{G}$, therefore, the flux density $S_{J}$ becomes 
\begin{align}
    S_{J} = \frac{2k T_{AJ}}{A_{e}}.
\end{align}
Hence, the intensity of Jupiter observed by a radio telescope would be
\begin{align}
    I_{J} = \frac{ S_{J} A_{e}}{\lambda^2}. \label{eq:IJ}
\end{align}
If we compare the intensities of two observations of Jupiter (i.e. $I_{J1}$ and $I_{J2}$) using different instruments and at different wavelengths we have 
\begin{align}
    \frac{I_{J1}}{I_{J2}} = \left( \frac{\lambda_{2}}{\lambda_{1}}\right)^2 \left(\frac{A_{e1}}{A_{e2}}\right).  \label{eq:IJcompare}
\end{align}

We consider an observation of Jupiter (Obs~$\#$1) with instrument 1 ($I_{J1}$) in a given frequency range (e.g. 16-26 MHz) and an observation of the sky background (Obs~$\#$2) with instrument 2 ($I_{S2}$) in an arbitrary frequency range (e.g. 50-60 MHz). The instruments and frequency ranges in these observations do not have to be the same. The goal is to synthesize a signal ($I_{sim}$) with the sky background ($I_{S2}$) from Obs~$\#$2 plus the Jupiter signal as it would have been observed with instrument 2 ($I_{J2}$) and attenuated by a factor $\alpha$
\begin{align}
    I_{sim} = I_{S2} + \alpha I_{J2}. \label{eq:Inew_Begin}
\end{align}
Therefore, we have 
\begin{align}
     I_{sim} =& I_{S2} \left[1 + \alpha \left(\frac{I_{J2}}{I_{S2}}\right) \right],\\
           =& I_{S2} \left[ 1+ \alpha \left(\frac{I_{J2}}{I_{J1}}\right) 
                                       \left(\frac{I_{J1}}{I_{S1}}\right) 
                                       \left(\frac{I_{S1}}{I_{S2}}\right)\right], \label{eq:Inew_1}
\end{align}
where $I_{S1}$ is the sky background in Obs~$\#$1. $I_{S1}$ has to be measured in an OFF-beam in Obs~$\#$1 since the Jupiter emission in the ON-beam is so immense. By using equation \eqref{eq:IScompare} for the sky background ratio and equation \eqref{eq:IJcompare} for the Jupiter signal ratio, we find that $I_{sim}$ is equal to 
\begin{align}
    I_{sim} =& I_{S2} \left[ 1+ \alpha  \left(\frac{I_{J1}}{I_{S1}}\right) \left(\frac{T_{G1} + T_{i1}}{A_{e1}}\right) \left( \frac{A_{e2}}{T_{G2} + T_{i2}}  \right) \right], \\
      I_{sim} =& I_{S2} \left[ 1+ \alpha  \left(\frac{I_{J1}}{I_{S1}}\right)  \left(\frac{S_{S1}}{ S_{S2}}\right) \right]. \label{eq:Inew_Final}
\end{align}
Jupiter's intensity $I_{J1}$ and the intensity of the sky $I_{S1}$ in Equation \eqref{eq:Inew_Final} can be measured directly from the data in Obs~$\#$1 (Section \ref{sec:JupSig}). Equation \eqref{eq:Inew_Final} is used as Equation \eqref{eq:transfer_I} in the main text.

\subsection{Scaling Jupiter's Signal in Polarization}

LOFAR measures directly the 4 Stokes parameters: Stokes-I, Stokes-Q, Stokes-U, and Stokes-V. The procedure for finding the synthesized Stokes-V signal ($V_{sim}$) is similar to the Stokes-I derivation shown above 
\begin{align}
    V_{sim} = V_{S2} + \alpha V_{J2}, \label{eq:Vnew_begin}
\end{align}
where $V_{S2}$ is the measured Stokes-V intensity in the sky background in Obs~$\#$2 and $V_{J2}$ is the Stokes-V Jupiter signal scaled as if it was observed in Obs~$\#$2. The polarized fraction in circular polarization ($v$) is 
\begin{align}
    v = \frac{V}{I}.
\end{align}

Throughout the analysis we assume that the polarized fraction is independent of frequency and instrument (i.e. $v_{J} = v_{J1} = v_{J2}$). Therefore, $V_{sim}$ in equation \eqref{eq:Vnew_begin} becomes 
\begin{align}
    V_{sim} = v_{S2} I_{S2} + \alpha v_{J2} I_{J2} = I_{S2}\left(v_{S2} + \alpha v_{J} \frac{I_{J2}}{I_{S2}}  \right). \label{eq:Vnew_step}
\end{align}

We can solve for the ratio in Equation \eqref{eq:Vnew_step} by using the derivations found in Equations \eqref{eq:IJcompare} and \eqref{eq:IScompare} 
\begin{align}
  \frac{I_{J2}}{I_{S2}}  =  \frac{I_{J1}}{I_{S1}}  \left(\frac{A_{e2}}{A_{e1}}\right) \left[ \frac{\left( T_{G2} + T_{i2}\right)}{\left(T_{G1} + T_{i1}\right)}\right] = \left(\frac{I_{J1}}{I_{S1}}\right) \left(\frac{S_{S1}}{S_{S2}}\right). 
\end{align}

With this, Equation \eqref{eq:Vnew_step} becomes
\begin{align}
    V_{sim} =& I_{S2}\left[v_{S2} + \alpha v_{J} \left(\frac{I_{J1}}{I_{S1}}\right) \left(\frac{S_{S1}}{S_{S2}}\right)   \right], \\
     V_{sim} =& V_{S2} + \alpha V_{J1} \left(\frac{I_{S2}}{I_{S1}}\right) \left(\frac{S_{S1}}{S_{S2}}\right).  \label{eq:Vnew}
\end{align}
Equation \eqref{eq:Vnew} is used as Equation \eqref{eq:transfer_V} in the main text.

The Stokes-Q, Stokes-U, and Stokes-V polarized signals are processed identically. Therefore, the $Q_{sim}$ and $U_{sim}$ polarization signals are very similar to Equation \eqref{eq:Vnew}
\begin{align}
    Q_{sim} =& Q_{S2} + \alpha Q_{J1} \left(\frac{I_{S2}}{I_{S1}}\right) \left(\frac{S_{S1}}{S_{S2}}\right), \label{eq:Qnew}\\
    U_{sim} =& U_{S2} + \alpha U_{J1} \left(\frac{I_{S2}}{I_{S1}}\right) \left(\frac{S_{S1}}{S_{S2}}\right), \label{eq:Unew}
\end{align}
where $Q_{J1}$ and $U_{J1}$ are the Stokes parameters in Obs$\#$1 of Jupiter and $Q_{S2}$ and $U_{S2}$ are Stokes parameters in the sky background of Obs~$\#$2.

\section{Elliptical correction}
\label{app:elliptical-correction}
Typically, the distribution of points is not 'circular' in the Q2 scatter plot 
(normalized high-pass filtered intensities, Fig. \ref{fig:Scatter_demo}a). 
This is an indication that left-over RFI and/or ionospheric fluctuations affect both the ON- and the OFF-beam simultaneously, leading to points close to the main diagonal. We are interested in signal only identified in the ON-beam (i.e. close to the x-axis), and, to be able to quantify the background of spurious events, the signal identified only in the OFF-beam (i.e. close to the y-axis).
In order to be able to detect points close to the x- and y-axis more easily, we circularize the ellipse in the following way:
\begin{itemize}
    \item We determine the eigenvalues and eigenvectors of the distribution of points.
    \item The eigenvector gives the angle $\alpha$ of the principal axis of the distribution. It is very close to 45$^\circ$, showing that there is indeed correlated signal in both beams.
    \item The eigenvalues give the width of the point distribution along the principal axis and in the direction perpendicular to it.
    \item We fit an ellipse (tilted at the angle $\alpha$) to the distribution of points. For each point on the ellipse, we calculate the distance of the point from the origin, $r_{ellipse}(\varphi)$. We normalize $r_{ellipse}(\varphi)$ by $r_{ellipse}(0)$.
    \item Going back to the initial point distribution, we convert the position of all points to polar coordinates ($r_{data}$, $\varphi$), and scale $r_{data}(\varphi)$ by $r_{ellipse}(\varphi)$.
    \item Finally, we go back to cartesian coordinates $x$ and $y$. We determine the standard deviation of the distribution of $x$ values, and normalize all $x$ values by this number. We proceed the same for $y$. Thus, final values of $x$ and $y$ are again in units of standard deviations.
\end{itemize}
Figs. \ref{fig:Scatter_demo}a and \ref{fig:Scatter_demo}b show a (simulated) point distribution before and after this elliptical correction. The red, filled squares show that data-points on the x- and y-axis are only weakly affected by this procedure; data-points close to the main diagonal are most strongly affected.

The black points in Figs. \ref{fig:Scatter_demo}a and \ref{fig:Scatter_demo}b represent what we expect from an observation, namely sky noise plus a few signal datapoints (visible at ON$\sim$2.0 and OFF$\sim$0.0 in panel a). Before elliptical correction, data-points on the x- and y-axis are difficult to pick out by automatic procedures. In particular, there are no points in the hatched orange and blue. This would not be labelled as a detection. After elliptical correction, outlying data-points on the x- and y-axis are much easier to locate. In particular, the injected signal on the x-axis in clearly in the orange hatched region, and would be picked up as a detection. In this way, the elliptical correction renders the pipeline more sensitive towards the expected signal.
\end{appendix}

\end{document}